\documentclass[lettersize,journal]{article}
\usepackage{amsmath,amsfonts}
\usepackage{algorithmic}
\usepackage{array}
\usepackage[caption=false,font=normalsize,labelfont=sf,textfont=sf]{subfig}
\usepackage{textcomp}
\usepackage{stfloats}
\usepackage{url}
\usepackage{verbatim}
\usepackage{graphicx}
\usepackage{geometry}

\usepackage{appendix}
\usepackage[switch]{lineno}  
\usepackage[export]{adjustbox}
\usepackage{algorithm}
\usepackage{cite}
\usepackage{amsthm,amsmath,amssymb}
\usepackage{mathrsfs}
\usepackage{pdfpages}
\usepackage{subfig}
\usepackage{geometry}
\usepackage{indentfirst}
\usepackage{multirow}
\usepackage{diagbox}
\usepackage{makecell}
\usepackage[section]{placeins}
\usepackage{authblk}
\usepackage{booktabs}
\usepackage{alphalph} 
\usepackage{caption}

\usepackage{subfig}
\usepackage{caption}

\begin{document}
\title{Structural Heterogeneity of the Drosophila Brain Network}
\author[1,2]{Xiaoyu Zhang}
\author[1]{Pengcheng Yang}
\author[1]{Yifei Zhang}
\author[2]{Bowei Qin}
\author[2]{Qiang Luo}

\author[2]{Wei Lin\thanks{Corresponding author: wlin@fudan.edu.cn}}
\author[1]{Xin Lu\thanks{Corresponding author: xin.lu.lab@outlook.com}}
\affil[1]{College of Systems Engineering, National University of Defense Technology, Changsha 410073, China}
\affil[2]{Research Institute of Intelligent Complex Systems, Fudan University, Shanghai 200433, China}

\maketitle

\begin{abstract}
Decoding the heterogeneity of biological neural systems is key to understanding the nervous system's complex dynamical behaviors. This study analyzes the comprehensive Drosophila brain connectome, which is the most recent data set, containing over 130,000 neurons and 50 million synapses. We conducted meticulous analyses of both network and spatial structure. Our findings reveal significant heterogeneity in network properties and distinct spatial clustering across functional regions. Besides, our analysis revealed a modular organizational pattern within the neural network, wherein regions with similar functions exhibited higher connection densities, forming distinct community structures. Moreover, we observed spatial clustering within functional regions but was not statistically significant. Additionally, we identify pervasive bilateral symmetry in network topology and spatial organization. Simulations based on the Kuramoto model demonstrate that the functional asymmetry between cerebral hemispheres arises from disparities in the intrinsic frequencies of neurons rather than from structural asymmetry within the neural network itself. Finally, we develop a 3D connectome visualization tool for detailed mapping of neuronal morphology. These insights advance our understanding of neural network organization and complexity in biological systems. 

\end{abstract}

\section{Introduction}

Neural heterogeneity has emerged as a fundamental concept in modern neuroscience research, profoundly shaping the understanding of the nervous system and therapeutic strategies for neurological disorders\cite{koch1999complexity,perez2021neural,segal2023regional}. From an evolutionary perspective, this heterogeneity represents the product of natural selection, enabling neural systems to cope with complex and dynamic environmental challenges while providing the foundation for species adaptability\cite{jerison2001evolution}. At the molecular and cellular levels, neuronal heterogeneity manifests in multiple aspects, including morphological characteristics, electrophysiological properties, neurotransmitter types, and gene expression patterns. Such diversity not only ensures the precision and reliability of neural circuit functions but also provides the material basis for neural plasticity\cite{zeng2017neuronal,cembrowski2018continuous,huang2019diversity}. Recent technological advances in single-cell sequencing and optogenetics have enabled researchers to reveal the intrinsic nature of neural system heterogeneity more thoroughly, revolutionizing our understanding of the organizational principles of the brain\cite{marder2011multiple}.  

Theoretical neuroscience has provided substantial insights into the computational advantages of neural heterogeneity. Gjorgjieva et al. \cite{gjorgjieva2016computational} systematically investigated how the biophysical diversity of neurons and synapses influences circuit performance, demonstrating that temporal scale diversity significantly enhances network information processing capacity. Specifically, different combinations of time constants generate richer dynamical behaviors, providing a theoretical foundation for understanding the computational advantages of neural diversity. Complementing this work, Amit et al.\cite{amit1997model} demonstrated that heterogeneity in neuronal parameters prevents networks from falling into pathological synchronization, thus increasing network resilience against noise.  

These theoretical predictions have been extensively validated through experimental studies. Marder et al.\cite{marder2006variability} discovered that identical network functions can be achieved through multiple distinct combinations of ion channels, revealing that parametric heterogeneity provides the foundation for homeostatic plasticity. Building on this, Prinz et al.\cite{prinz2004similar} found that parameters controlling network activity may exhibit considerable inter-animal heterogeneity, with various combinations of synaptic strengths and intrinsic membrane properties supporting proper network performance. At the cellular level, Bean et al.\cite{bean2007action} revealed that individual neurons can express up to 20 different voltage-gated ion channels, enabling rich dynamical behaviors and explaining the mechanisms underlying diverse neuronal electrical activity. Nusser et al. \cite{nusser2009variability} combined electron microscopy with immunolabeling techniques to discover that ion channel subcellular localization displays precise spatial patterns, directly influencing synaptic integration and information processing.  

A particularly significant manifestation of neural heterogeneity is bilateral lateralization, which has profound implications for specialized biological functions across diverse species. This lateralization—the asymmetrical distribution of neural structures and functions between the left and right hemispheres—represents a specialized form of heterogeneity that enhances neural processing efficiency through functional specialization\cite{Rogers2024,corballis2017evolution,schmitz2019building}.Drosophila has emerged as a powerful model organism for studying neural lateralization due to its relatively simple yet functionally sophisticated nervous system. Pascual et al. \cite{pascual2004brain} documented in detail the Asymmetrical Body (AB) within the central complex, confirming that approximately 95\% of wild-type Drosophila display a consistent lateralization pattern where the right AB is significantly larger than the left AB. This high population consistency suggests functional significance rather than random variation. Expanding on this structural foundation, Turner-Evans et al. \cite{turner2016insect} characterized the microstructural features of the asymmetrical body, revealing specialized neuronal groups with differential connectivity patterns between hemispheres. Beyond the central complex, Linneweber et al. \cite{linneweber2020neurodevelopmental} identified structural asymmetries in the Drosophila visual system, particularly in the R7/R8 photoreceptor projection regions, demonstrating these asymmetries result from programmed developmental processes.  

Structure determines function\cite{konieczny2014structure,kohn2018connecting,10786305}. Despite the widespread theoretical acceptance of neural heterogeneity as a concept, systematic investigations of neural structural heterogeneity based on authentic biological connectomes remain severely limited. Particularly regarding connectivity pattern heterogeneity, existing research is largely confined to specific brain regions or simplified models\cite{demirtacs2019hierarchical,segal2023regional}, lacking comprehensive analyses of heterogeneity characteristics across complete neural networks. For example, while pronounced left-right asymmetries have been discovered in localized structures such as the Asymmetrical Body (AB) and visual system\cite{pascual2004brain}, analyses of heterogeneity properties at the whole-brain network level—especially regarding symmetry and asymmetry patterns in global inter-hemispheric connectivity and spatial heterogeneity—remain virtually unexplored. This research gap primarily stems from the extreme difficulty in obtaining complete connectome data from biological organisms, with comprehensive connectomes currently available for only a few species\cite{cook2019whole}. This limitation has significantly impeded the ability to systematically analyze structural heterogeneity features embedded within biological connectomes and understand how these heterogeneity characteristics contribute to the functional diversity of neural systems. 

\begin{figure*}[!tbp]  
	\centering  
	\includegraphics[width=\textwidth, ]{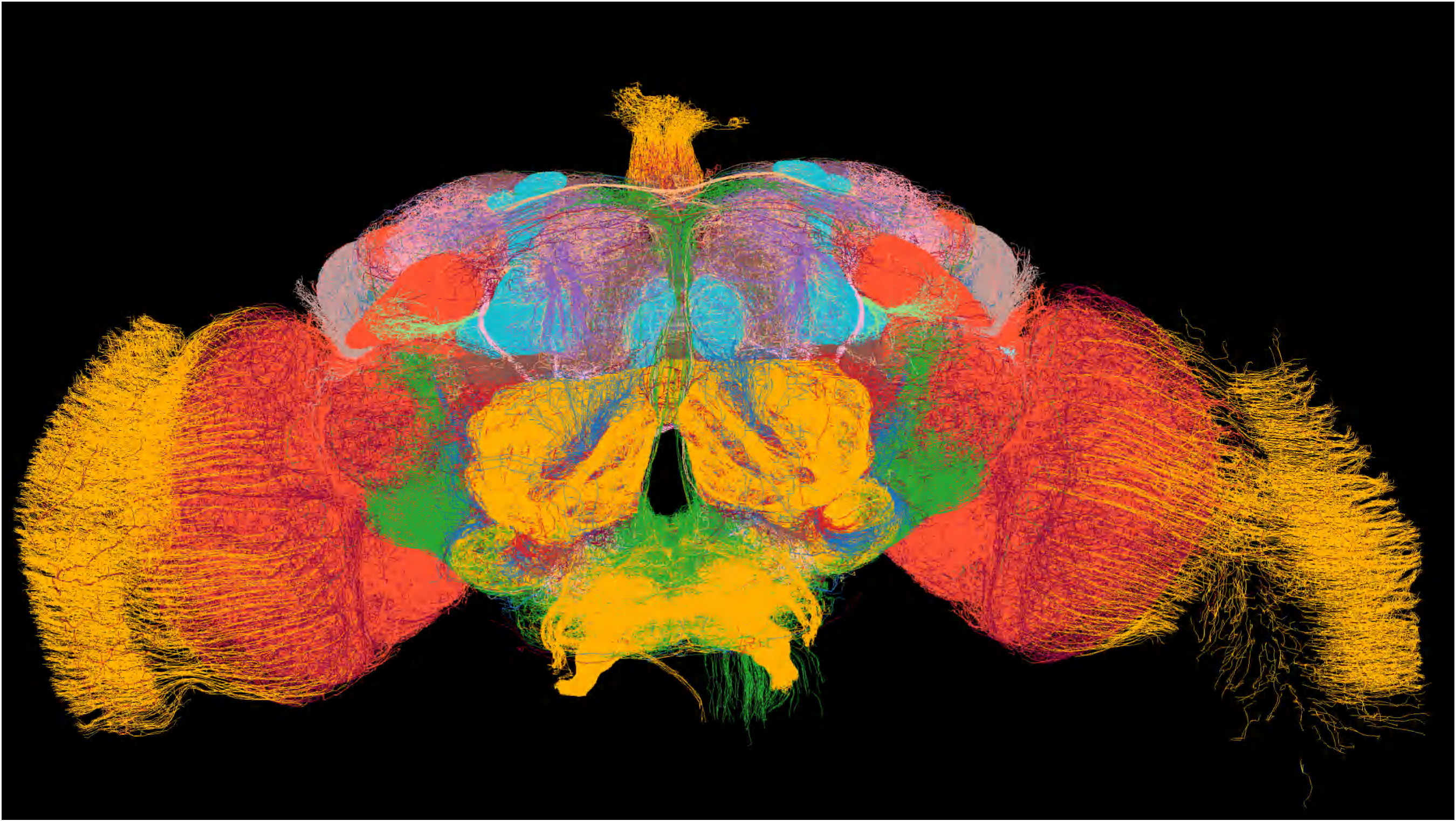}  
	\caption{Visualization of the Drosophila brain network.}  
	\label{fig0}  
\end{figure*}

With recent advances in connectomics technology, particularly the completion of the Drosophila connectome, researchers have the unprecedented opportunity to systematically investigate structural heterogeneity features and their potential functional significance in a complex biological neural system. Dorkenwald et al.\cite{dorkenwald2023neuronal} has mapped the adult Drosophila brain connectome with unprecedented detail, cataloging approximately 130,000 neurons, over 50 million synapses, and the positional morphology and functional classification of most neurons. This comprehensive data set provides an exceptional foundation for exploring the complexity of biological neural systems. Building on this resource, our study systematically investigates structural heterogeneities in the Drosophila brain from complementary perspectives. While previous research\cite{schlegel2023consensus,lin2023network} has primarily focused on statistical properties of the Drosophila brain network, we analyze heterogeneity through both network structural and spatial perspectives. Our approach examines global network heterogeneity while also characterizing region-specific patterns. Furthermore, as a crucial aspect of heterogeneity, we specifically investigate symmetry and asymmetry of the Drosophila brain network. To extend our analysis beyond structural features, we employ simulation models to study the functional asymmetry under both input-free and input-driven conditions. Additionally, to facilitate enhanced analysis and visualization of such complex networks, we have developed a specialized software platform for ultra-large-scale neural systems, with its efficacy demonstrated in Fig. \ref{fig0}.

\section{Results}

\subsection{Structure Heterogeneity}
The Drosophila brain consists of a vast network of neurons engaged in complex interactions, which collectively defines its structural complexity. In this section, we characterize the statistical properties of the Drosophila brain network at both global and regional scales, demonstrating the heterogeneous nature of network nodes and their associating connectivity patterns.

\subsubsection{Multi-Dimensional Heterogeneity in the Global Structure}

\begin{figure*}[!tbp]  
\centering  

\subfloat[{\tiny}]{  
	\begin{minipage}{4cm}  
		\centering  
		\includegraphics[width=\linewidth,height=3.11cm]{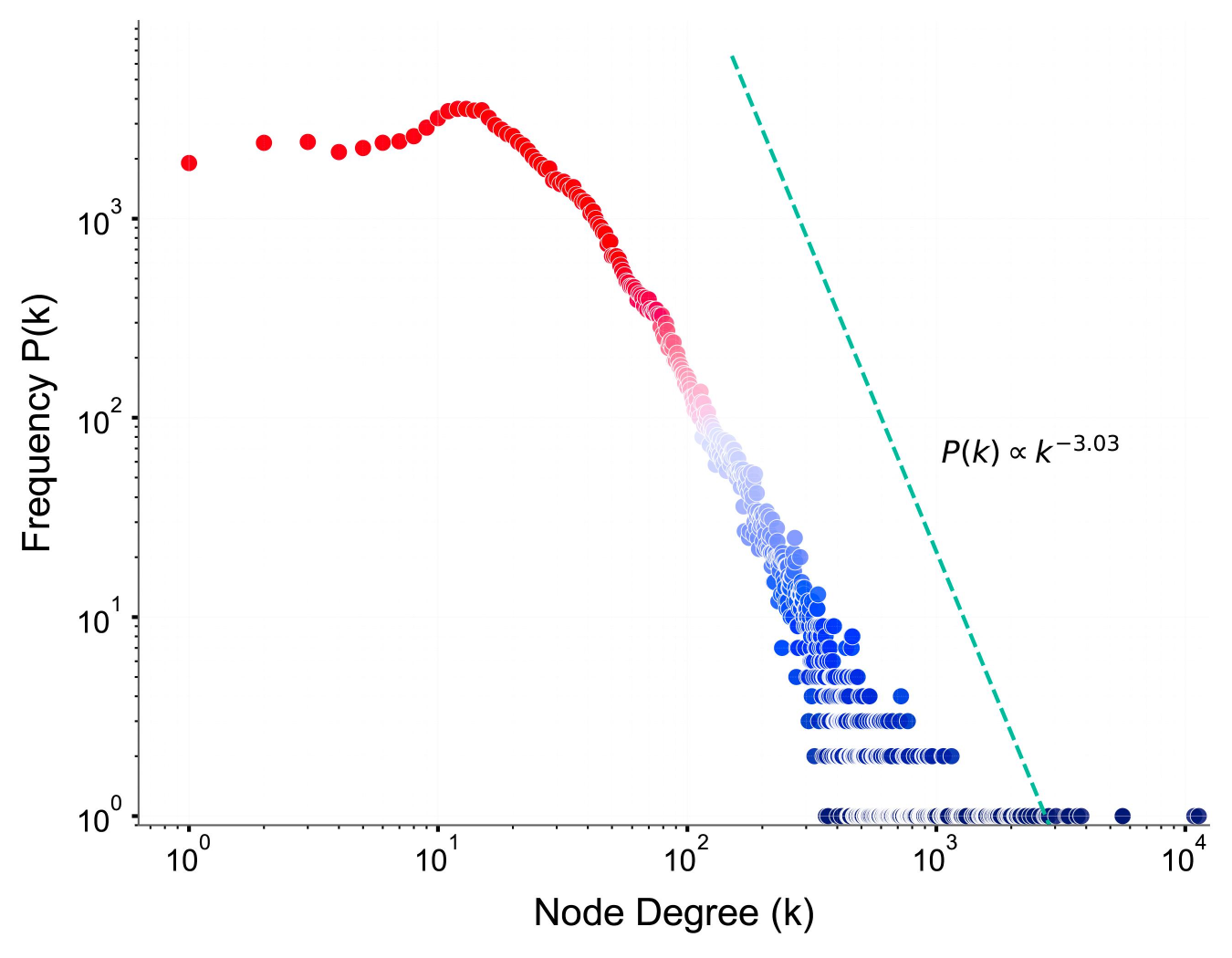}  
	\end{minipage}  
}  
\subfloat[\tiny]{  
	\begin{minipage}{4cm}  
		\centering  
		\includegraphics[width=\linewidth,height=3.11cm]{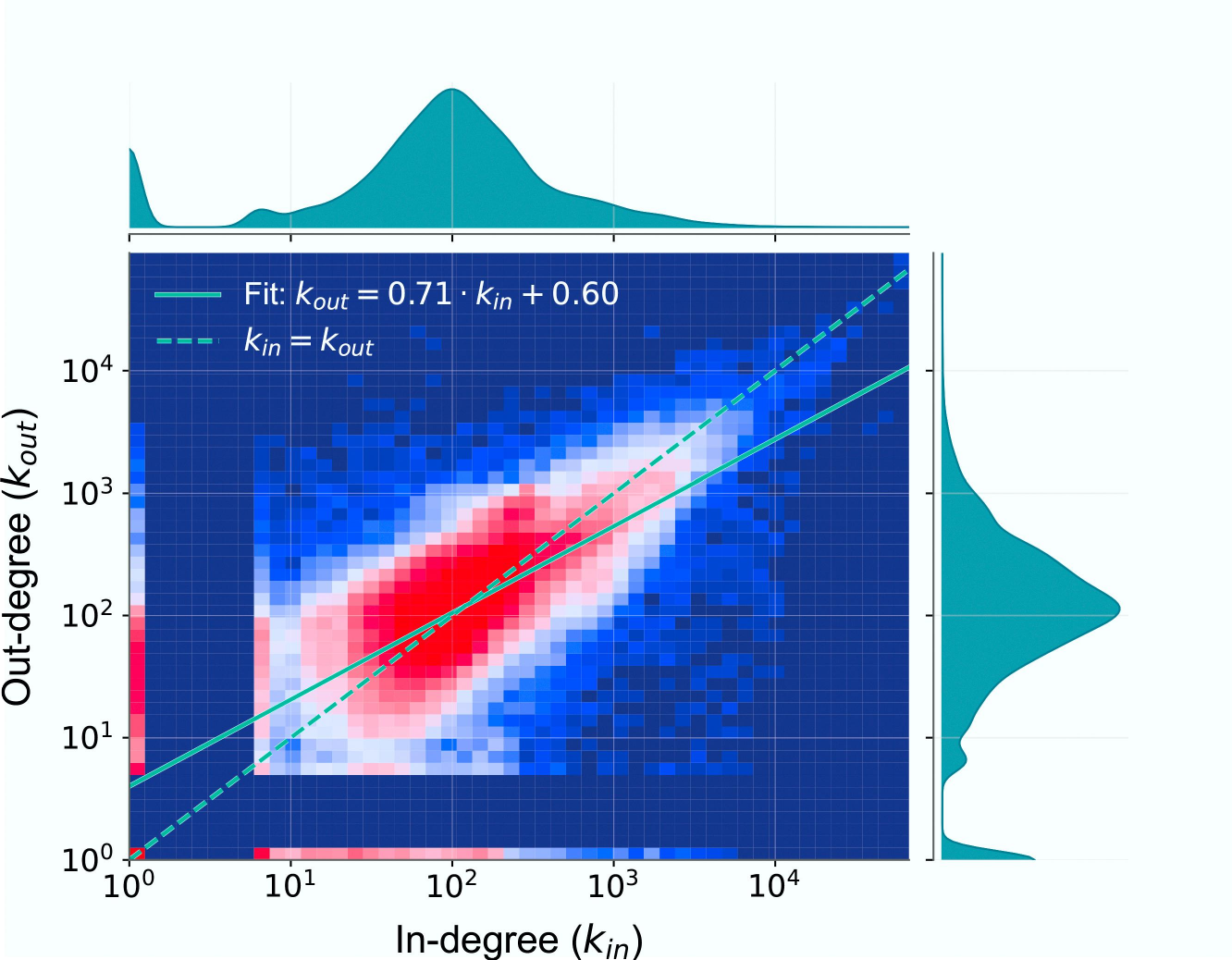}  
	\end{minipage}  
}  
\subfloat[\tiny]{  
	\begin{minipage}{4cm}  
		\centering  
		\includegraphics[width=\linewidth,height=3.11cm]{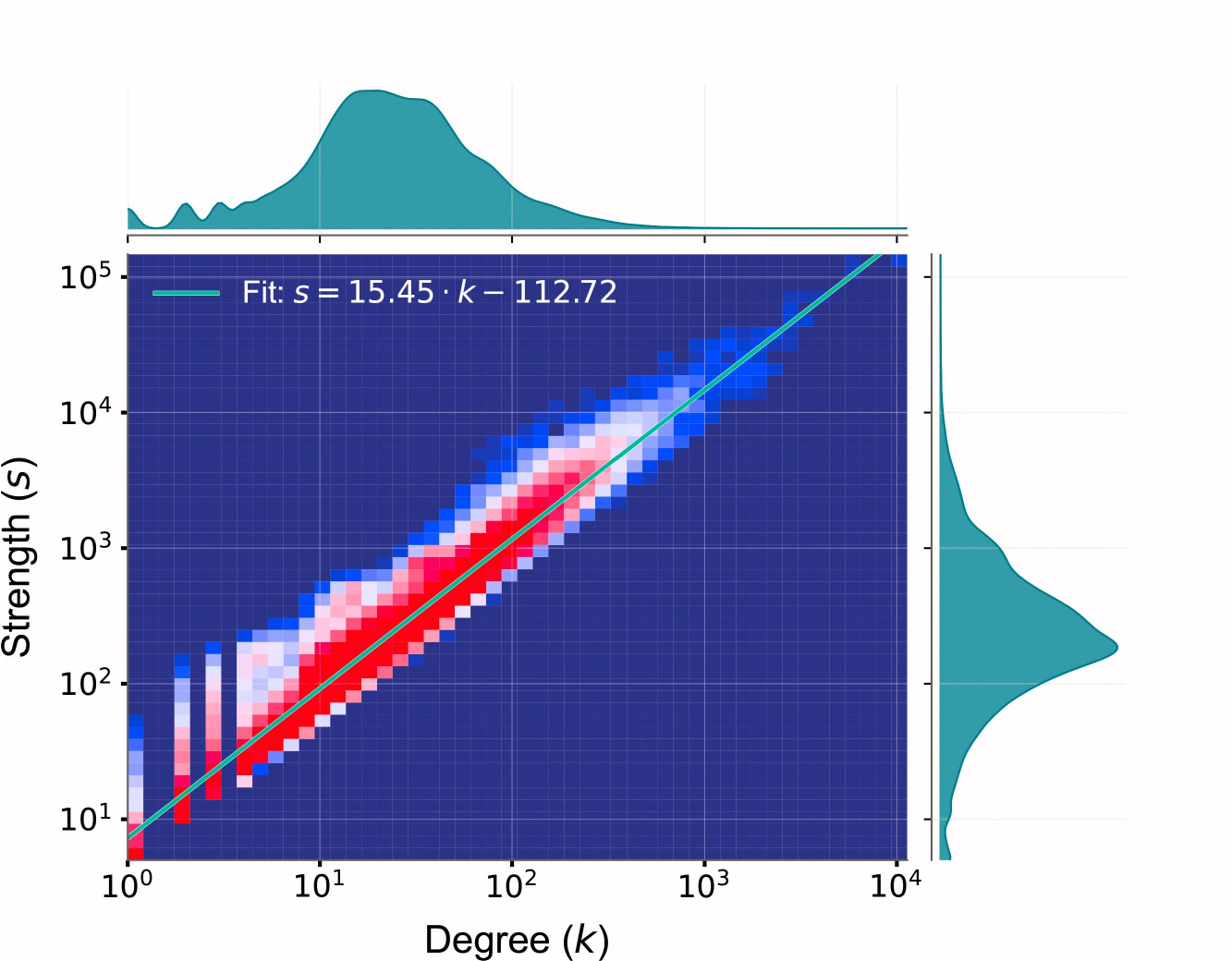}  
	\end{minipage}  
}  
\subfloat[]{  
	\begin{minipage}{4cm}  
		\centering  
		\includegraphics[width=\linewidth,height=3.11cm]{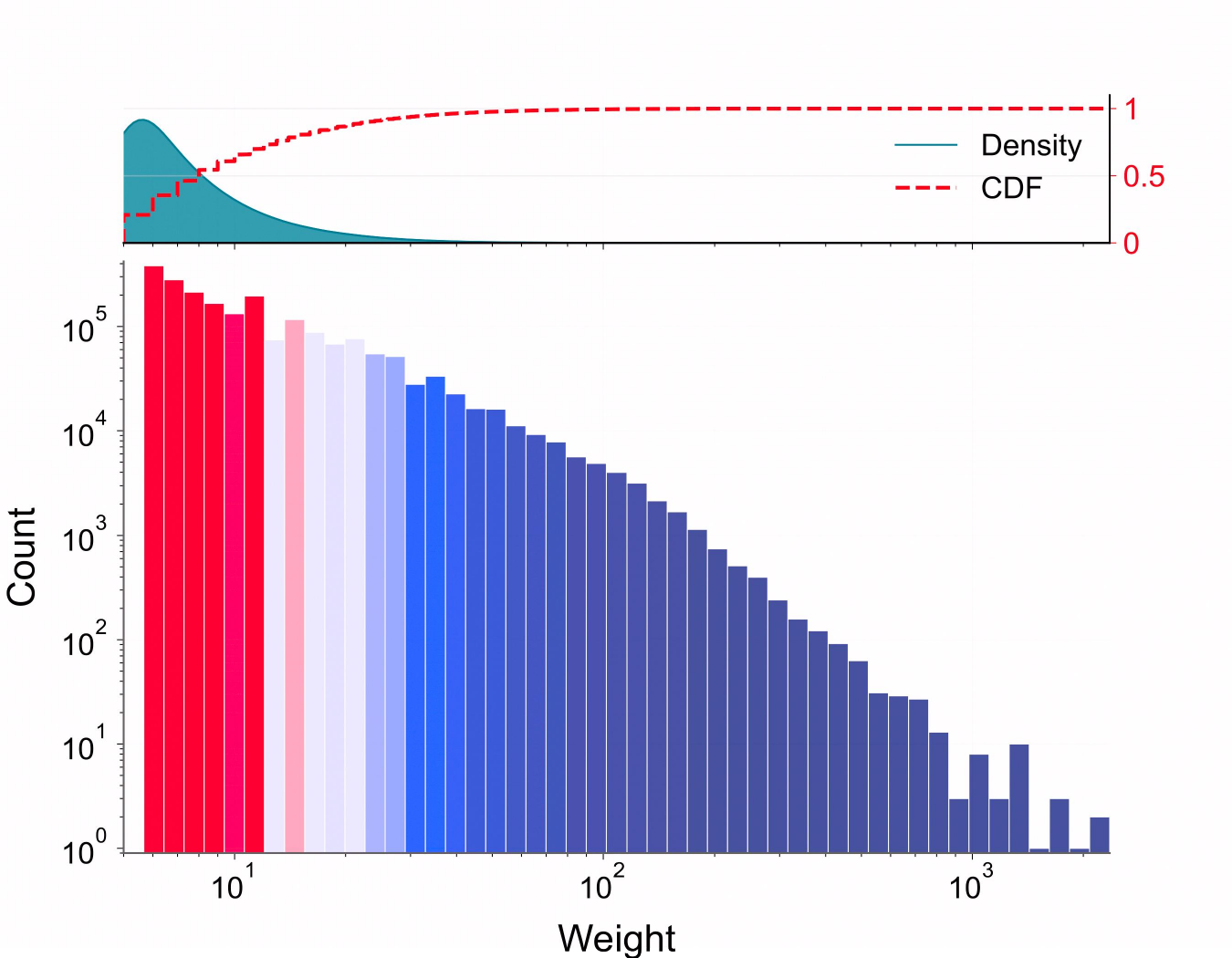}  
	\end{minipage}  
} 
 
\caption{Statistic Properties of the Drosophila brain network. (a) Degree distribution of Drosophila brain network. (b) Correlation between in-degree and out-degree. The red line denotes the reference values, whereas the green line represents the fitted curve. (c) Relationship between node's degree and its strength. The distribution of connection weight. }  
\label{fig1}  
\end{figure*}

The degree distribution of neurons in the Drosophila brain network follows a power-law pattern (Fig. \ref{fig1}(a)). The median and mean degrees are 20 and 37.01, respectively. Notably, only 7.35\% of neurons have more than 100 synaptic connections, and a mere 0.076\% exhibit degrees exceeding 1000. In general, higher in-degree—defined as the number of afferent synaptic inputs received from presynaptic partners—is positively correlated with higher out-degree, i.e., the number of efferent projections sent to postsynaptic targets (Pearson's r = 0.766; Fig. \ref{fig1}(b)). Moreover, node degree and node strength are strongly correlated (r = 0.95, p < 0.001), following a power-law relationship: as the degree increases, the node strength—representing the total number of synapses—also increases accordingly (Fig. \ref{fig1}(c)). Similarly, the distribution of connection weights follows a power-law, indicating that while a small subset of neuron pairs are connected by a large number of synapses, most exhibit relatively few connections (Fig. \ref{fig1}(d)). The mean connection weight is 12.6, suggesting that, on average, connected neuron pairs share approximately 12.6 synapses. Together, these results highlight the pronounced heterogeneity of the network at the global level.


\renewcommand\floatpagefraction{.99}
\renewcommand\topfraction{.99}
\renewcommand\bottomfraction{.99}
\renewcommand\textfraction{.1}
\setcounter{totalnumber}{50}
\setcounter{topnumber}{50}
\setcounter{bottomnumber}{50}

\subsubsection{Structural Heterogeneity and Bilateral Symmetry in Functional Regions}

The nervous systems of most organisms, including Drosophila, are divided into functional regions, each with specialized characteristics in structures and functions. Investigating the heterogeneity of these regions—along with their similarities and differences—offers valuable insights into the organizational principles and complexity of the nervous systems. In this section, we quantify the neurons' numbers, synaptic densities, and inter-region synaptic connectivity within dhistinct anatomical areas to investigate the heterogeneity among different functional regions.

\begin{figure*}
	\subfloat[]{  
		\begin{minipage}[t]{0.33\textwidth}  
			\centering  
			\includegraphics[width=\textwidth]{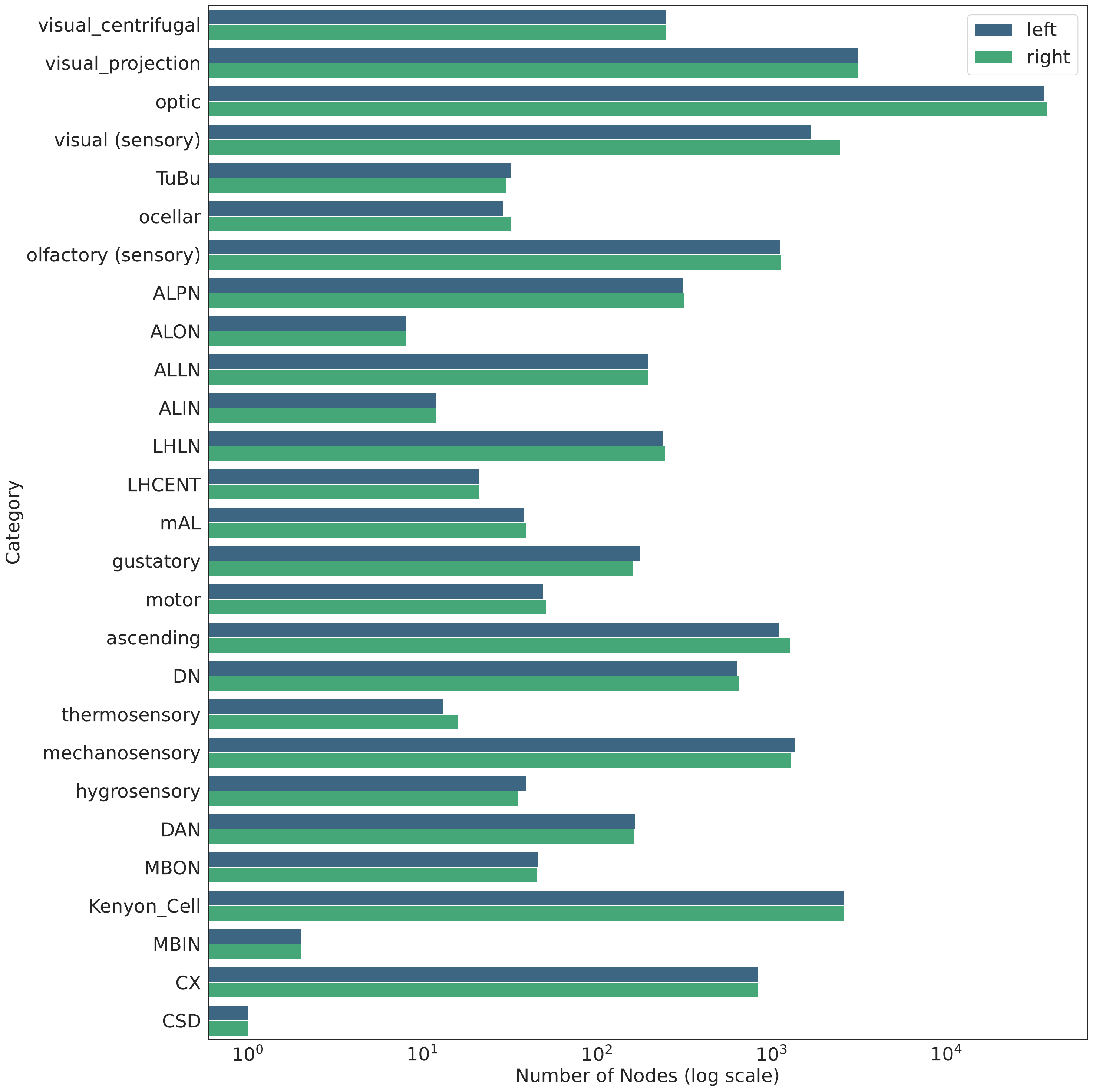}  \end{minipage} } 
	\subfloat[]{ 
		\begin{minipage}[t]{0.33\textwidth}  
			\centering  
			\includegraphics[width=\textwidth]{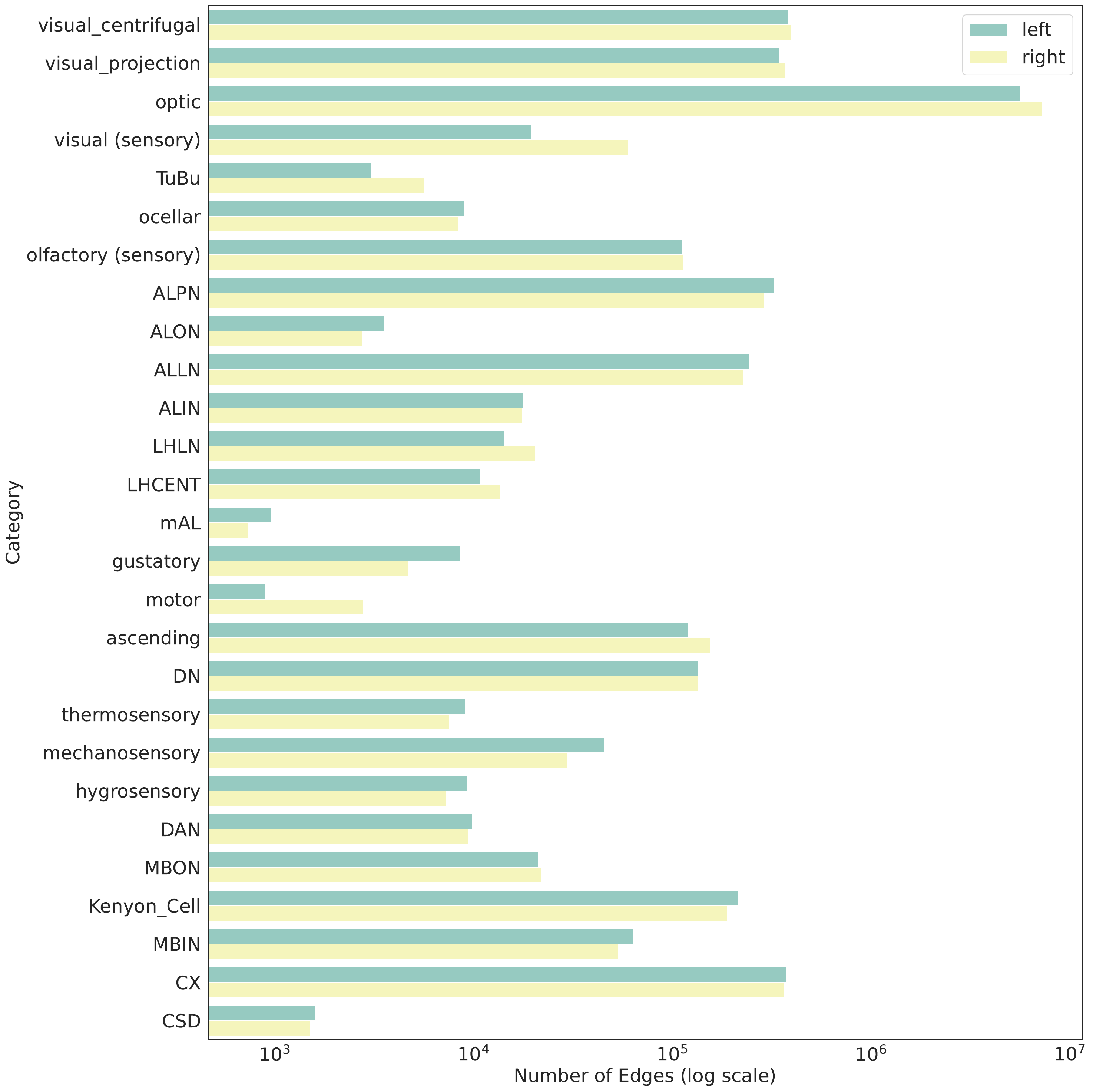}  \end{minipage}}  
	\subfloat[]{  
		\begin{minipage}[t]{0.3\textwidth}  
			\centering  
			\includegraphics[width=\textwidth]{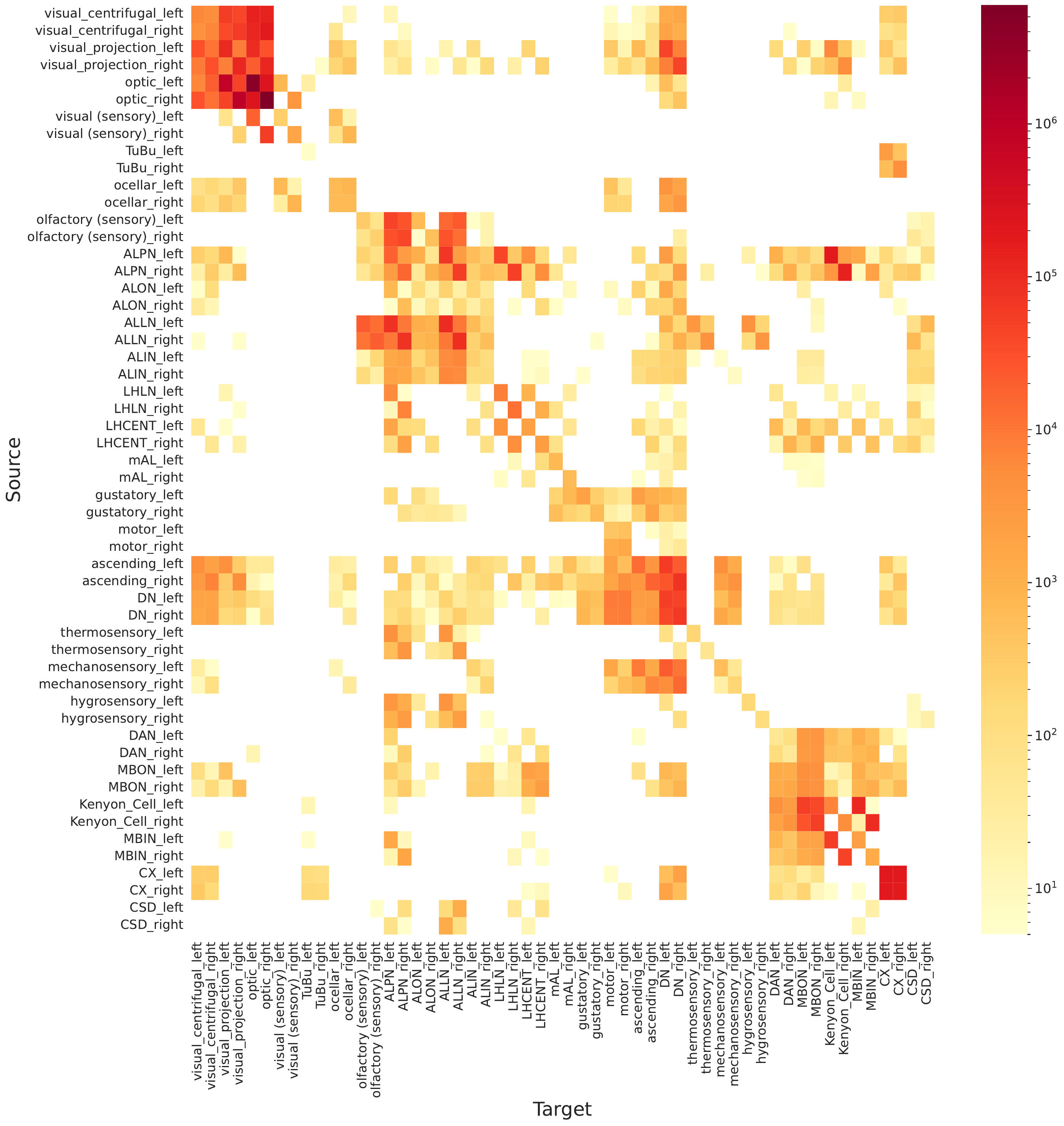}  \end{minipage}}  
	\caption{The neurons' numbers and synaptic densities within distinct anatomical regions. (a) The number of neurons across distinct regions of the Drosophila connectome. (b) The synaptic density within each region. (c) Heat map of the inter-regional synaptic connectivity patterns. } 
	\label{num}
\end{figure*}

The neuronal distribution across various regions, revealing that vision-related neurons constitute the largest proportion (Fig. \ref{num}(a)). We initially report the heterogeneity across various functional regions with respect to the number of neurons they contain. Specifically, neurons associated with optic and visual projection areas collectively represent 71.06\% of the total number of neurons. Furthermore, the neuronal counts exhibit a relatively symmetrical distribution between the left and right hemispheres across most regions. The most pronounced hemispheric heterogeneity is observed in the visual (sensory) region, where the neuronal population differs by approximately 37.50\% between the left and right hemispheres. 

In addition, we analyzed the quantity of connections within each region. Fig. \ref{num}(b) depicts the histogram of synaptic connections, defined as all neuronal interconnections within the boundaries of each analyzed region. Consistent with the neuronal distributions, vision-related regions exhibit the highest synaptic counts, accounting for 80.11\% of the total intra-regional synapses. Notably, the number of synapses maintain a high degree of bilateral symmetry across regions. The most pronounced heterogeneity is observed in the gustatory region, where the inter-hemispheric difference in synaptic density approximates 29.41\% of the total synapses within this region. Furthermore, we analyzed the number of connections between regions. It is evident that numerous regions exhibit a pronounced small-world phenomenon, exemplified by notable clustering effects among areas associated with visual (visual\_centrifugal, visual\_projection, optic, visual (sensory), TuBu, and ocellar), olfactory (olfactory (sensory), ALPN, ALON, ALLN, and ALIN), and learning \& memory functions (DAN, MBON, Kenyon Cell, and MBIN) (Fig. \ref{num}(c)). Concurrently, we calculated the ratio of intra-regional to inter-regional connections, finding that the magnitude of intra-regional connections is approximately 23.48 times greater than that of inter-regional connections. 
The results demonstrate pronounced heterogeneity in neuronal connectivity patterns both within and between functional regions. Specifically, intra-regional connections exhibit remarkably high density, whereas inter-regional connections are notably sparse, which is consistent with previous studies\cite{zhang2024simple,ZHANG2025}. Furthermore, regions with similar functions display a higher number of synaptic connections between them, indicating significant heterogeneity in inter-regional connection strength. Moreover, this study observed functional segregation between many functionally distinct areas. For instance, internal connections within visual processing regions constitute 98.96\% of the total connections in these areas. Concurrently, internal connections within olfactory processing regions represent 73.98\% of the total connections in those regions.

\begin{figure*}[b]
	\centering
	\subfloat[]{  
		\begin{minipage}{0.3\textwidth}  
			\centering  
			\includegraphics[width=\textwidth]{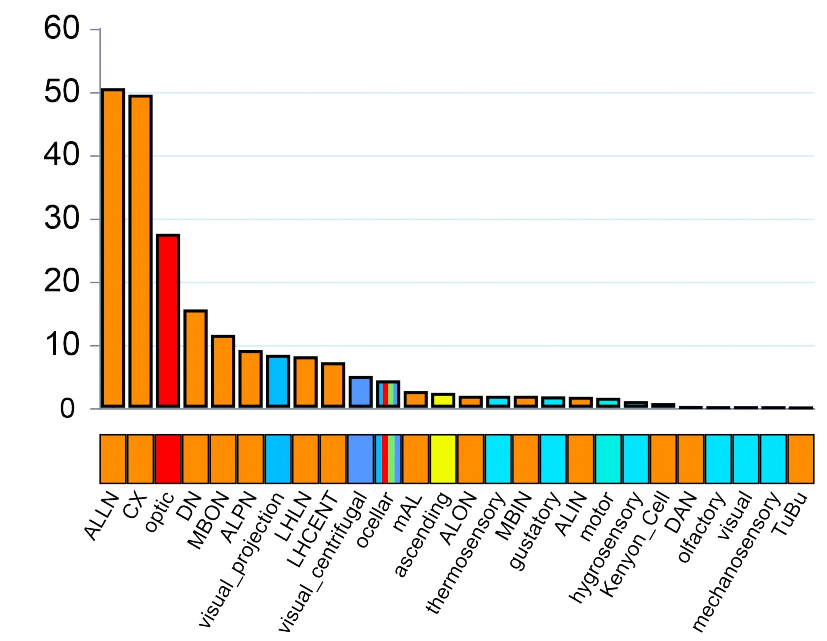}  \end{minipage}} 
	\subfloat[]{ 		
		\begin{minipage}{0.3\textwidth}  
			\centering  
			\includegraphics[width=\textwidth]{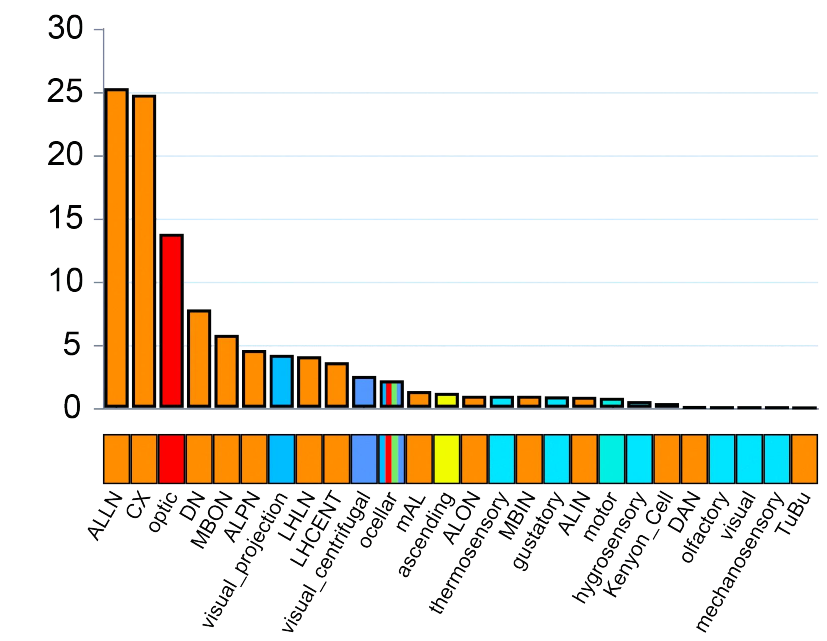}  \end{minipage}}  
	\subfloat[]{ 
		\begin{minipage}{0.3\textwidth}  
			\centering  
			\includegraphics[width=\textwidth]{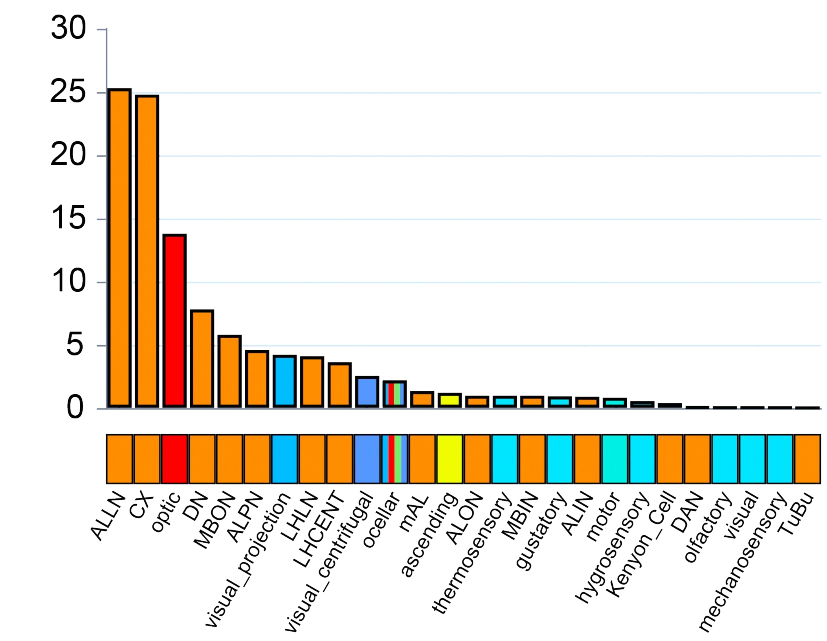}  \end{minipage}}
	
	\subfloat[]{ 
		\begin{minipage}{0.3\textwidth}  
			\centering  
			\includegraphics[width=\textwidth]{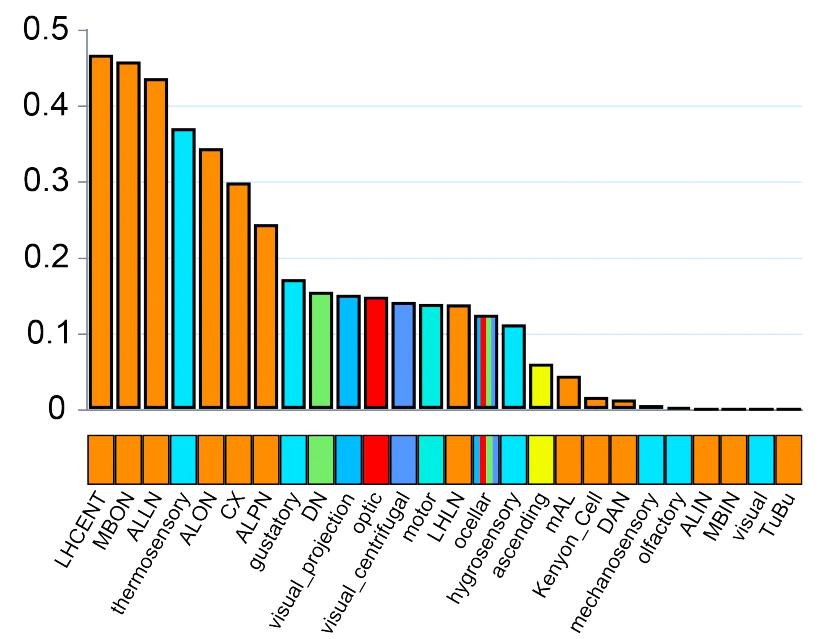}  \end{minipage}}
	\subfloat[]{ 
		\begin{minipage}{0.3\textwidth}  
			\centering  
			\includegraphics[width=\textwidth]{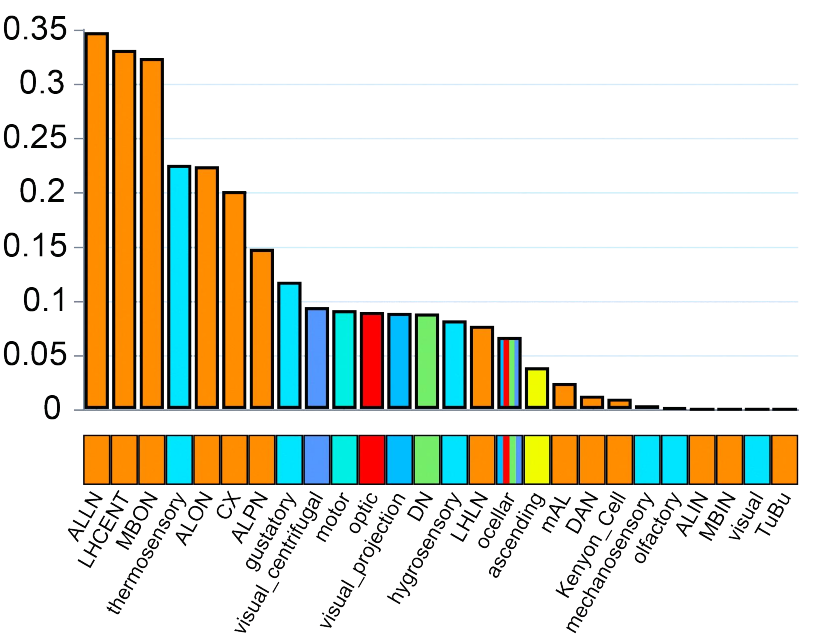}  \end{minipage}}
	\subfloat[]{   
		\begin{minipage}{0.3\textwidth}  
			\centering  
			\includegraphics[width=\textwidth]{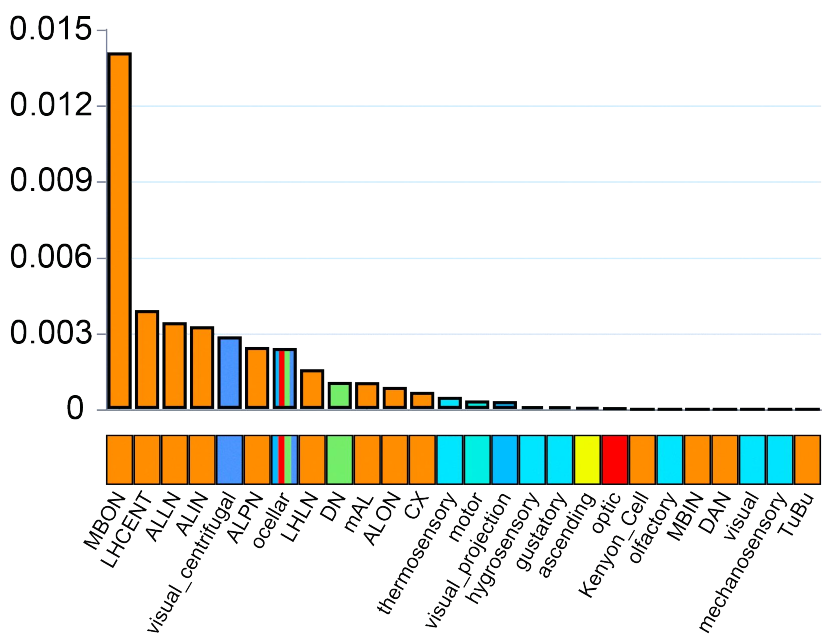}  \end{minipage}}
	
	\caption{Statistical distribution of networks composed of regions. (a) illustrates the average degree across various regions. (b) depicts the clustering coefficient for each region (undirected graph). (c) represents the average in-degree for the respective regions. (d) shows the average out-degree for each region. (e) demonstrates the average directed clustering coefficient across regions. (f) portrays the average betweenness centrality coefficient for the regions.}
	\label{cat}
\end{figure*}

To further investigate the heterogeneity of inter-regional statistical characteristics, we conceptualized regions as subgraphs and analyzed the statistical distributions of these subgraphs (Fig. \ref{cat}), including degree distribution, clustering coefficient and betweenness centrality. The regions with the highest average degree are ALLN and CX (with an average degree of approximately 50), while the remaining regions exhibit average degrees not exceeding 30 (Fig. \ref{cat}(a)). 
In addition, the internal in-degree and out-degree distributions within these regions are approximately equivalent (Fig. \ref{cat}(b) and (c)). Fig. \ref{cat}(d) and (e) illustrate the clustering coefficients for directed and undirected networks among regions, respectively. The results show minimal disparity between the clustering coefficients of directed and undirected networks, suggesting a predominance of bidirectional connections among neurons (i.e., neuron A can transmit signals to neuron B, and vice versa). The mean betweenness centrality coefficients for each region are generally low, consistent with the overall network distribution (Fig. \ref{cat}(f)). However, the MBON (Mushroom Body Output Neuron) region stands out with a significantly higher mean betweenness centrality. MBONs are the primary efferent neurons of the Mushroom Body, which is critical for learning, memory, and behavioral output in Drosophila. The elevated betweenness centrality of MBONs suggests that they serve as key nodes linking multiple functional areas, likely playing a crucial role in transmitting processed information to other regions. Their high centrality may also reflect their involvement in translating learning and memory outcomes into behavioral responses, underlining their importance in coordinating complex behavioral patterns. 


\begin{figure*}[t]  
	\centering  
	\subfloat[ALIN]{\includegraphics[width=0.19\textwidth]{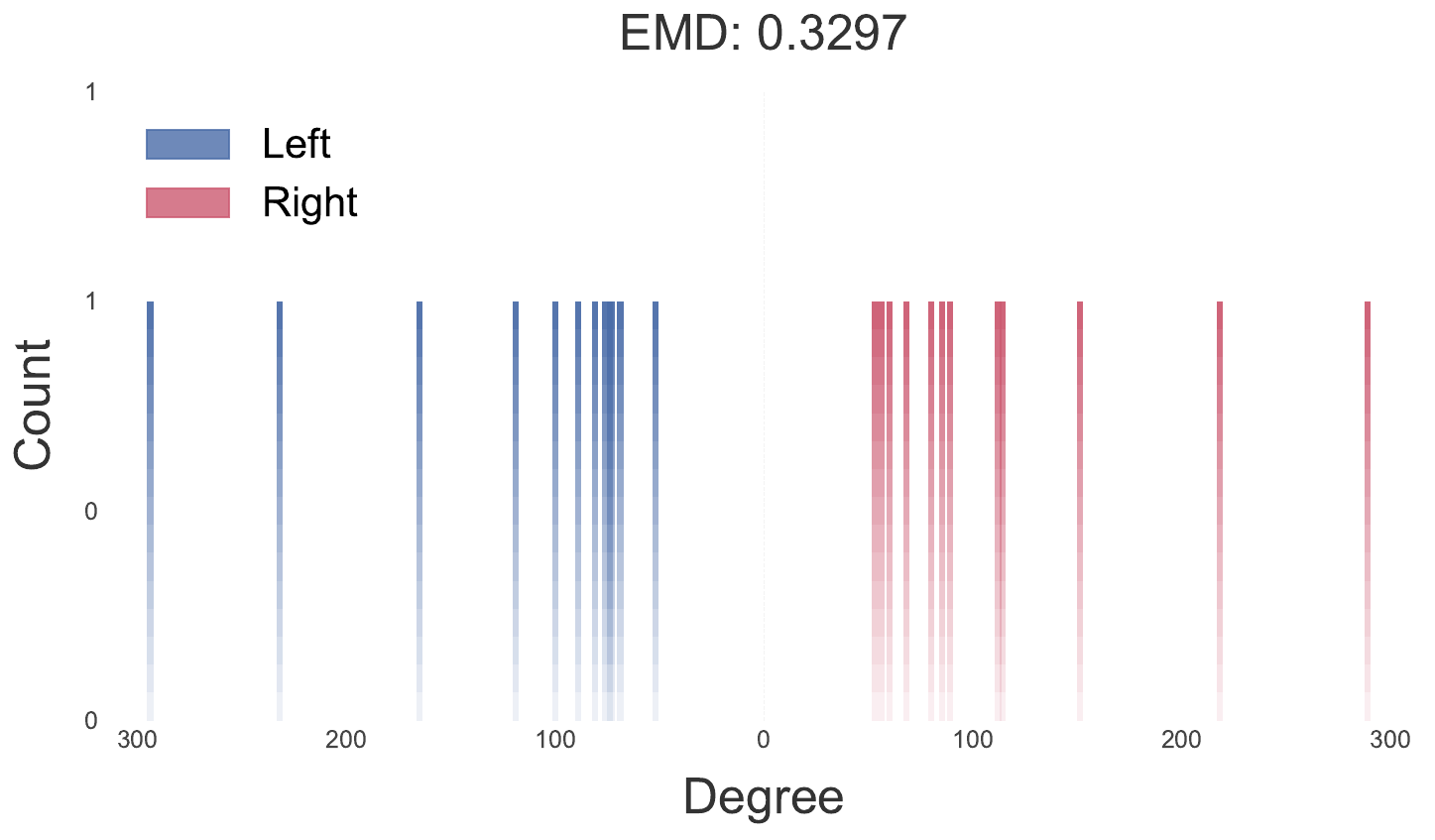}}  
	\hspace{0.1em}  
	\subfloat[ALON]{\includegraphics[width=0.19\textwidth]{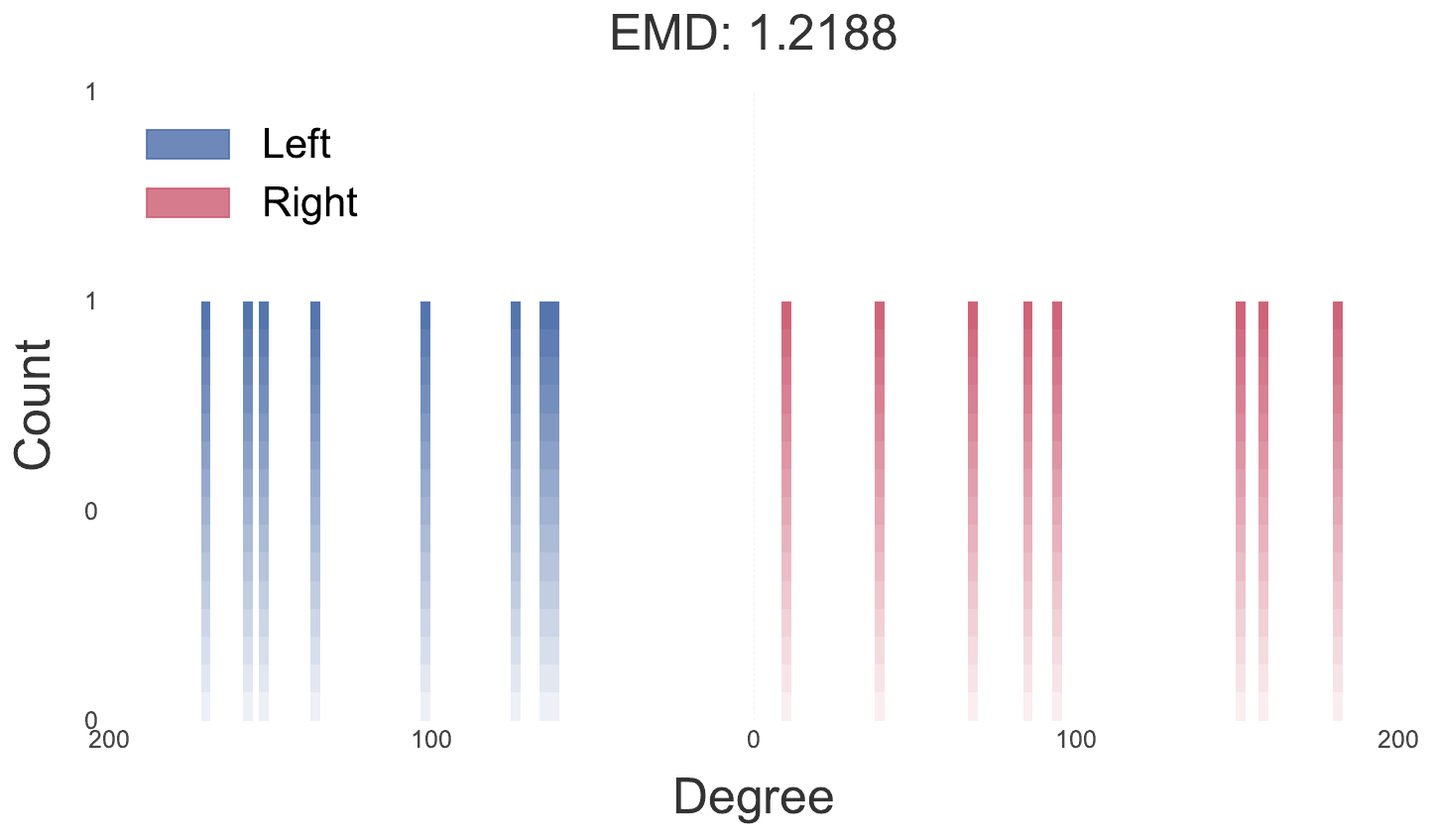}}  
	\hspace{0.1em}  
	\subfloat[ALLN]{\includegraphics[width=0.19\textwidth]{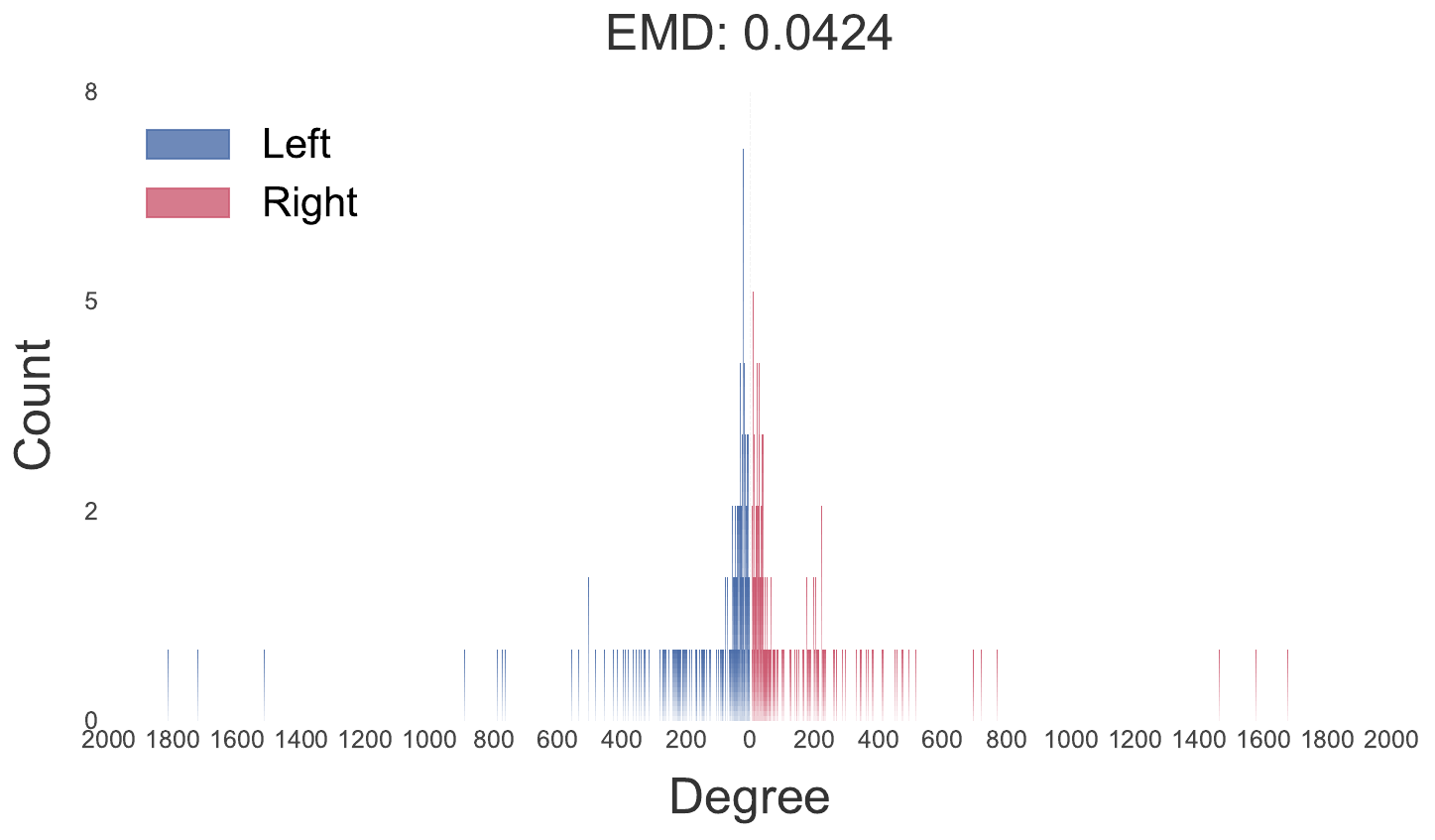}}  
	\hspace{0.1em}   
	\subfloat[ALPN]{\includegraphics[width=0.19\textwidth]{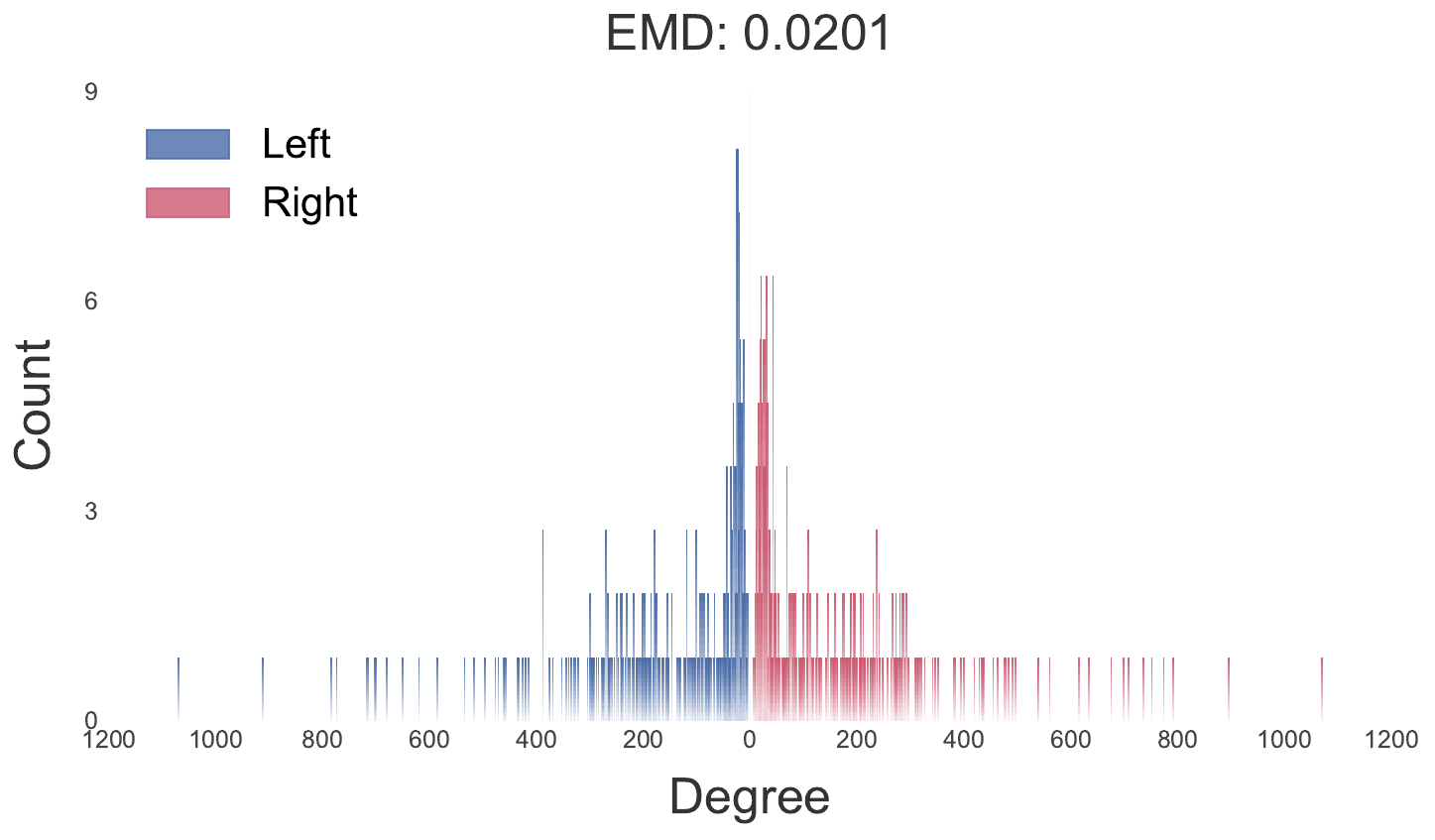}}  
	\hspace{0.1em}  
	\subfloat[ascending]{\includegraphics[width=0.19\textwidth]{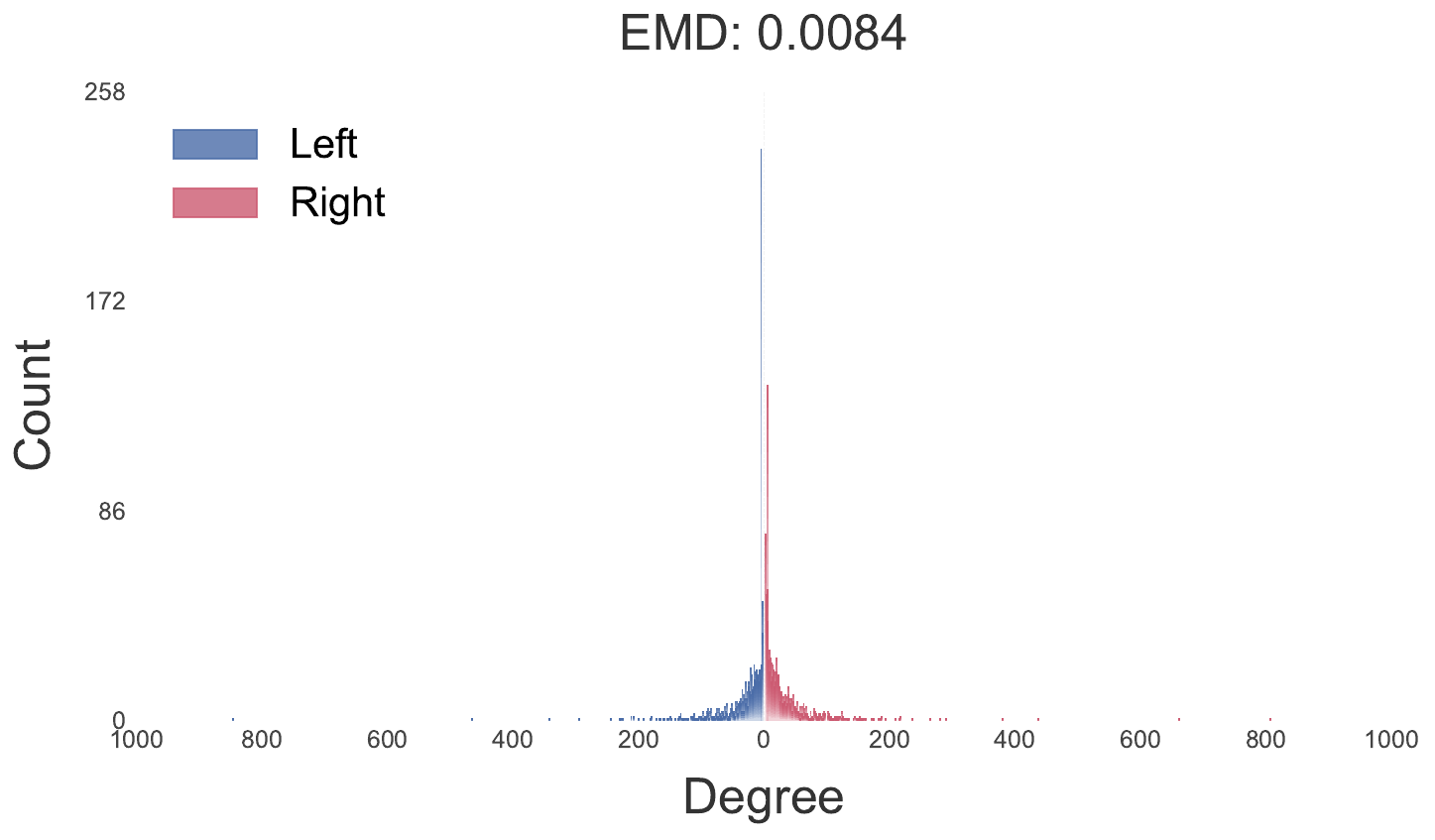}}  
	
	\vspace{-1em}  
	\subfloat[visual projection]{\includegraphics[width=0.19\textwidth]{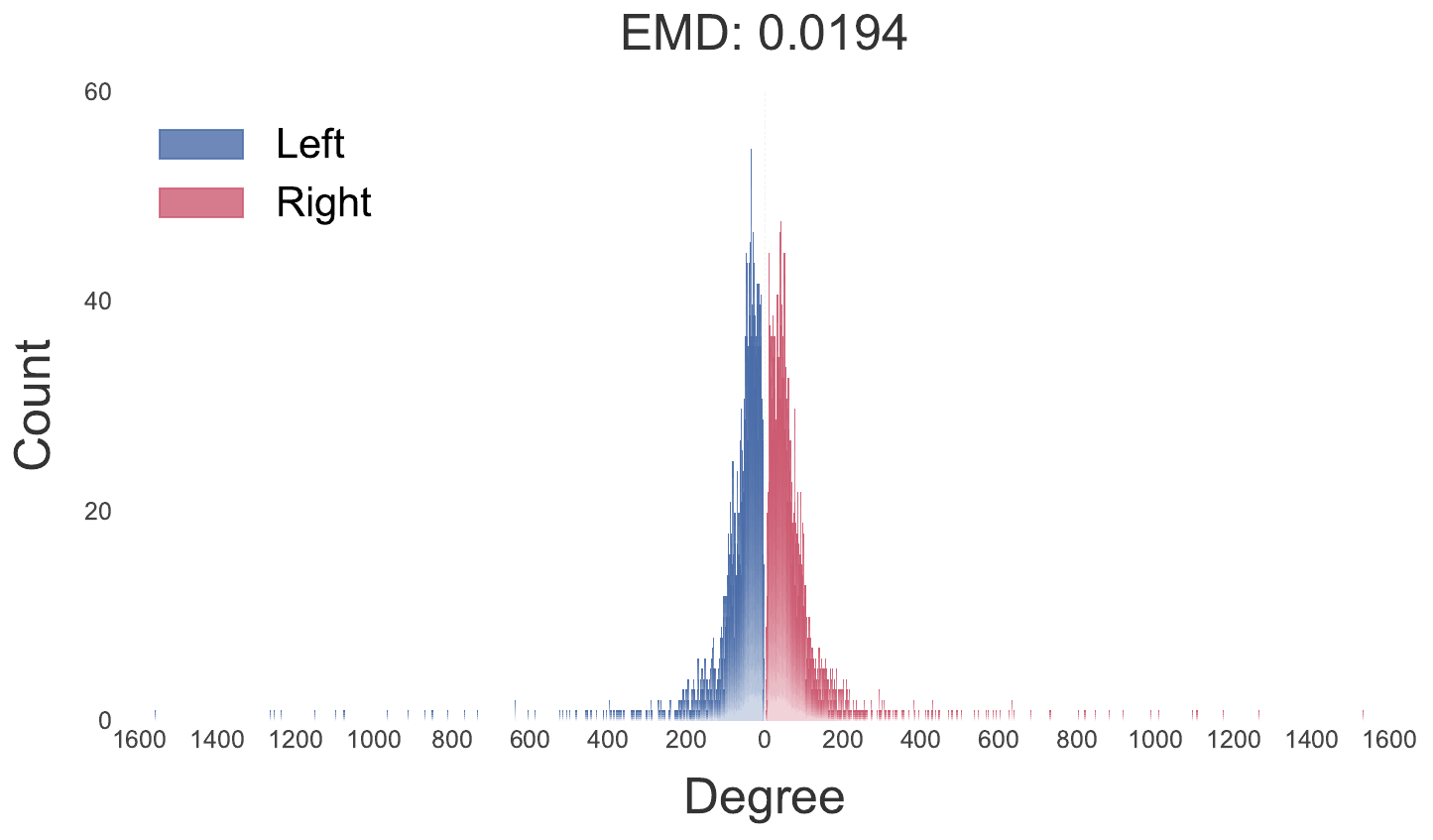}}  
	\hspace{0.1em}  
	\subfloat[CX]{\includegraphics[width=0.19\textwidth]{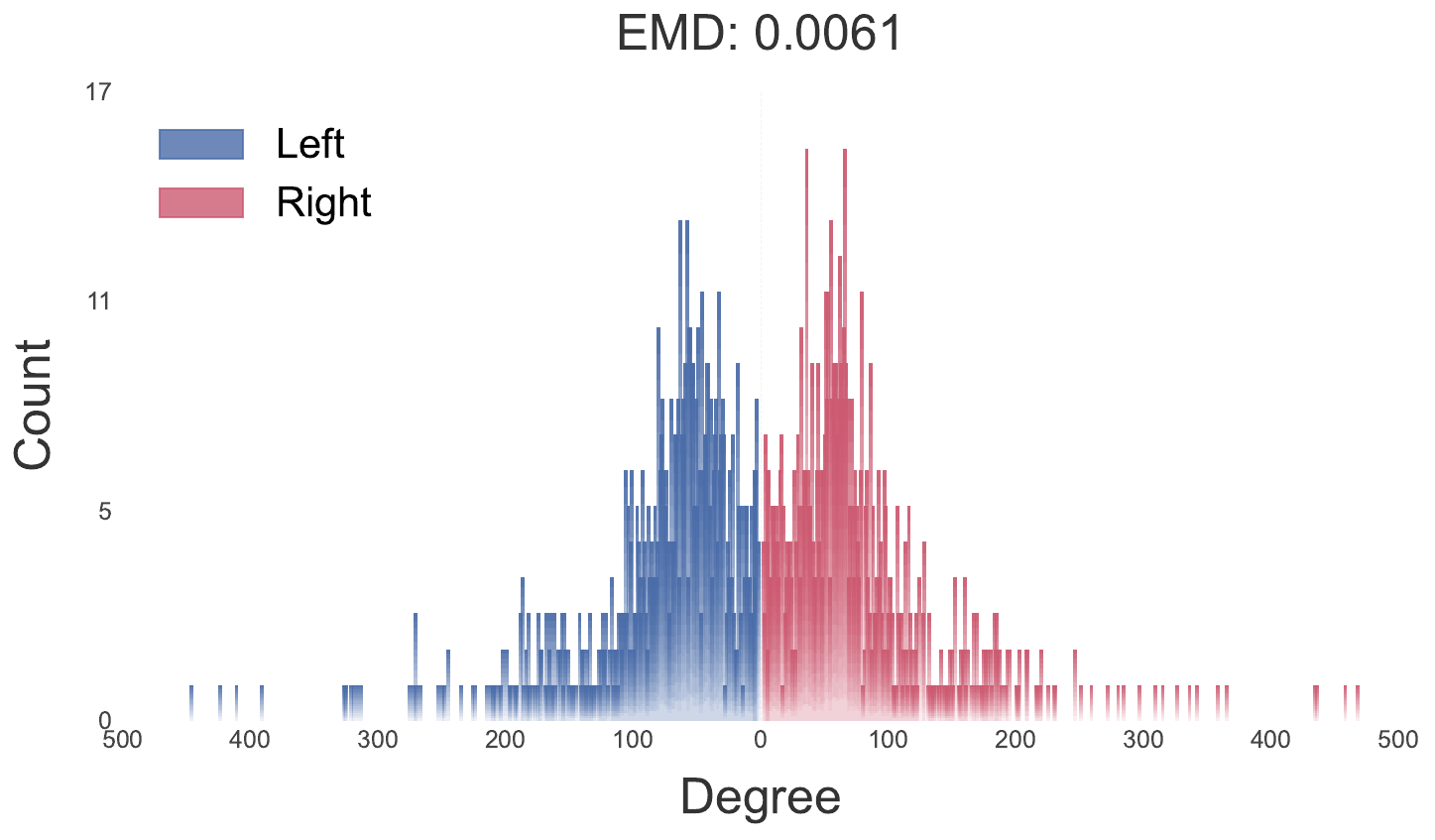}}  
	\hspace{0.1em}  
	\subfloat[DAN]{\includegraphics[width=0.19\textwidth]{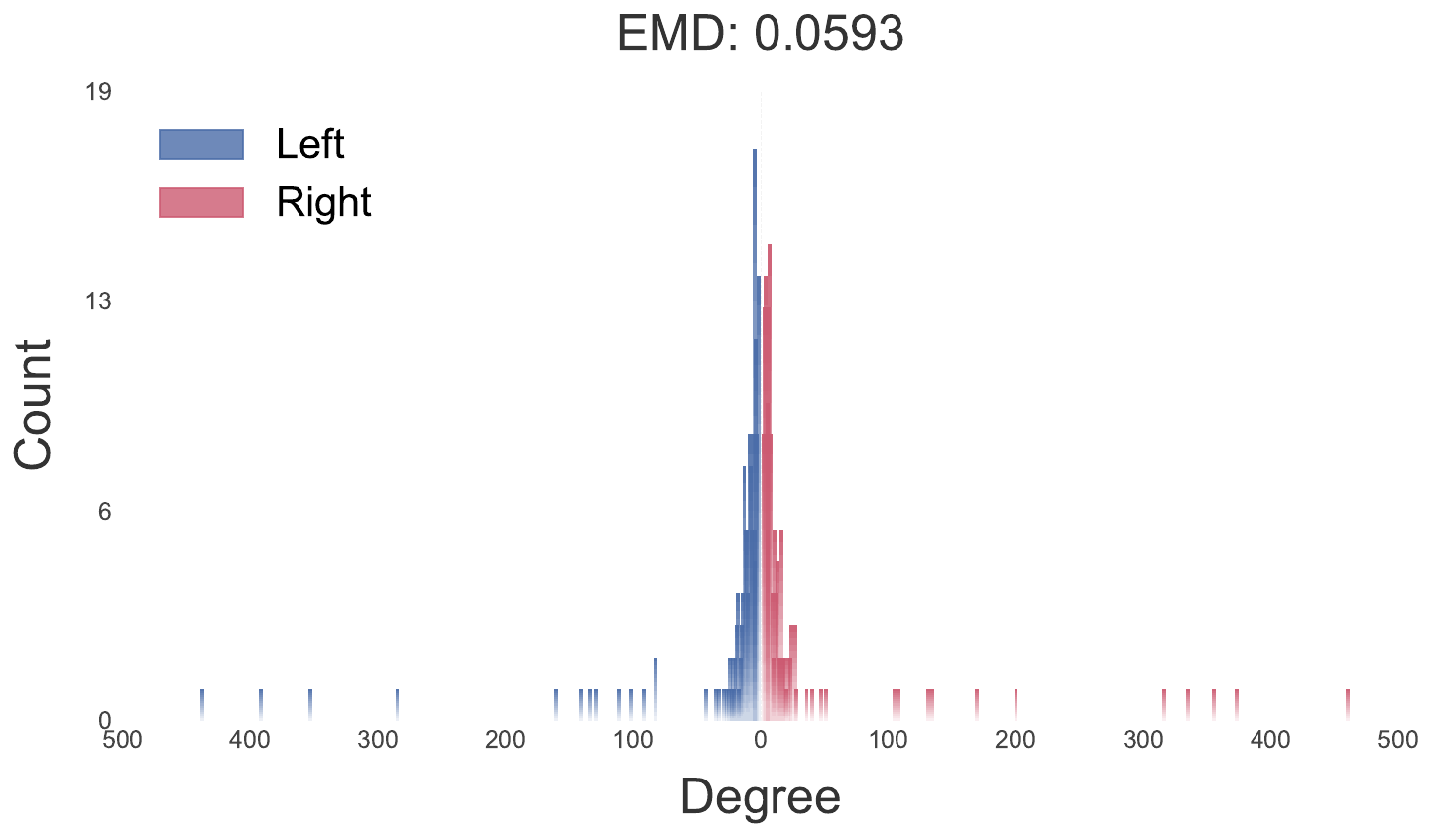}}  
	\hspace{0.1em}  
	\subfloat[DN]{\includegraphics[width=0.19\textwidth]{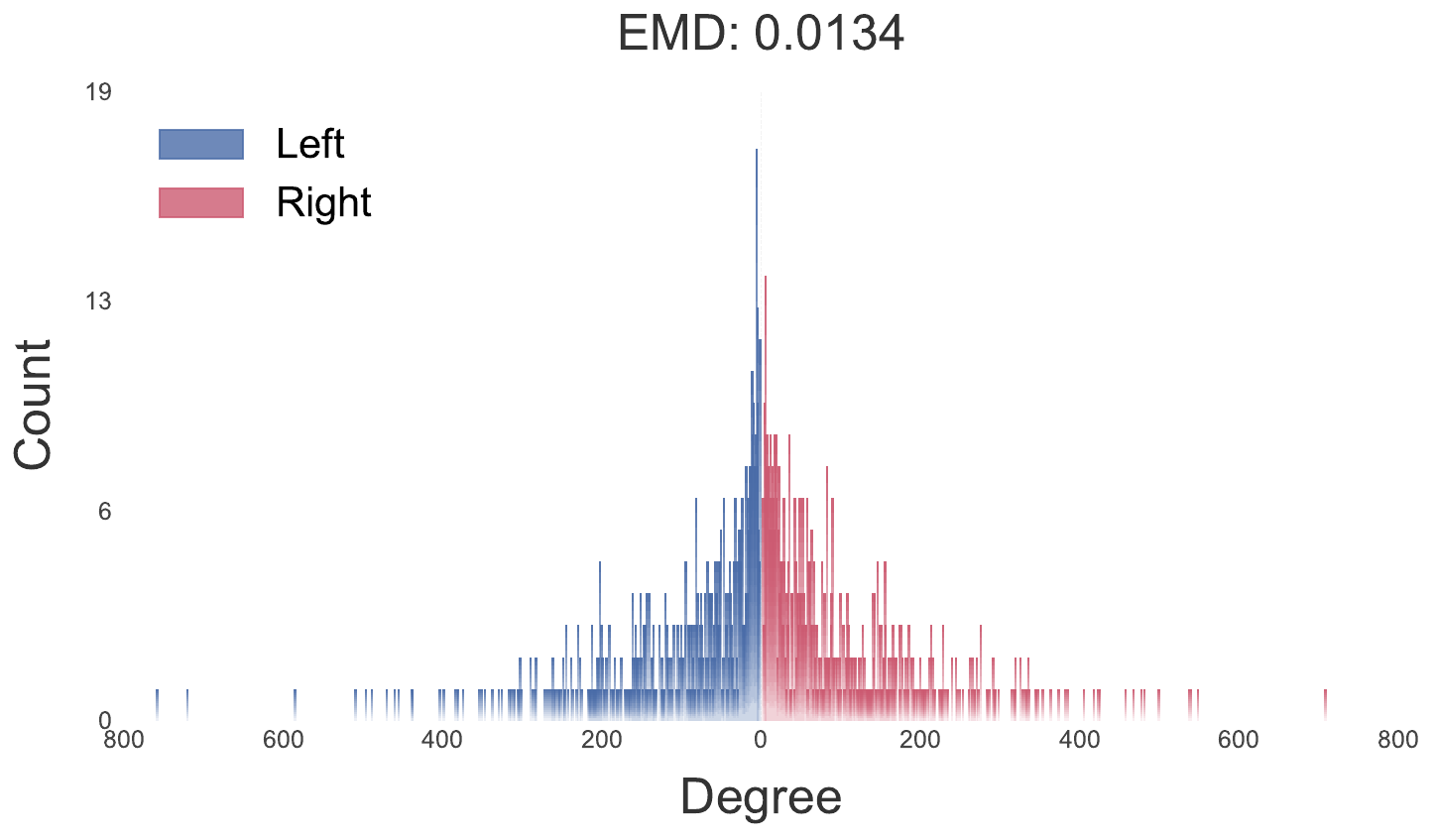}}  
	\hspace{0.1em}  
	\subfloat[gustatory]{\includegraphics[width=0.19\textwidth]{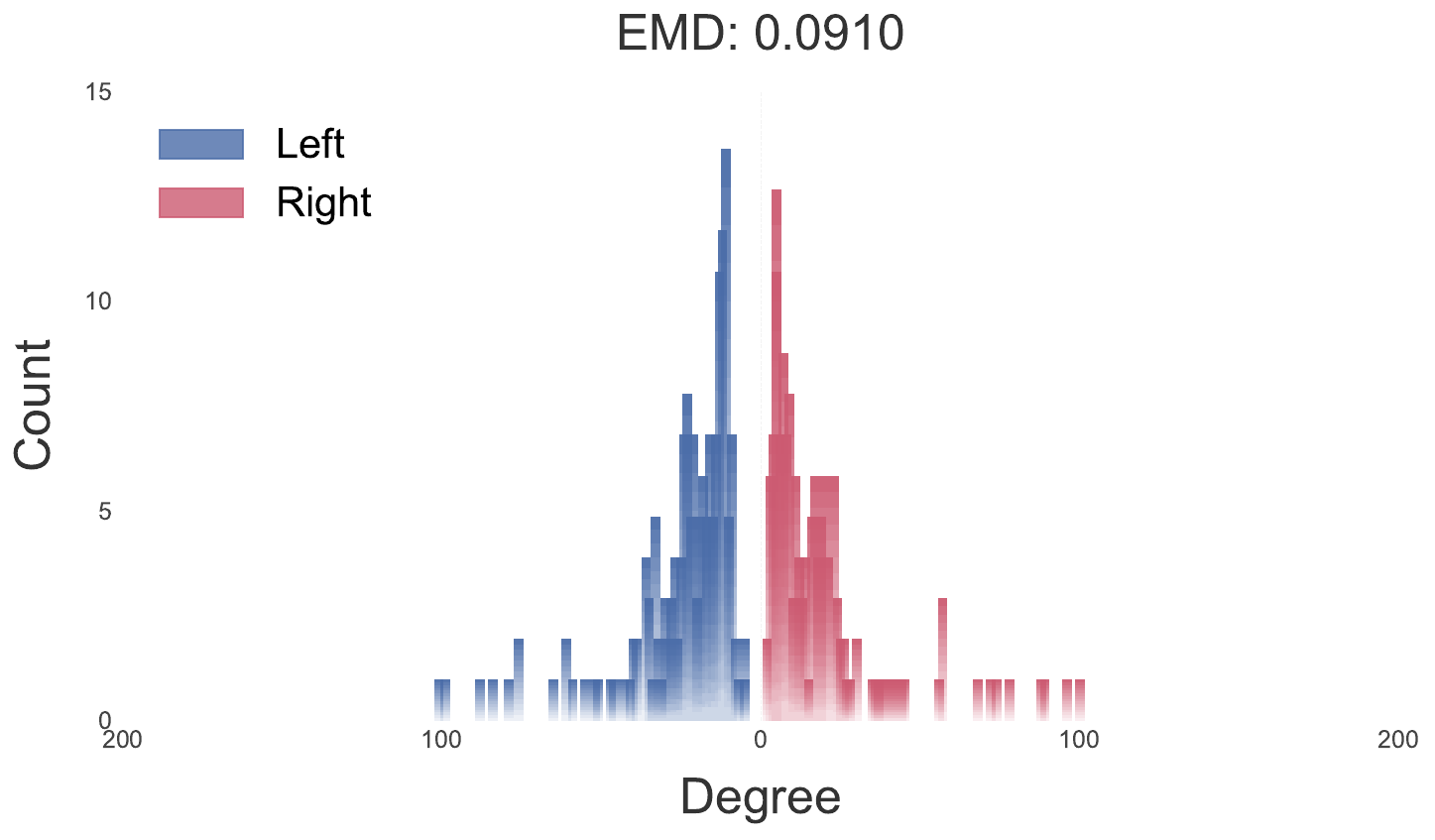}}  
	
	\vspace{-1em}  
	\subfloat[hygrosensory]{\includegraphics[width=0.19\textwidth]{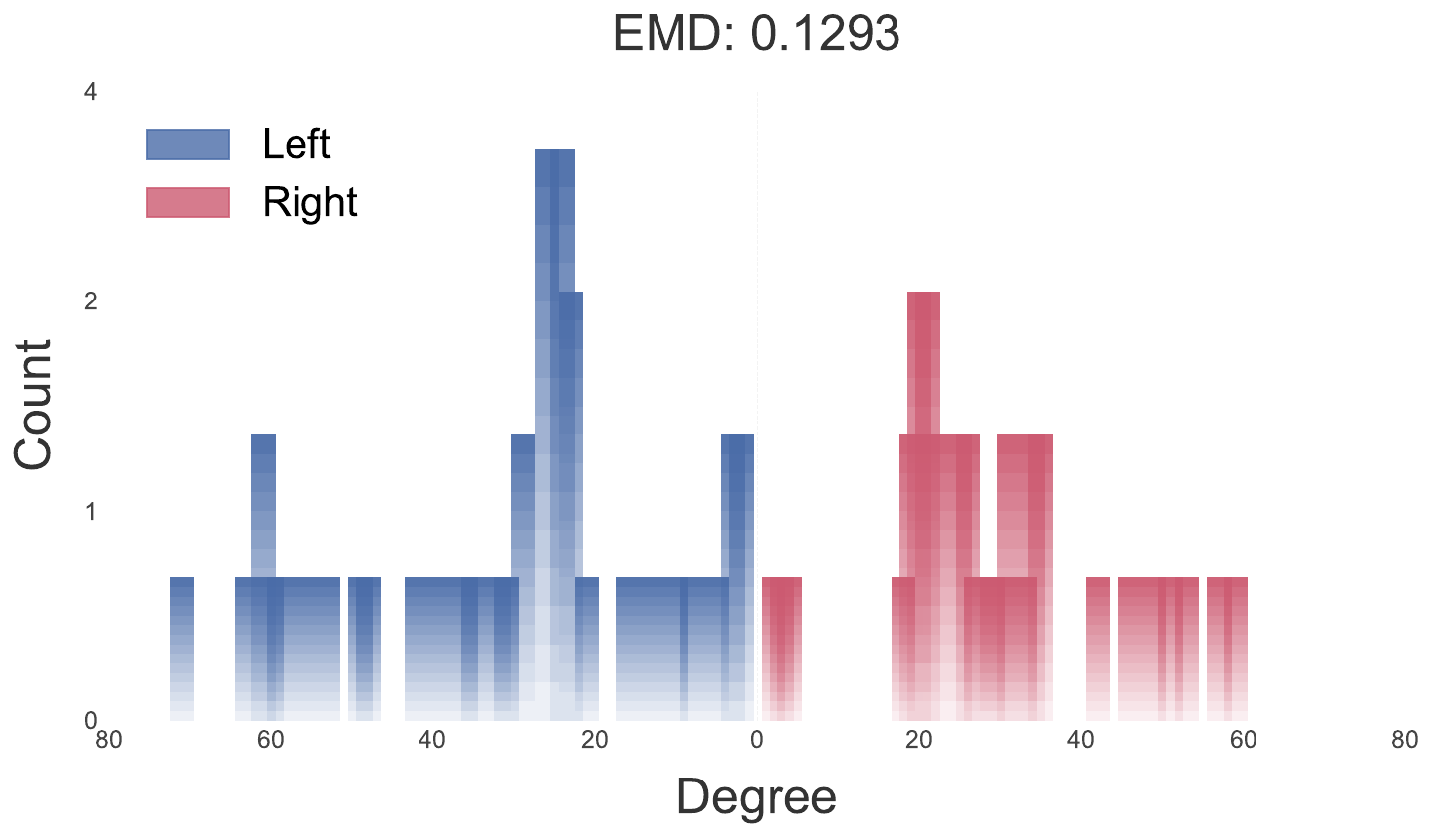}}  
	\hspace{0.1em}  
	\subfloat[Kenyon Cell]{\includegraphics[width=0.19\textwidth]{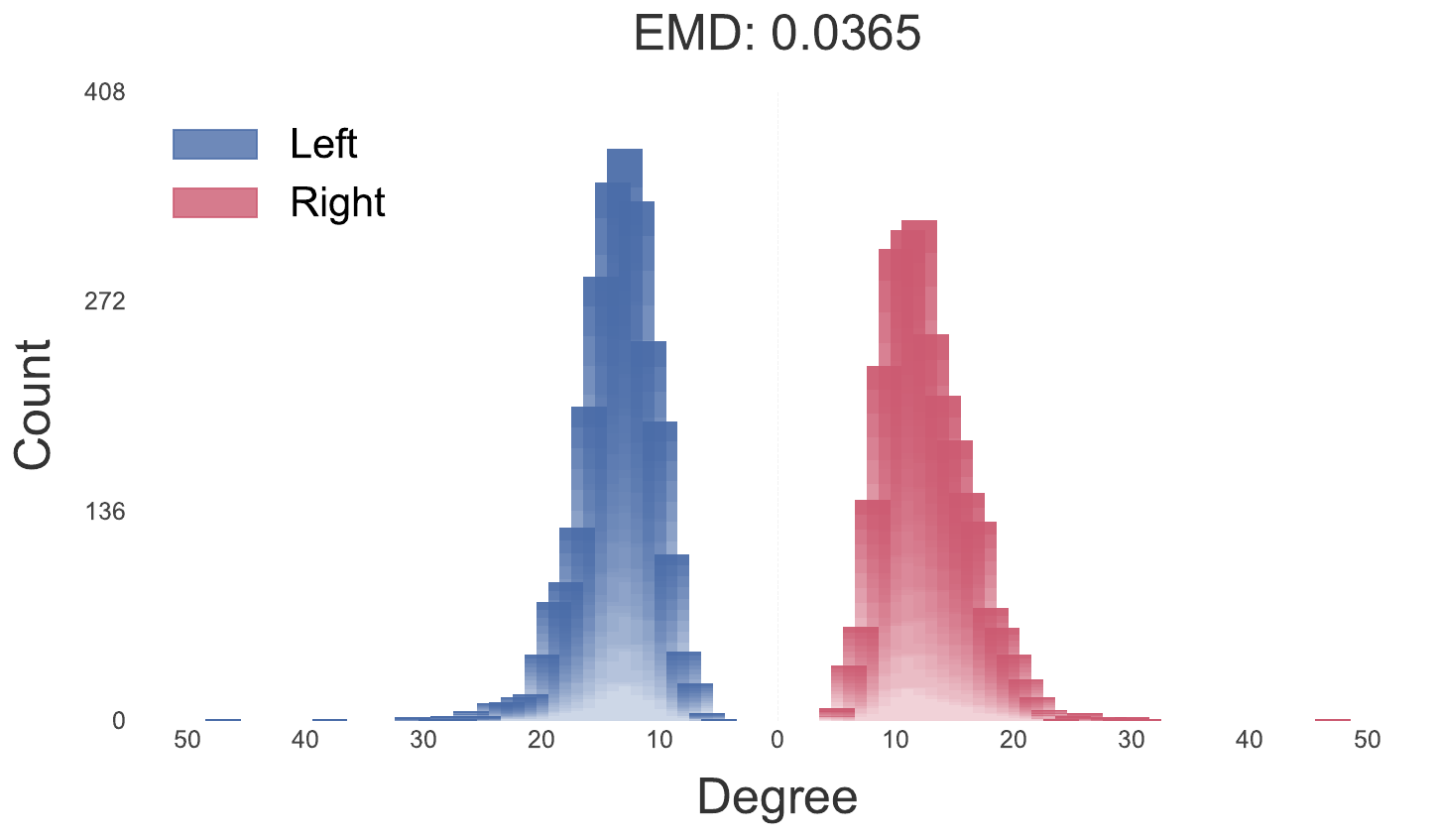}}  
	\hspace{0.1em}   
	\subfloat[LHCENT]{\includegraphics[width=0.19\textwidth]{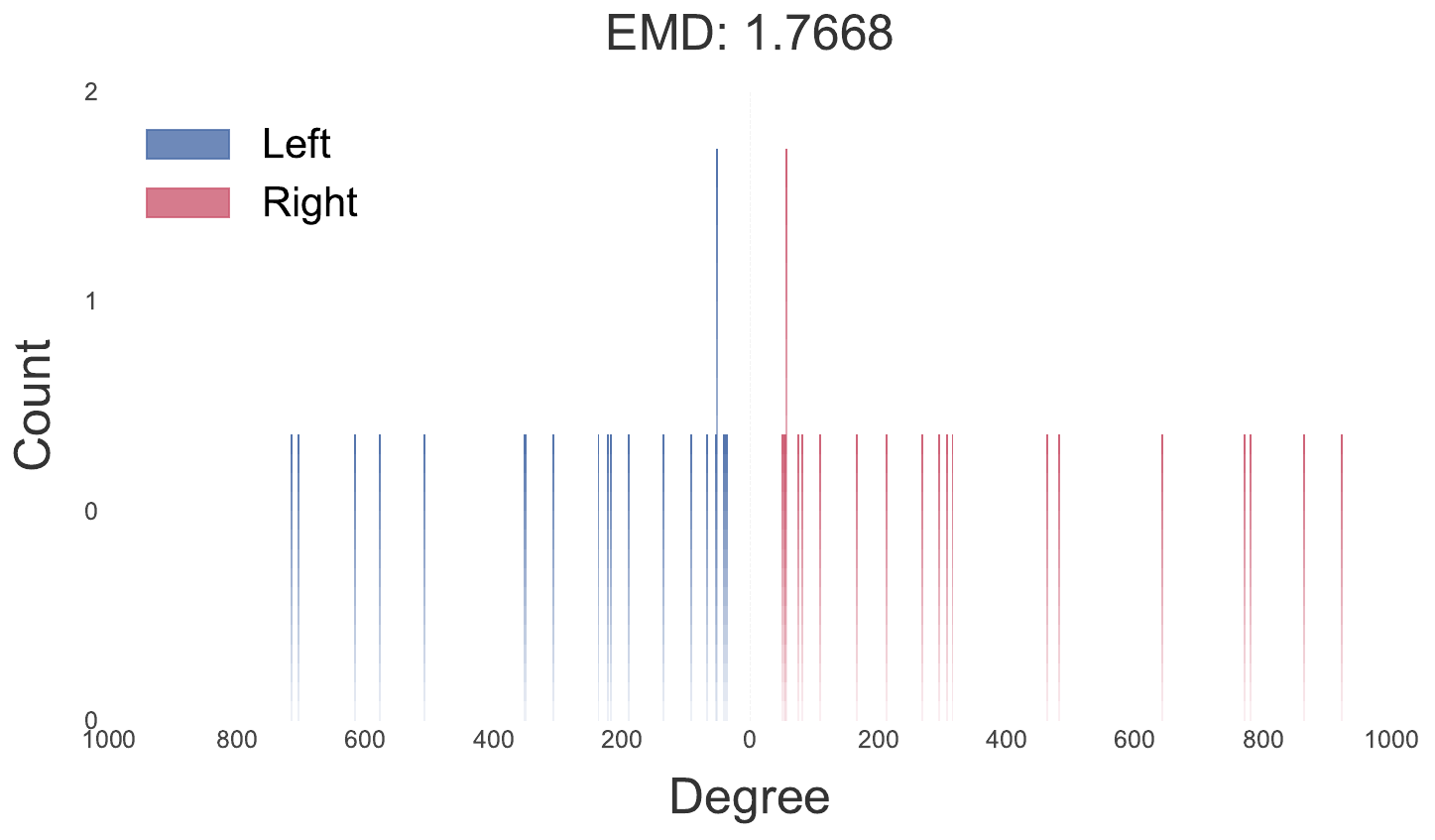}}  
	\hspace{0.1em}  
	\subfloat[LHLN]{\includegraphics[width=0.19\textwidth]{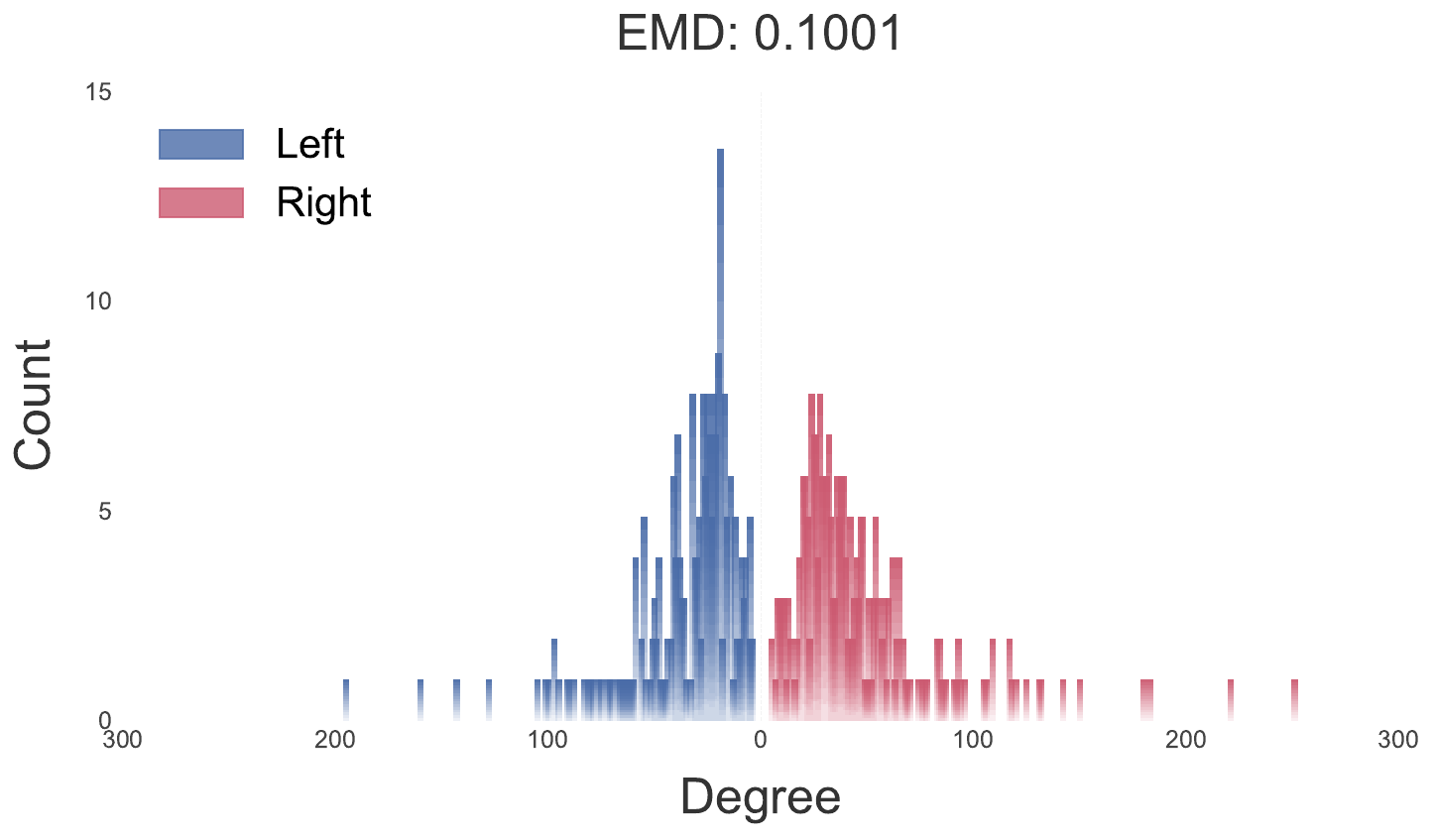}}  
	\hspace{0.1em}    
	\subfloat[mAL]{\includegraphics[width=0.19\textwidth]{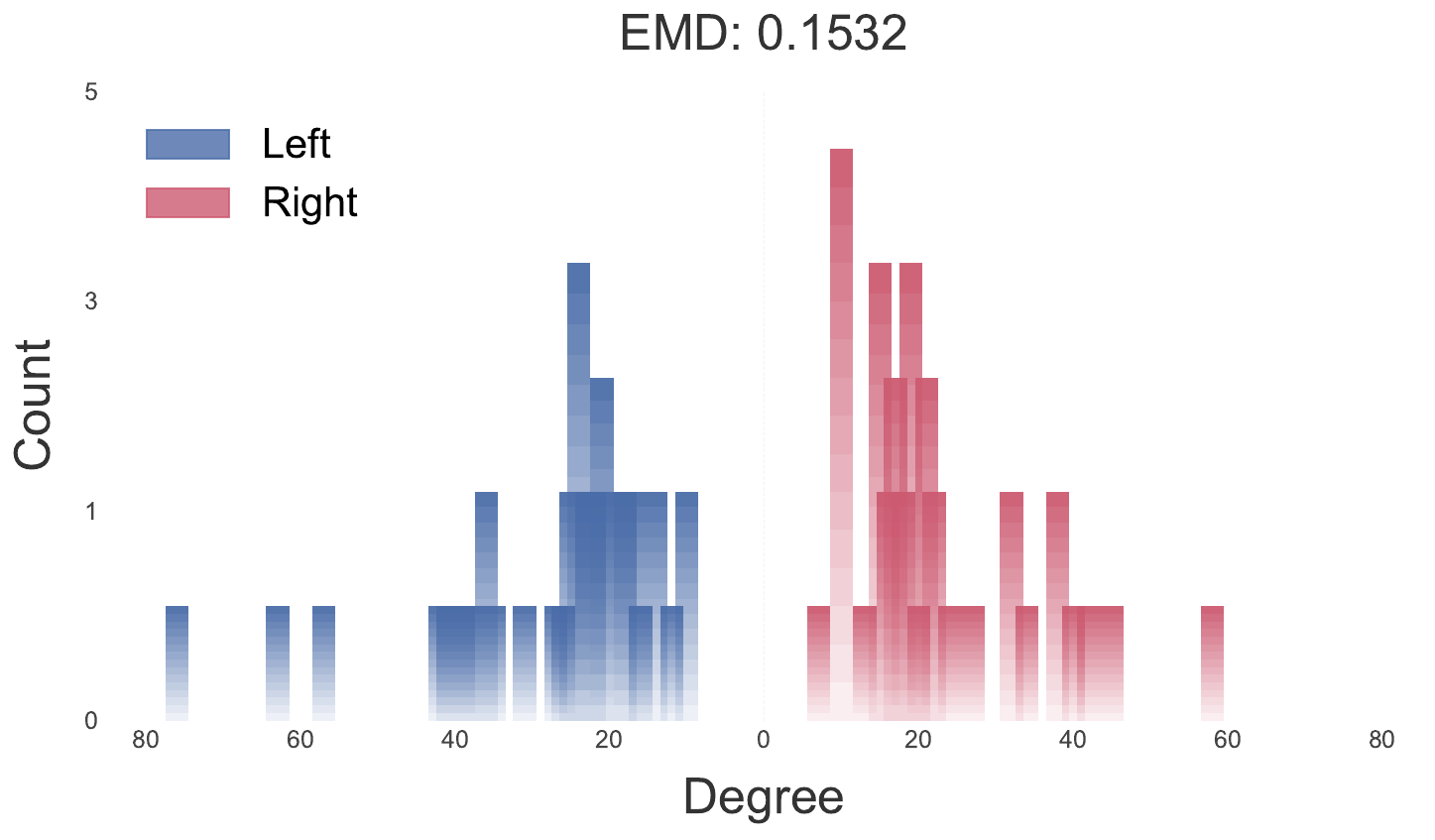}}  
	
	\vspace{-1em}  
	\subfloat[MBIN]{\includegraphics[width=0.19\textwidth]{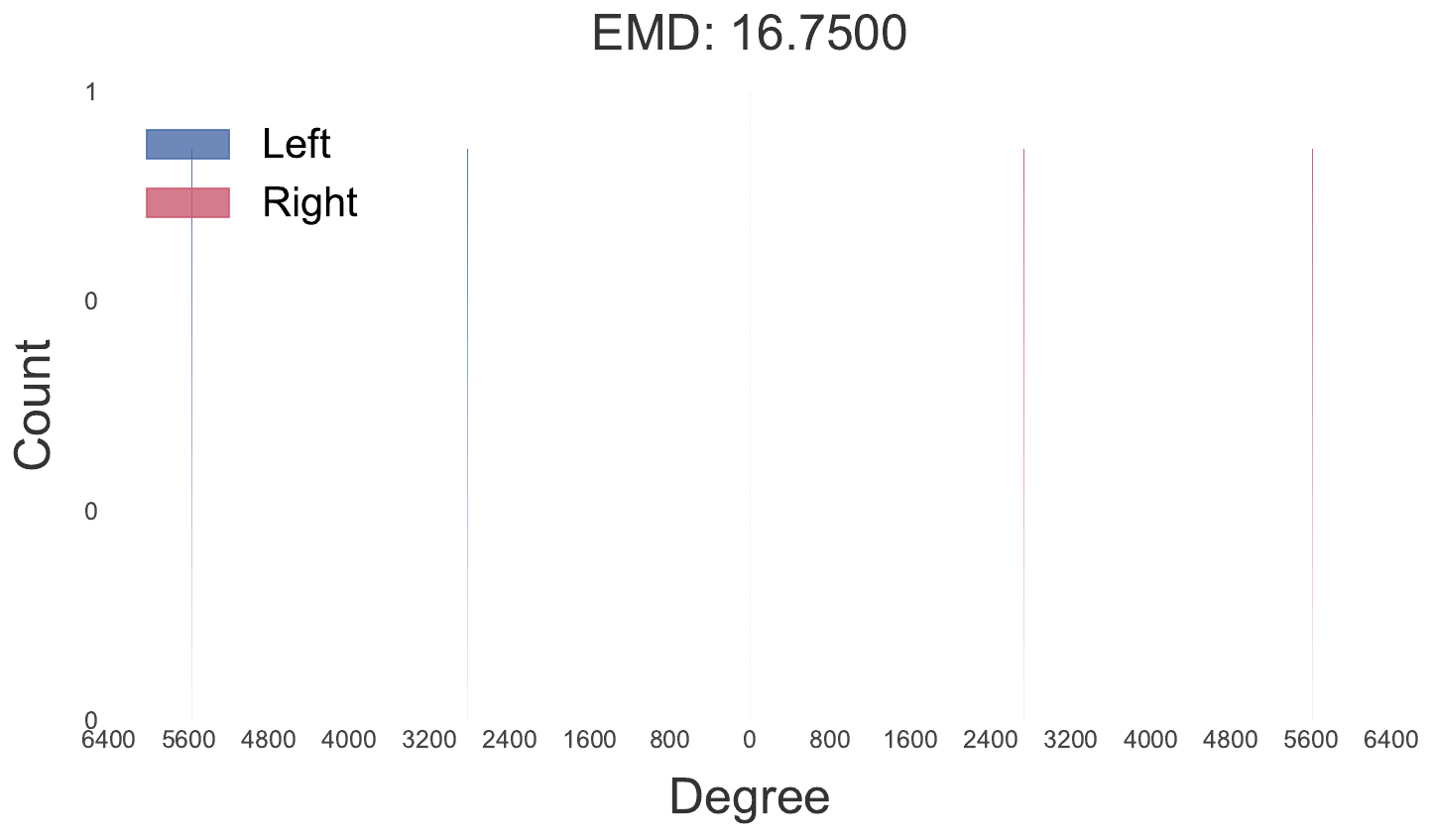}}  
	\hspace{0.1em}   
	\subfloat[MBON]{\includegraphics[width=0.19\textwidth]{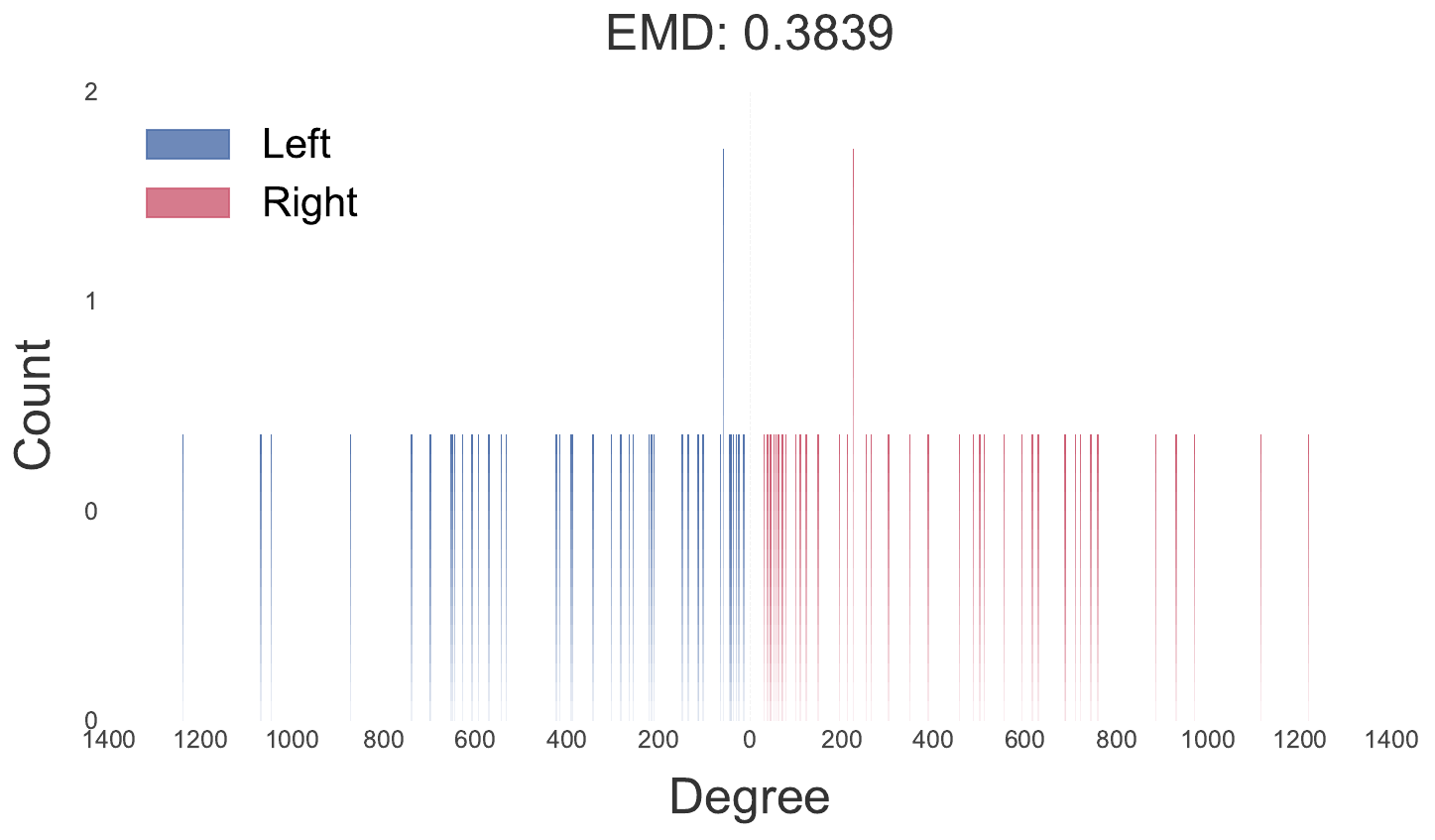}}  
	\hspace{0.1em}  
	\subfloat[mechanosensory]{\includegraphics[width=0.19\textwidth]{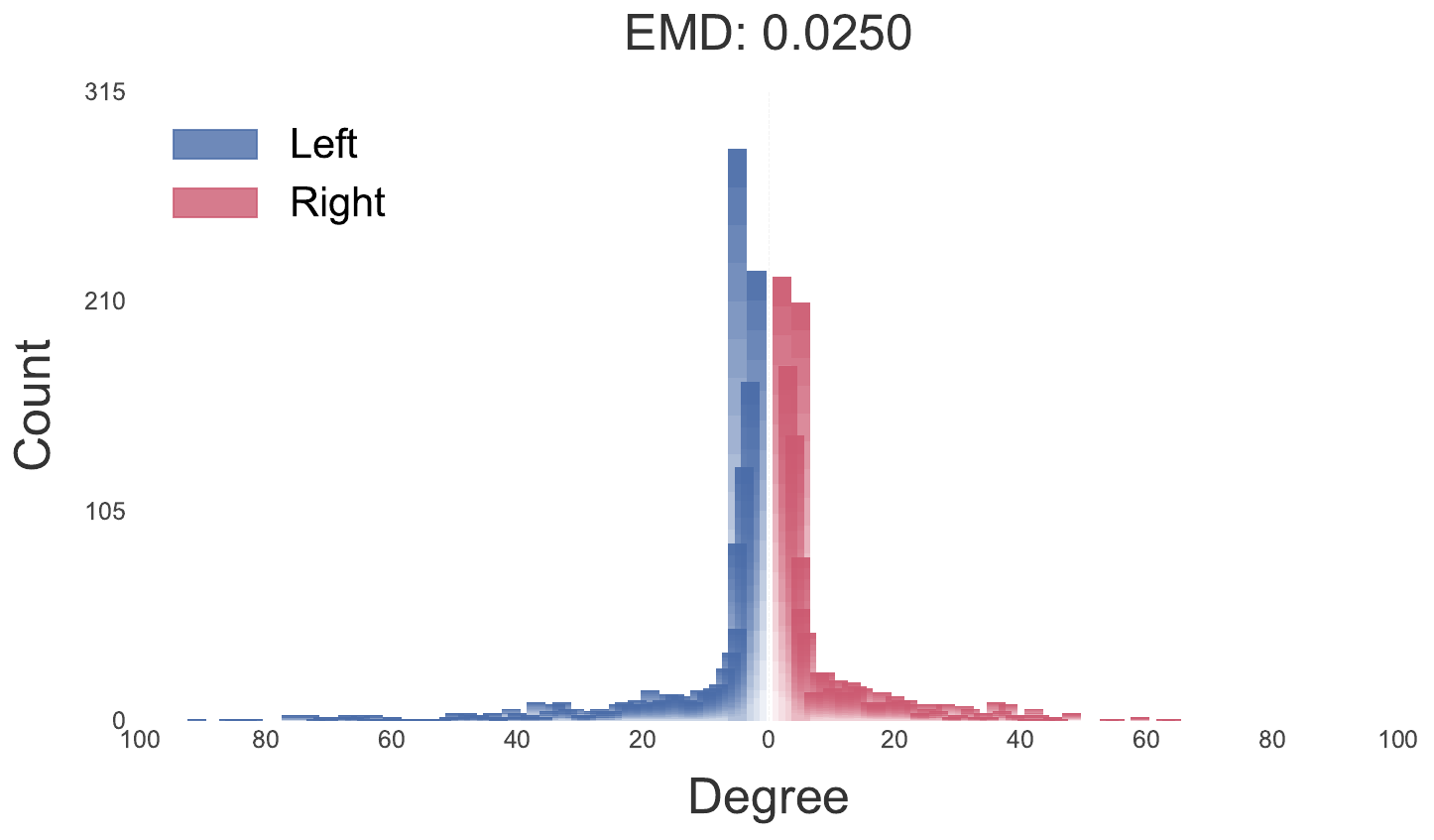}}  
	\hspace{0.1em}   
	\subfloat[motor]{\includegraphics[width=0.19\textwidth]{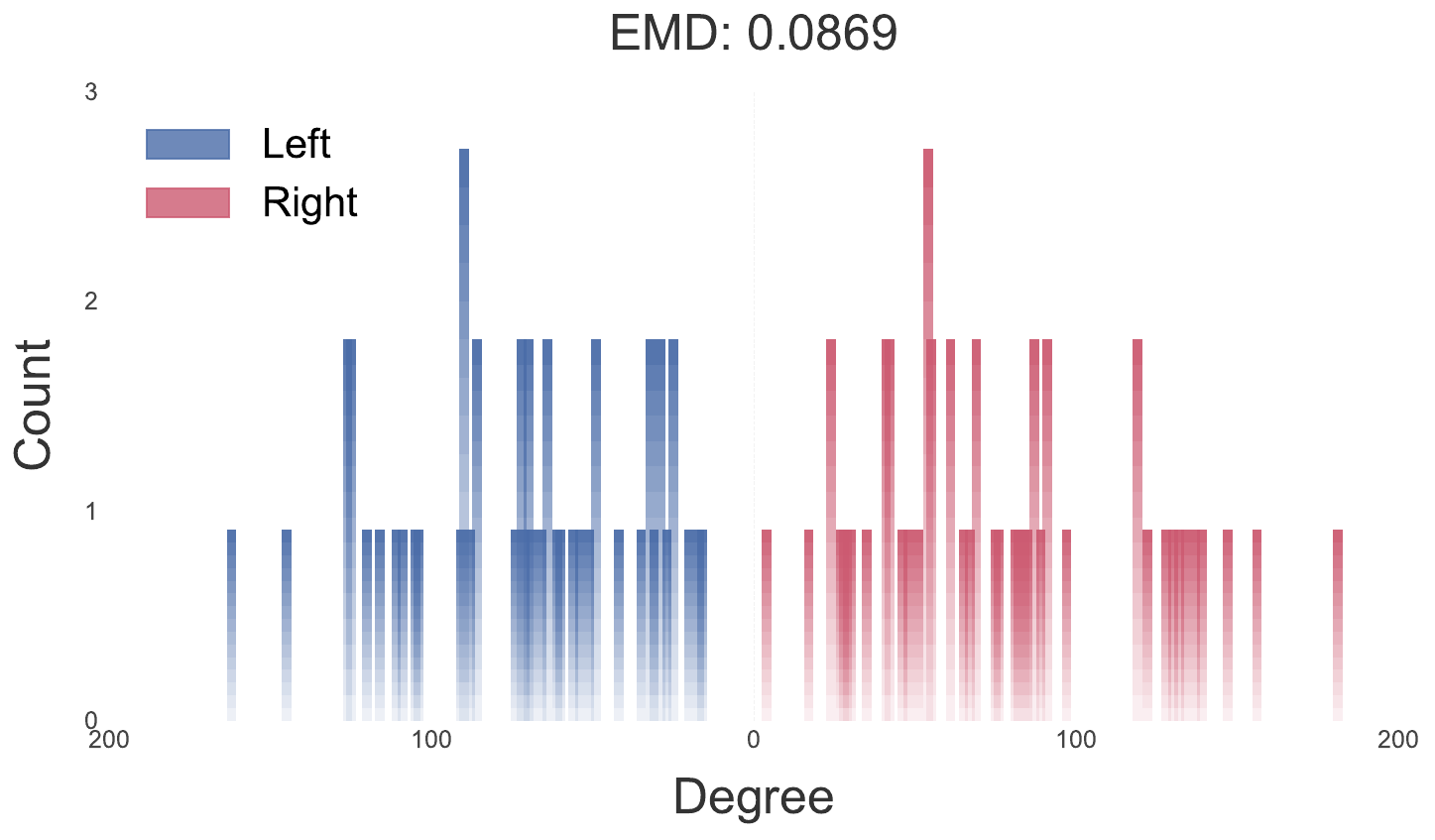}}  
	\hspace{0.1em}  
	\subfloat[olfactory sensory]{\includegraphics[width=0.19\textwidth]{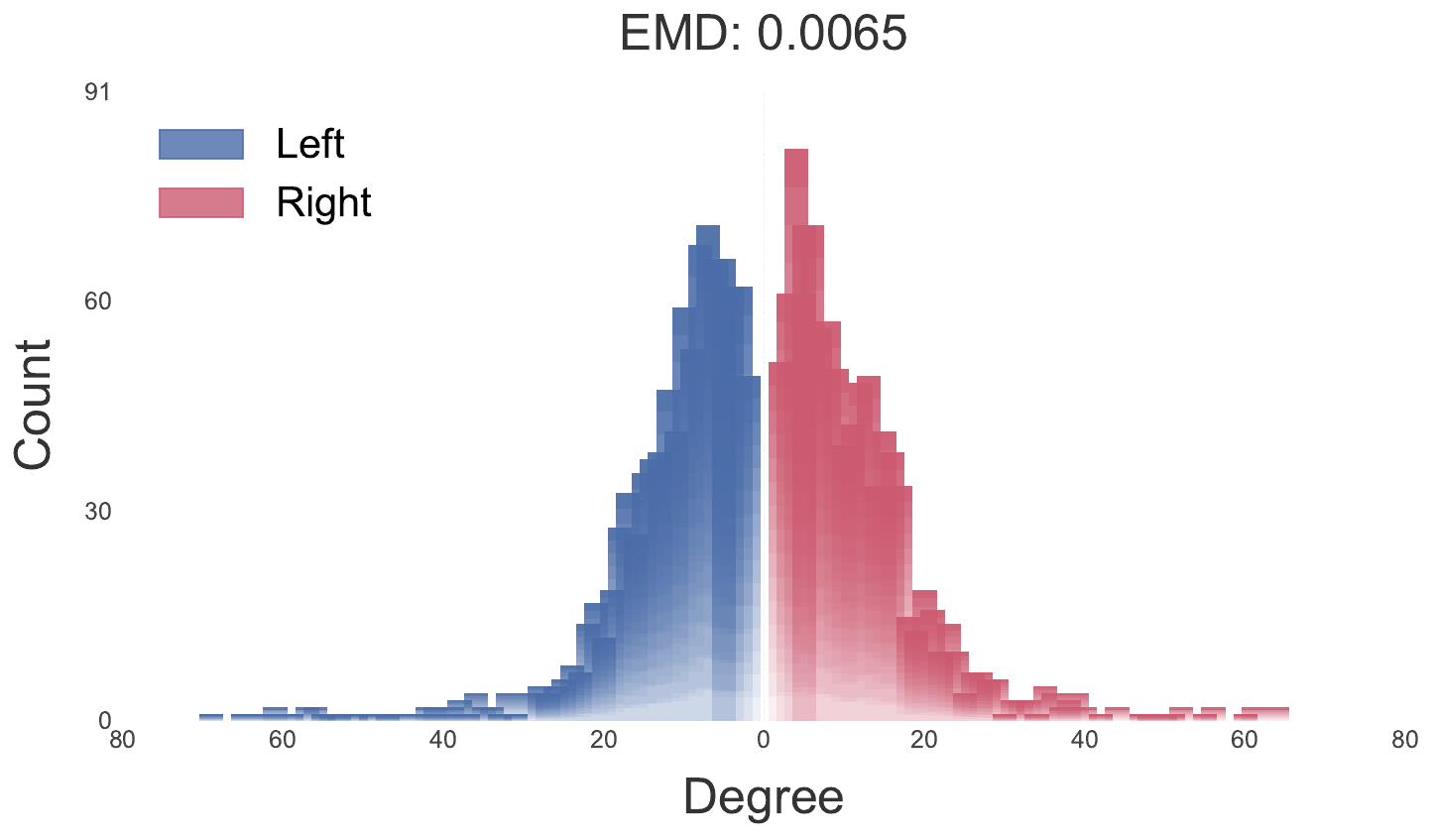}}  
	
	\vspace{-1em}  
	\subfloat[ocellar]{\includegraphics[width=0.19\textwidth]{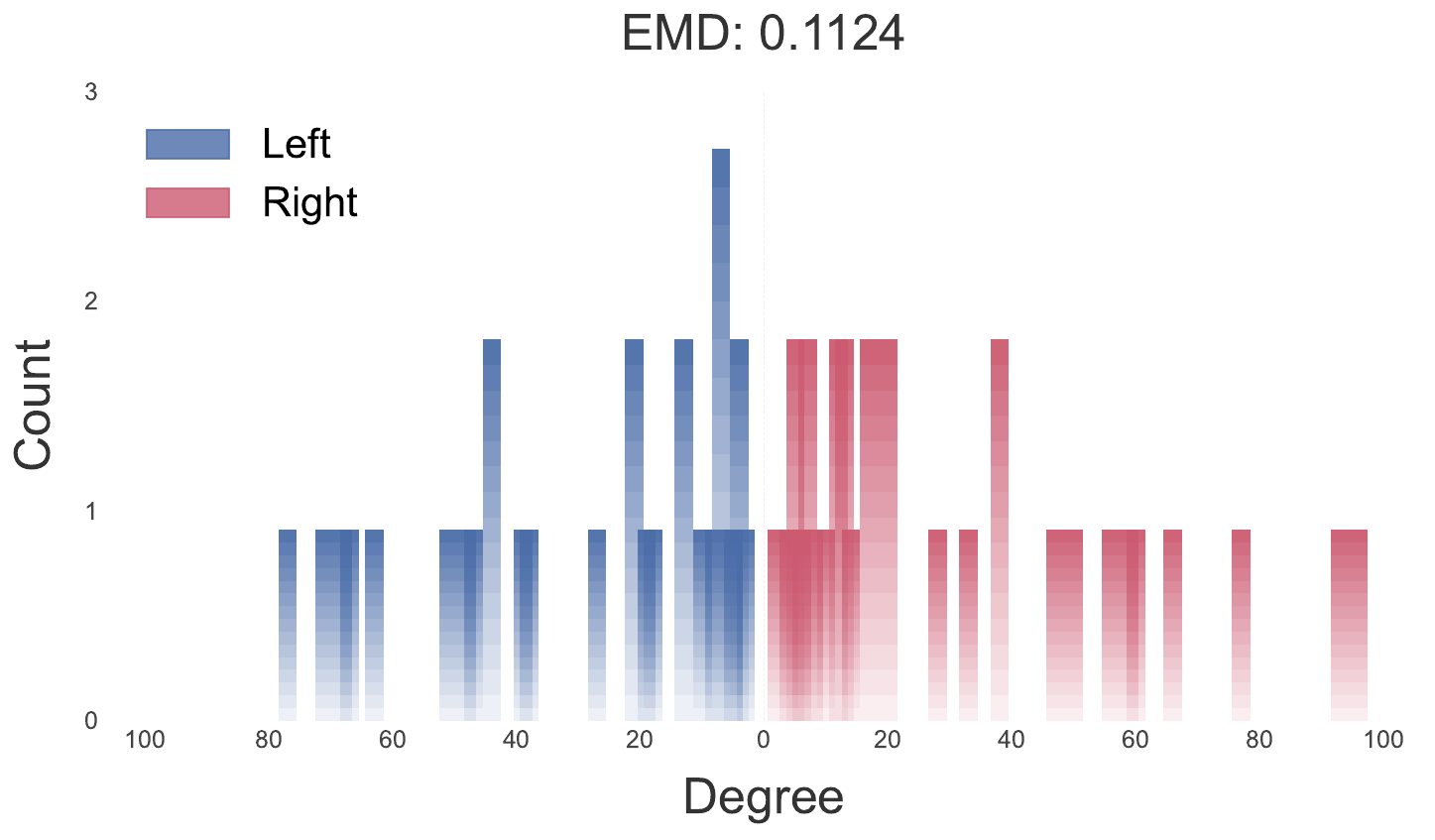}}  
	\hspace{0.1em}  
	\subfloat[optic]{\includegraphics[width=0.19\textwidth]{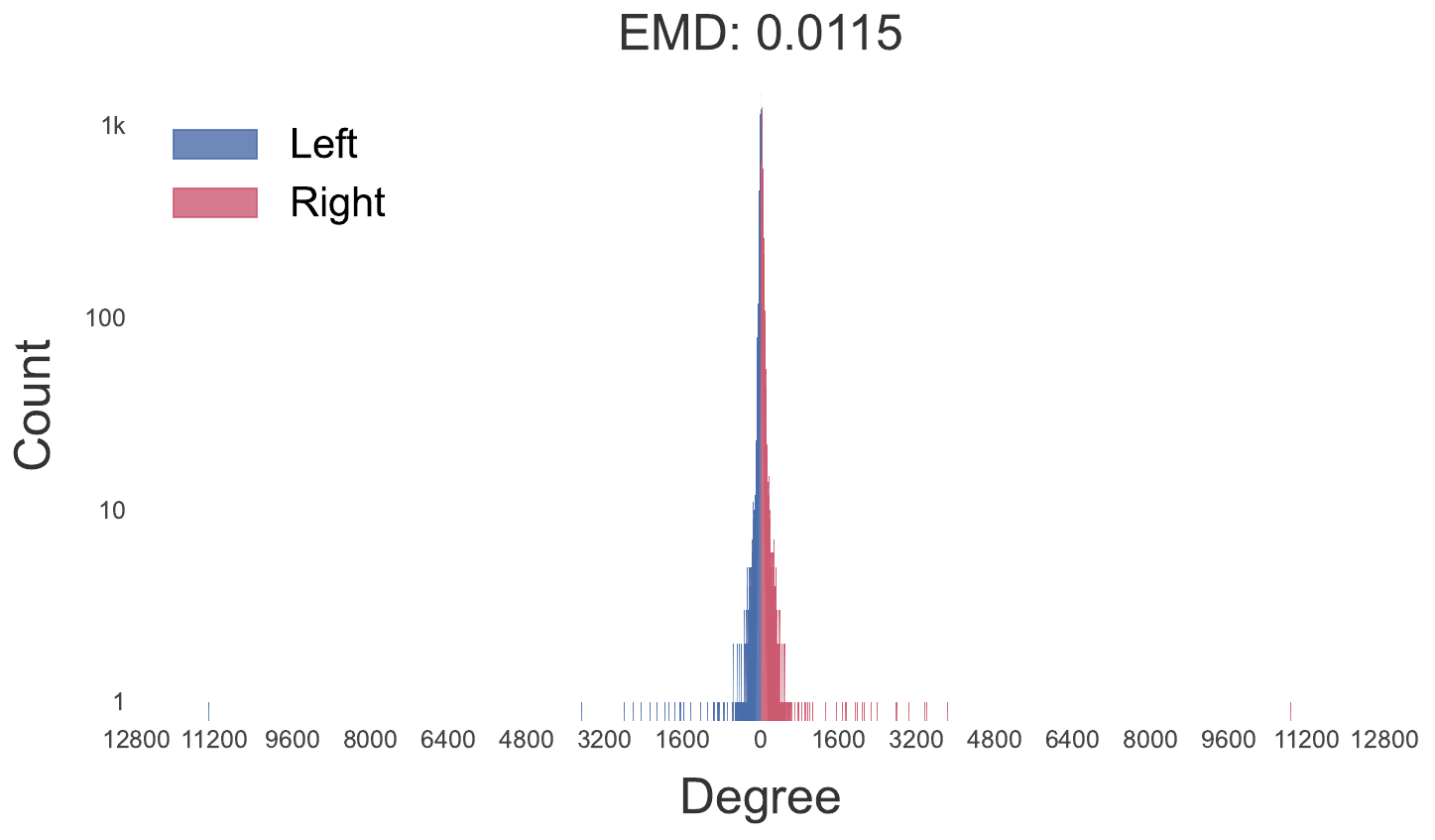}}  
	\hspace{0.1em}   
	\subfloat[thermosensory]{\includegraphics[width=0.19\textwidth]{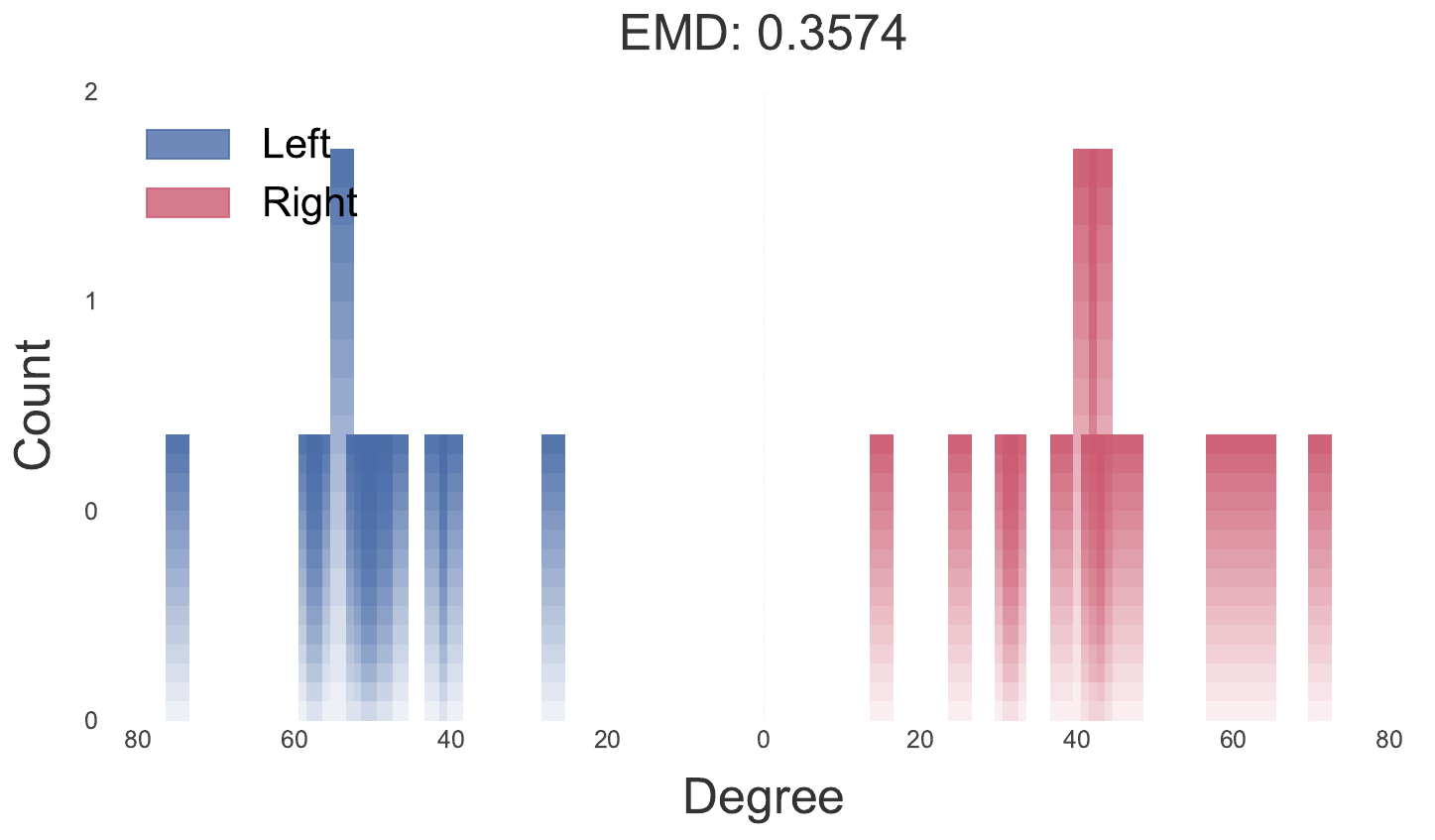}}  
	\hspace{0.1em}  
	\subfloat[TuBu]{\includegraphics[width=0.19\textwidth]{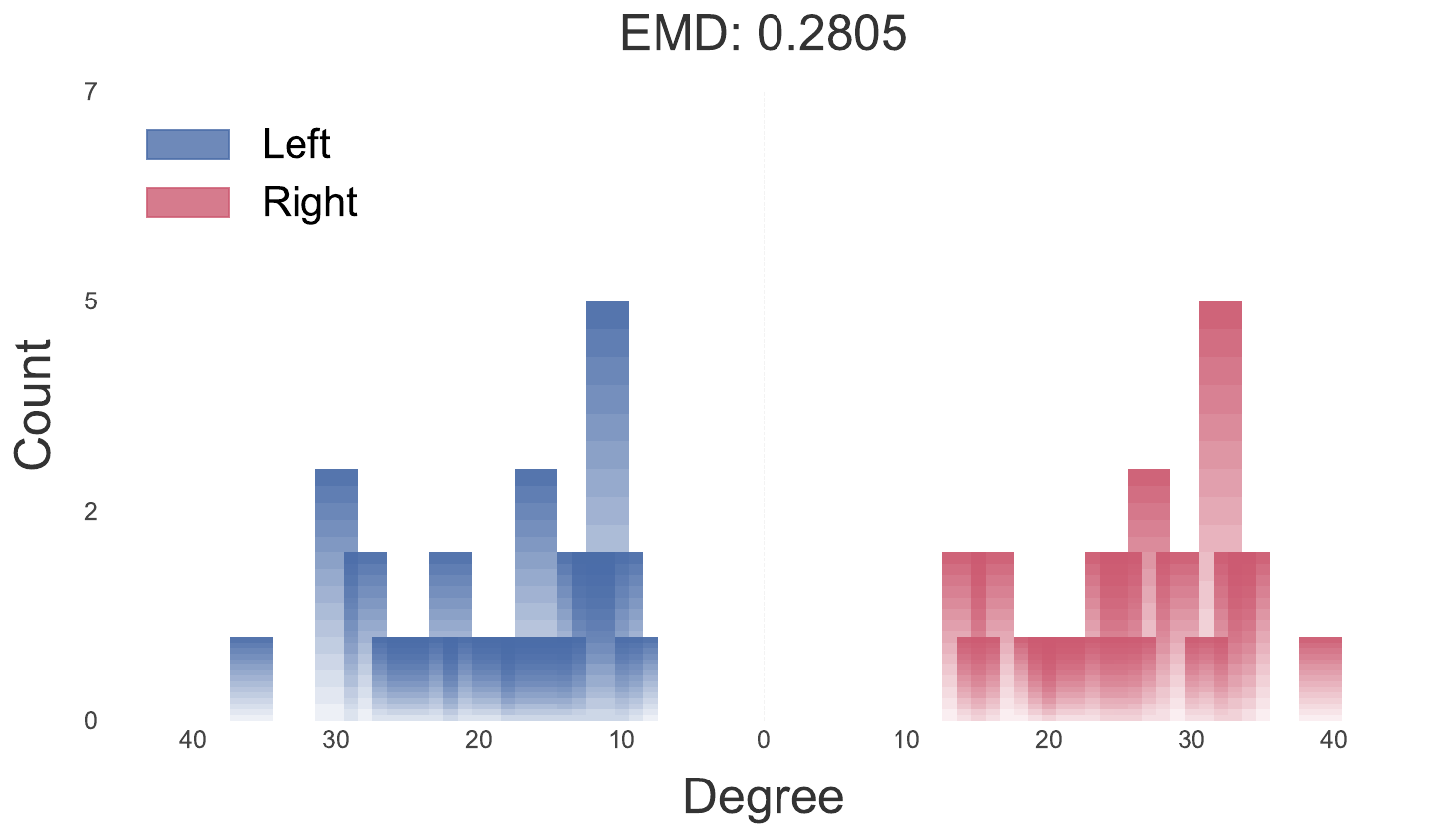}}  
	\hspace{0.1em}  
	\subfloat[visual centrifugal]{\includegraphics[width=0.19\textwidth]{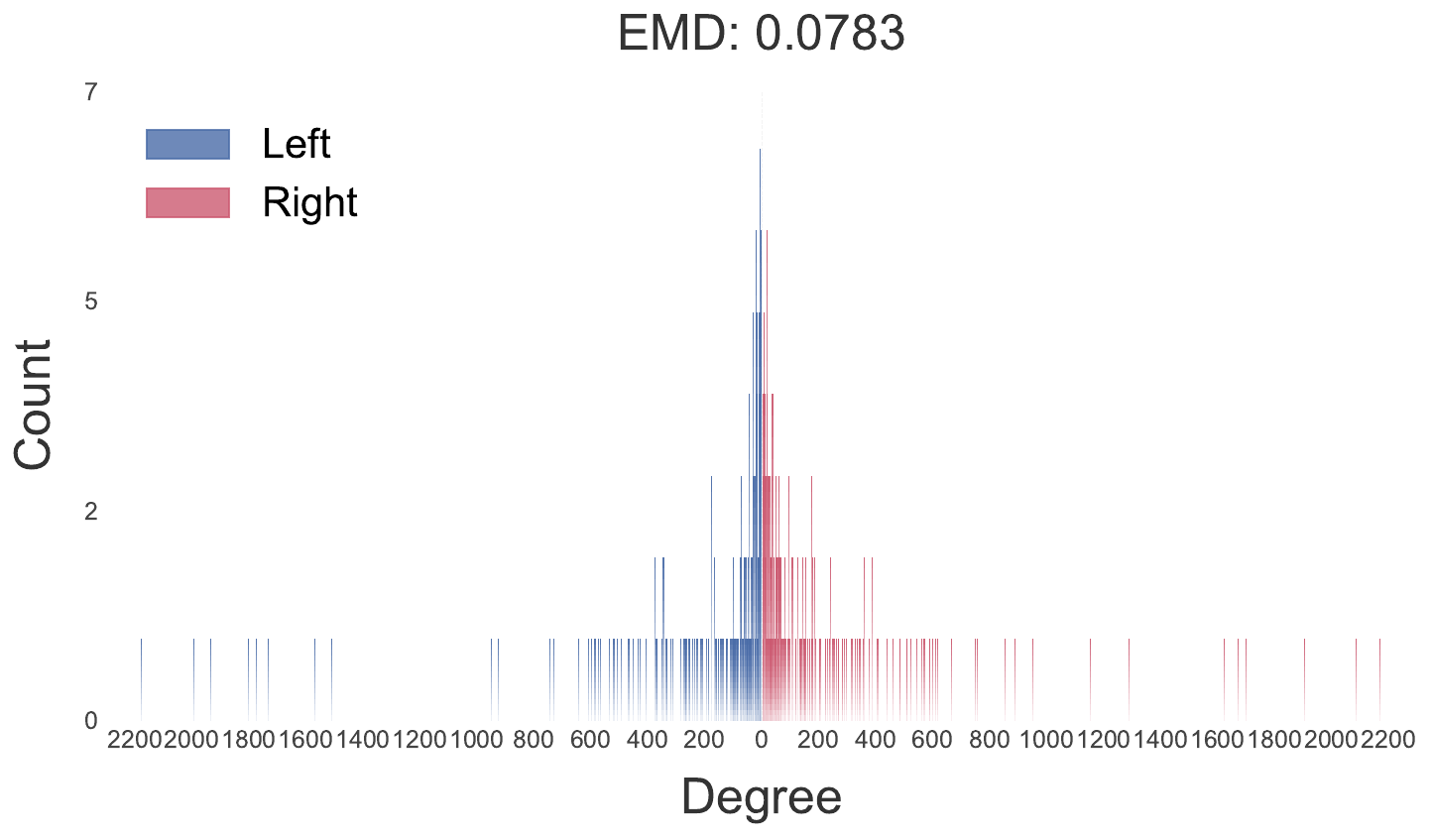}}  
	
	\vspace{-1em}  
	\subfloat[visual sensory]{\includegraphics[width=0.19\textwidth]{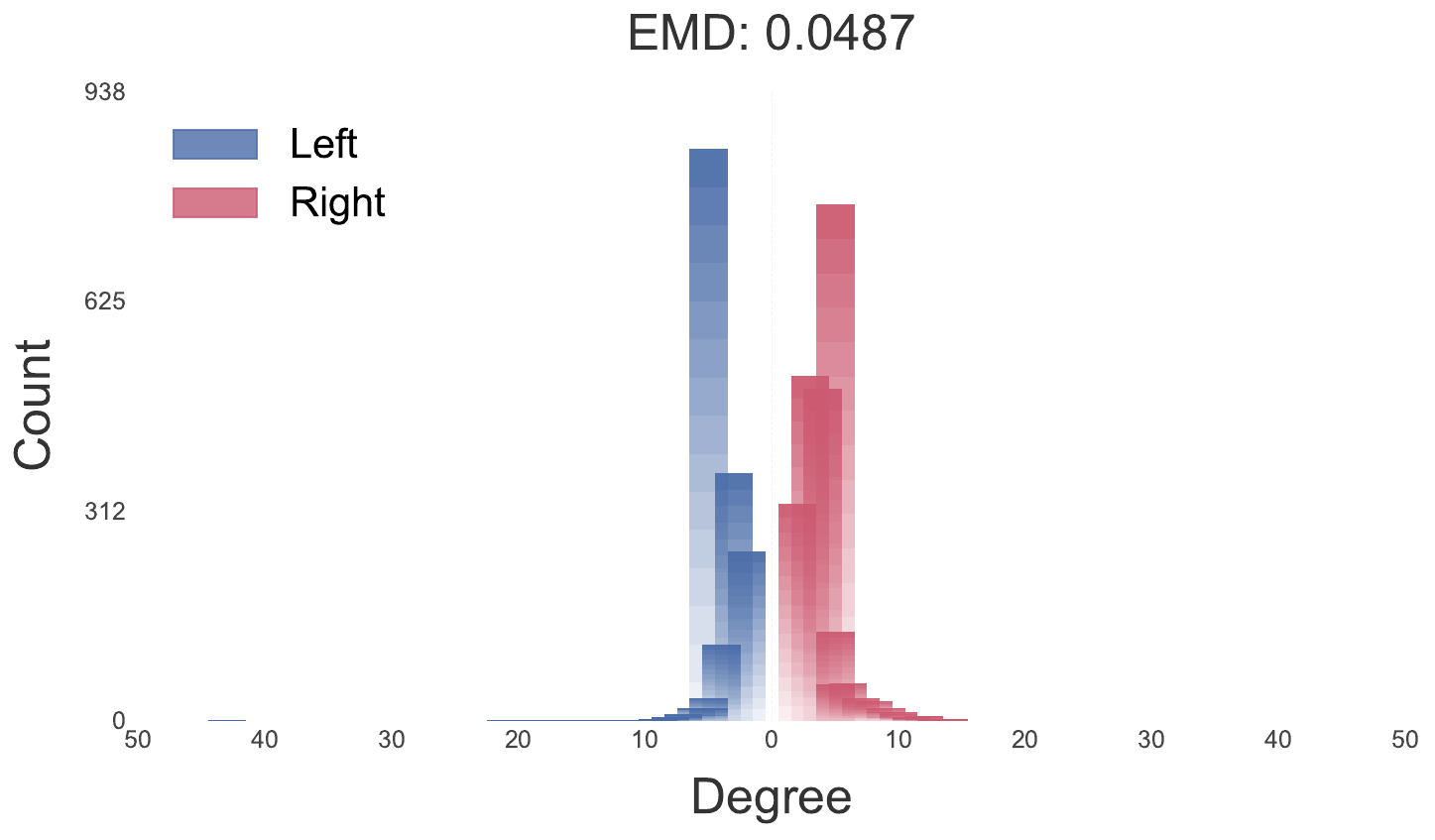}}  
	\hspace{1.0em}  
		\subfloat[gustatory]{\includegraphics[width=0.19\textwidth]{./figures/degree_distribution/gustatory_degree_distribution.pdf}}  
	\caption{Degree distributions of left and right hemispheres across diverse regions}  
	\label{degree_region}  
\end{figure*}

In order to further examine the similarity between homologous subregions across the left and right hemispheres, we computed and visualized the degree distributions for each cortical region (Fig. \ref{degree_region}). To quantify the interhemispheric similarity of these distributions, we employed the Earth Mover's Distance (EMD, see Data and Methods) metric, normalized to account for potential variations in distribution scale. Specifically, we calculated the normalized EMD similarity coefficient between the left and right hemisphere distributions for each region. Our results demonstrate that an overwhelming majority of regions exhibit remarkably low EMD similarity coefficients. This finding strongly suggests a high degree of concordance in the degree distributions between homologous regions in the left and right hemispheres. Such bilateral symmetry in network topology may have significant implications for our understanding of brain organization and function. Note that the visual (sensory) and MBIN region exhibits a relatively high EMD coefficient, indicating substantial asymmetry in the degree distribution of visual input neurons between the left and right hemispheres of the brain. 
\subsection{Spatial Clustering and Functional Organization}
In addition to its network structure, the nervous system exhibits distinctive spatial organization, as exemplified by the spatial distribution patterns of neurons and synaptic connections. These distribution patterns are predominantly established through developmental processes. Investigation of such spatial heterogeneity provides crucial empirical evidence for understanding the development of the nervous system.

\begin{figure*}[t]
	\centering
	\subfloat[]{  
		\begin{minipage}{0.45\textwidth}  
			\centering  
			\includegraphics[width=\textwidth]{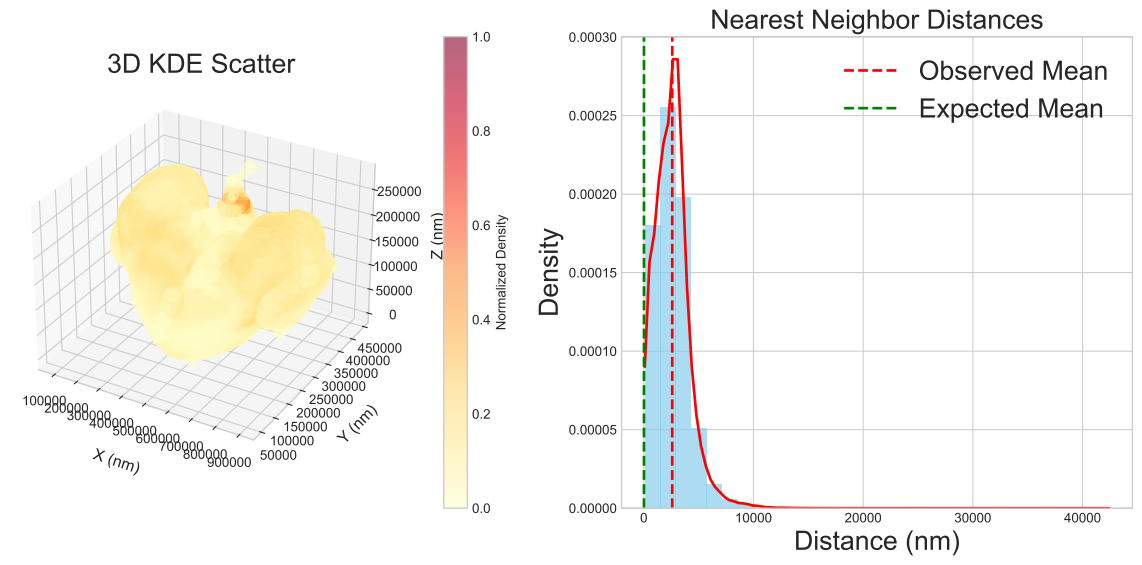}  \end{minipage}  	} 
				\subfloat[]{ 		
					\begin{minipage}{0.45\textwidth}  
				\centering  
				\includegraphics[width=\textwidth]{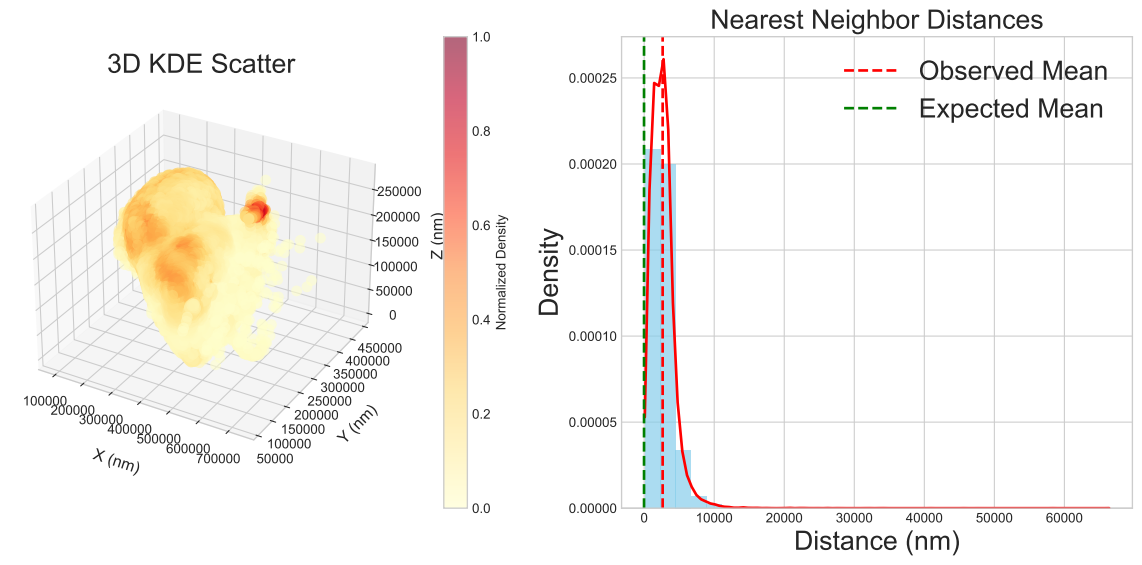}  \end{minipage}}  
				\\
					\subfloat[]{ 
						\begin{minipage}{0.45\textwidth}  
					\centering  
					\includegraphics[width=\textwidth]{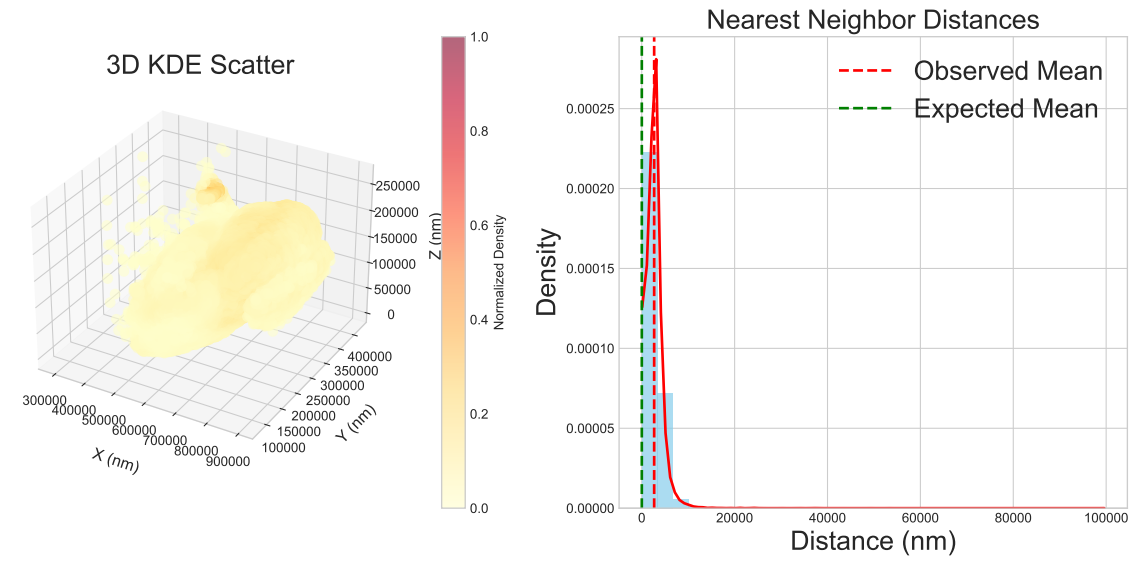}  \end{minipage}}
					\subfloat[]{   
							\begin{minipage}{0.45\textwidth}  
						\centering  
						\includegraphics[width=\textwidth]{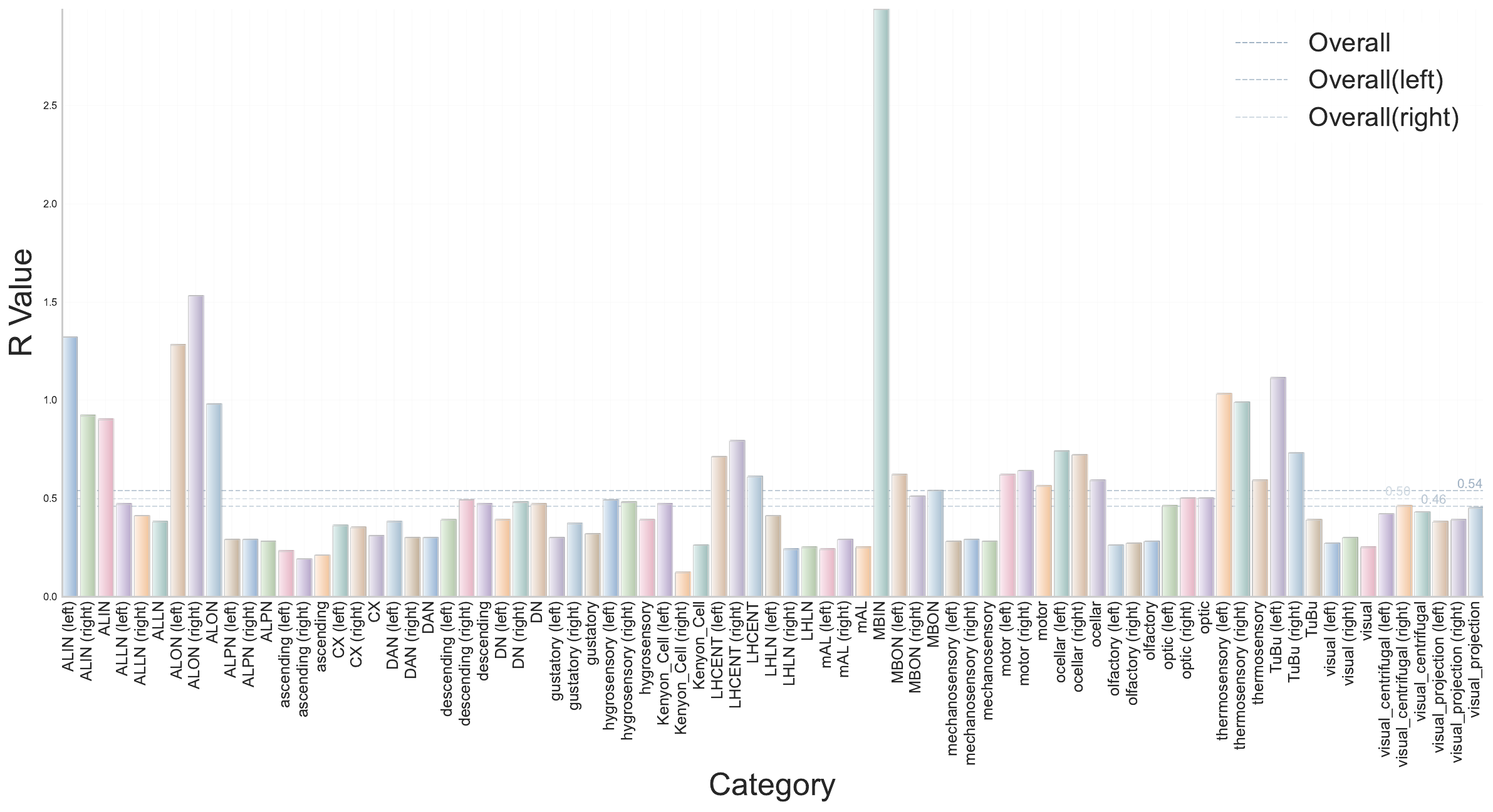}  \end{minipage}  }

 \caption{Spatial distribution and clustering analysis of neuronal somata. Panels (a), (b), and (c) illustrate the spatial distribution of neurons and the distribution of nearest neighbor distances for the entire brain, left hemisphere, and right hemisphere, respectively. Panel (d) presents the nearest neighbor index $R$ for regions. Each bar represents a specific brain area, with the bar height indicating the $R$ value for that region. }
 		\label{spatial}
\end{figure*}

The spatial distribution and clustering patterns of neurons are analyzed and visualized in Fig. \ref{spatial}. As shown in Fig. \ref{spatial}(a), the 3D kernel density estimation (KDE) scatter plot and the nearest neighbor distance histogram reveal a nearest neighbor index (R = 0.54), indicating a significant deviation from random spatial distribution and reflecting an overall agglomerative pattern. In the left and right hemispheres, 3D KDE scatter plots (Fig. \ref{spatial}(b) and Fig. \ref{spatial}(c)) reveal R values of 0.46 and 0.54, respectively, suggesting notable hemispheric differences in neuronal clustering. Regional analysis, as depicted in Fig. \ref{spatial}(d), shows that most brain regions have R values lower than the whole-brain average, implying that neurons with similar functions tend to cluster within specific anatomical areas.

Interestingly, Mushroom Body Input Neurons (MBINs) exhibit a markedly dispersed distribution, with an R value of 2.99. This unusual spatial pattern may be attributed to their functional role in transmitting sensory information and internal state signals to the Mushroom Body. Such a function requires widespread anatomical coverage, enabling MBINs to integrate inputs from multiple sensory modalities and distributed brain regions. This dispersed arrangement likely facilitates the efficient collection and integration of information necessary for the modulation of complex behaviors governed by the Mushroom Body.

Moreover, significant disparities in the distribution of neurons between the left and right hemispheres in specific regions, notably ALIN, ALON, Kenyon cells, and TuBu (Tubercle Bulb) neurons (Fig. \ref{spatial}(d)). This hemispheric asymmetry may be attributed to several factors: First, certain brain regions may have distinct functional roles in the left and right hemispheres, resulting in differential neuronal distributions. This functional lateralization could be associated with specific cognitive processes or behavioral outputs\cite{geschwind1968human}. Second, the observed asymmetry in neuronal distribution between hemispheres may reflect divergent information processing strategies or optimizations in computational efficiency. Such differences could potentially enhance the overall processing capacity and flexibility of the Drosophila brain\cite{ivry1998two}.

\begin{figure*}[t]  
	\centering  
	\subfloat[ALIN]{\includegraphics[width=0.19\textwidth]{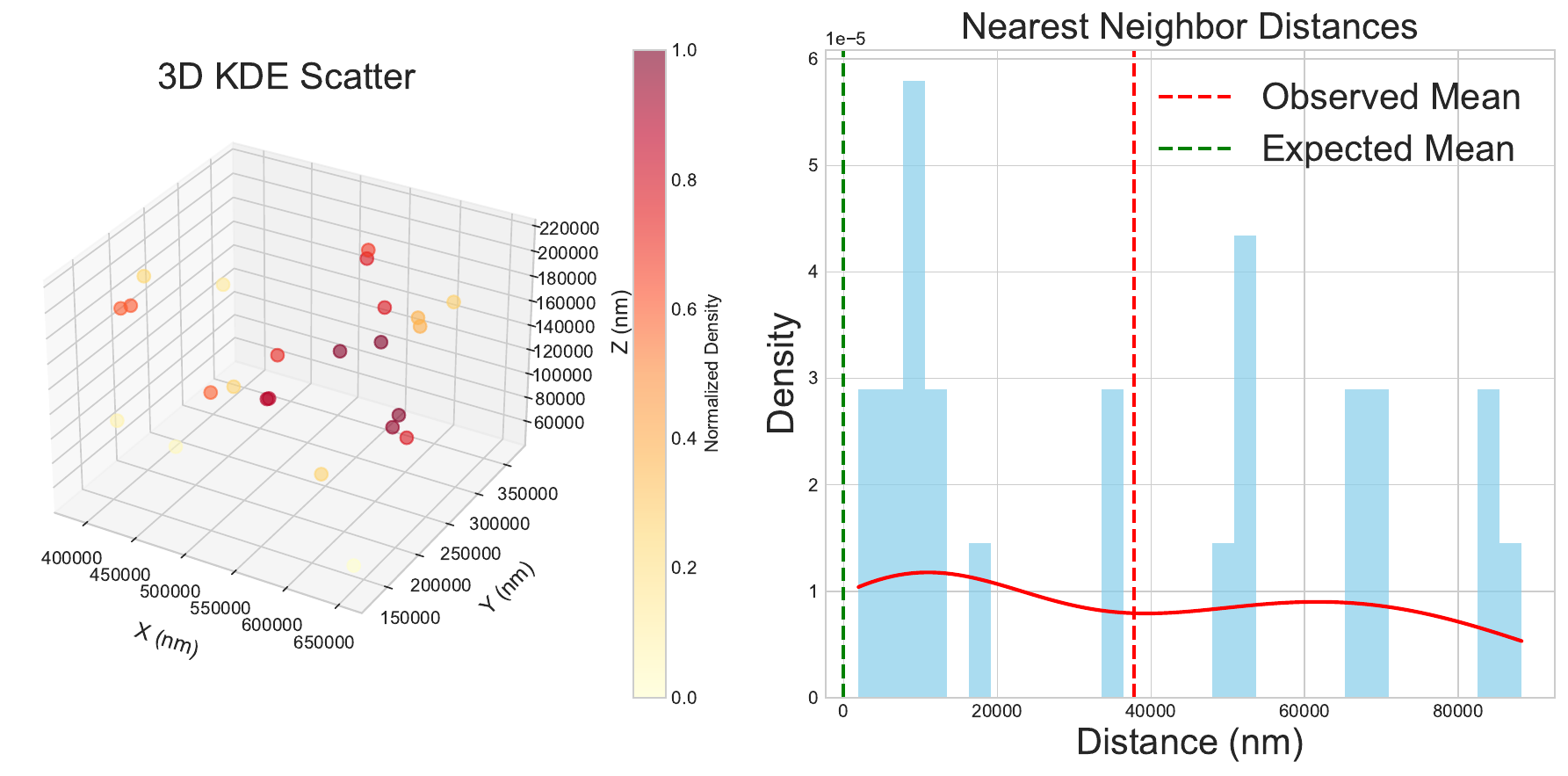}}  
	\hspace{-0.5em}  
	\subfloat[ALON]{\includegraphics[width=0.19\textwidth]{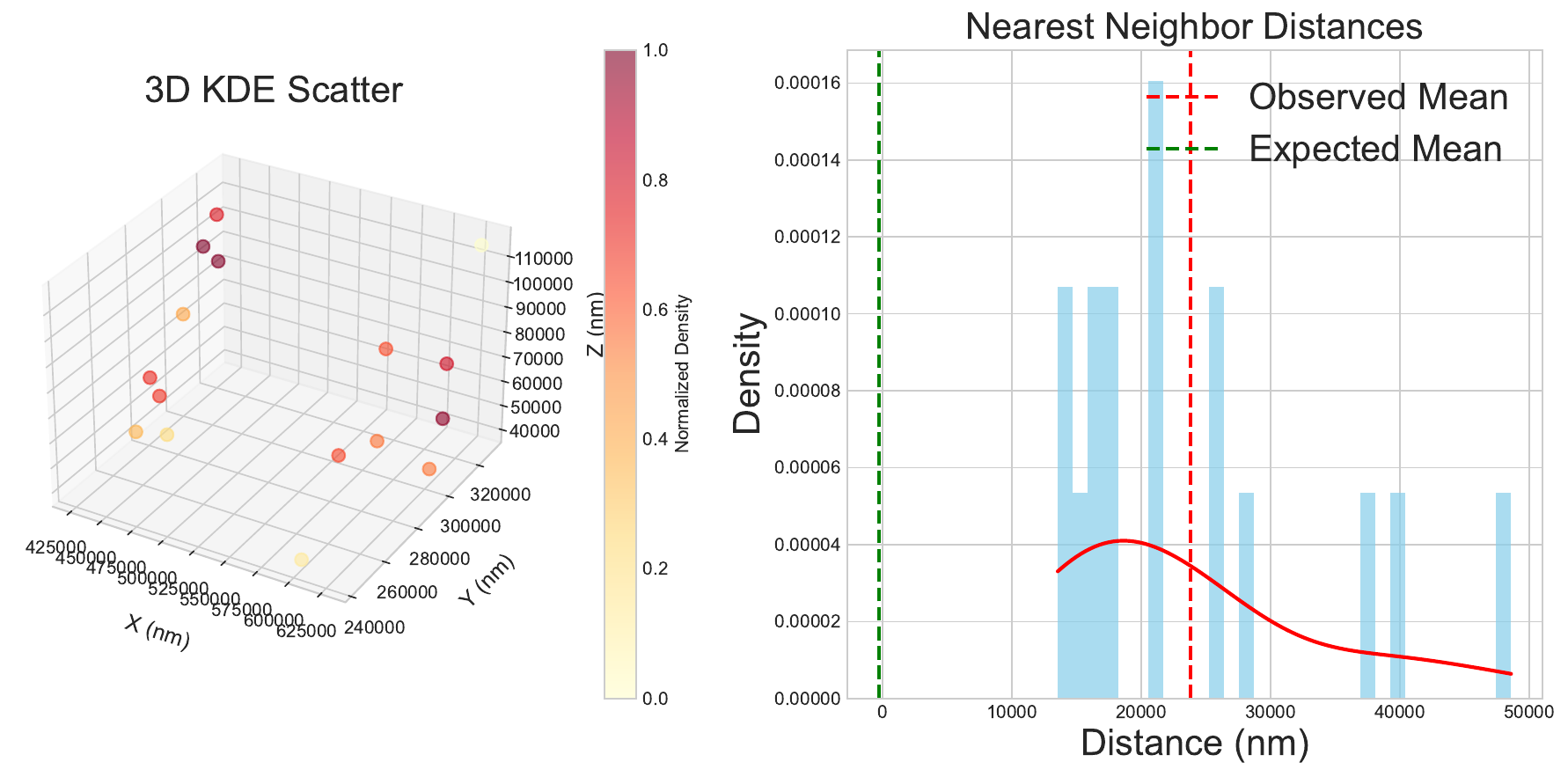}}  
	\hspace{-0.5em}  
	\subfloat[ALLN]{\includegraphics[width=0.19\textwidth]{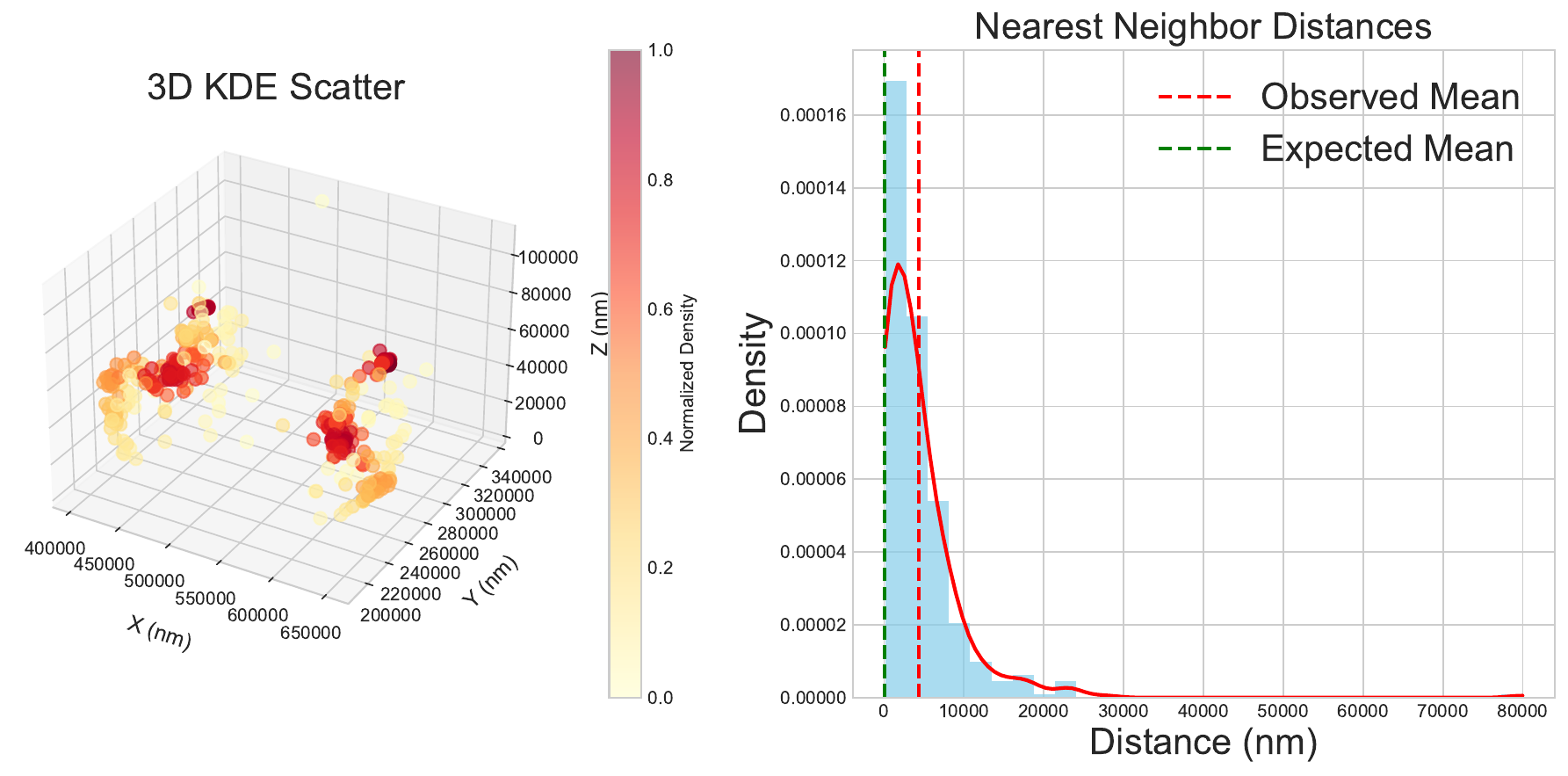}}  
	\hspace{-0.5em}  
	\subfloat[ALPN]{\includegraphics[width=0.19\textwidth]{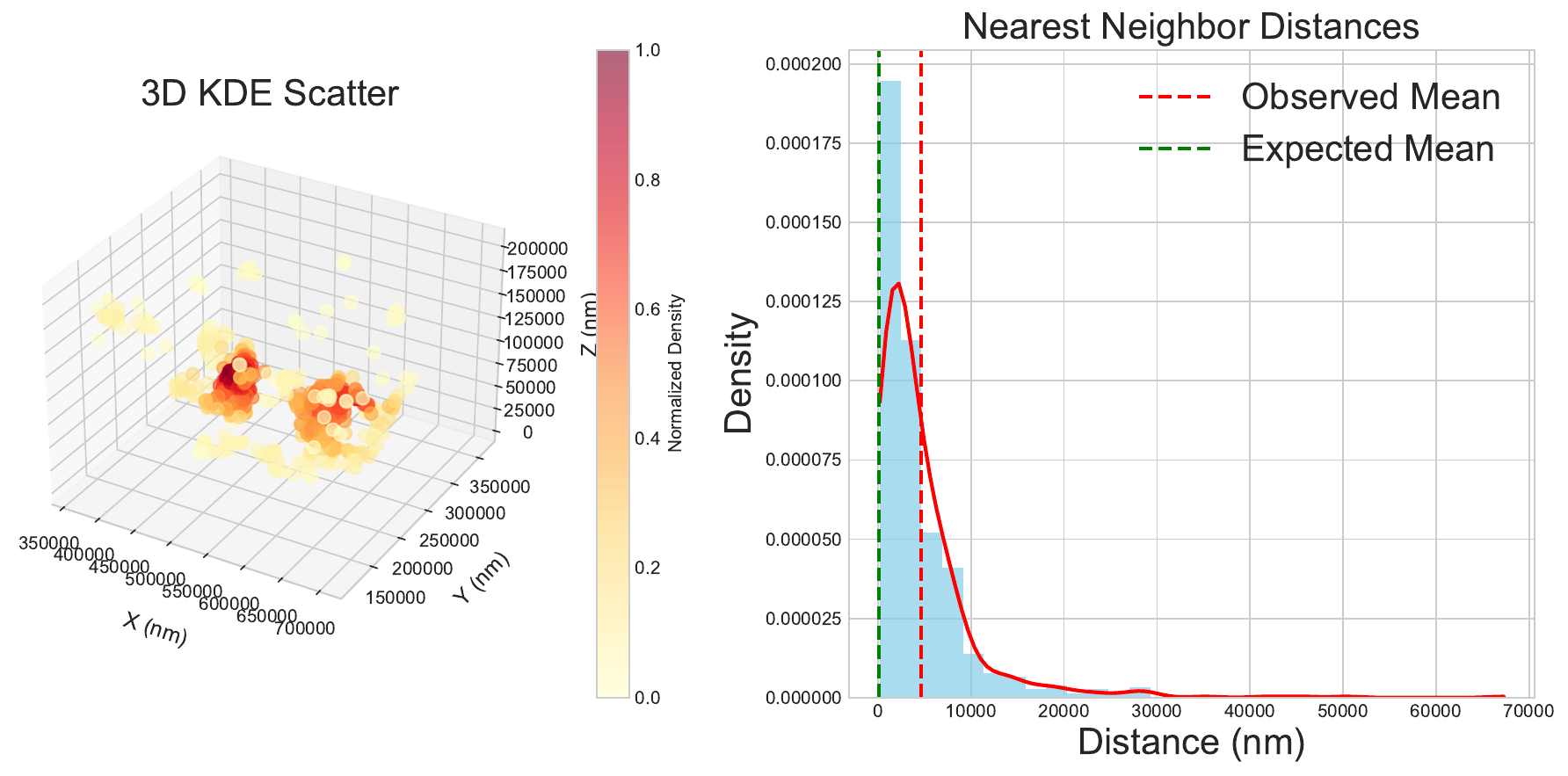}}  
	\hspace{-0.5em}  
	\subfloat[ascending]{\includegraphics[width=0.19\textwidth]{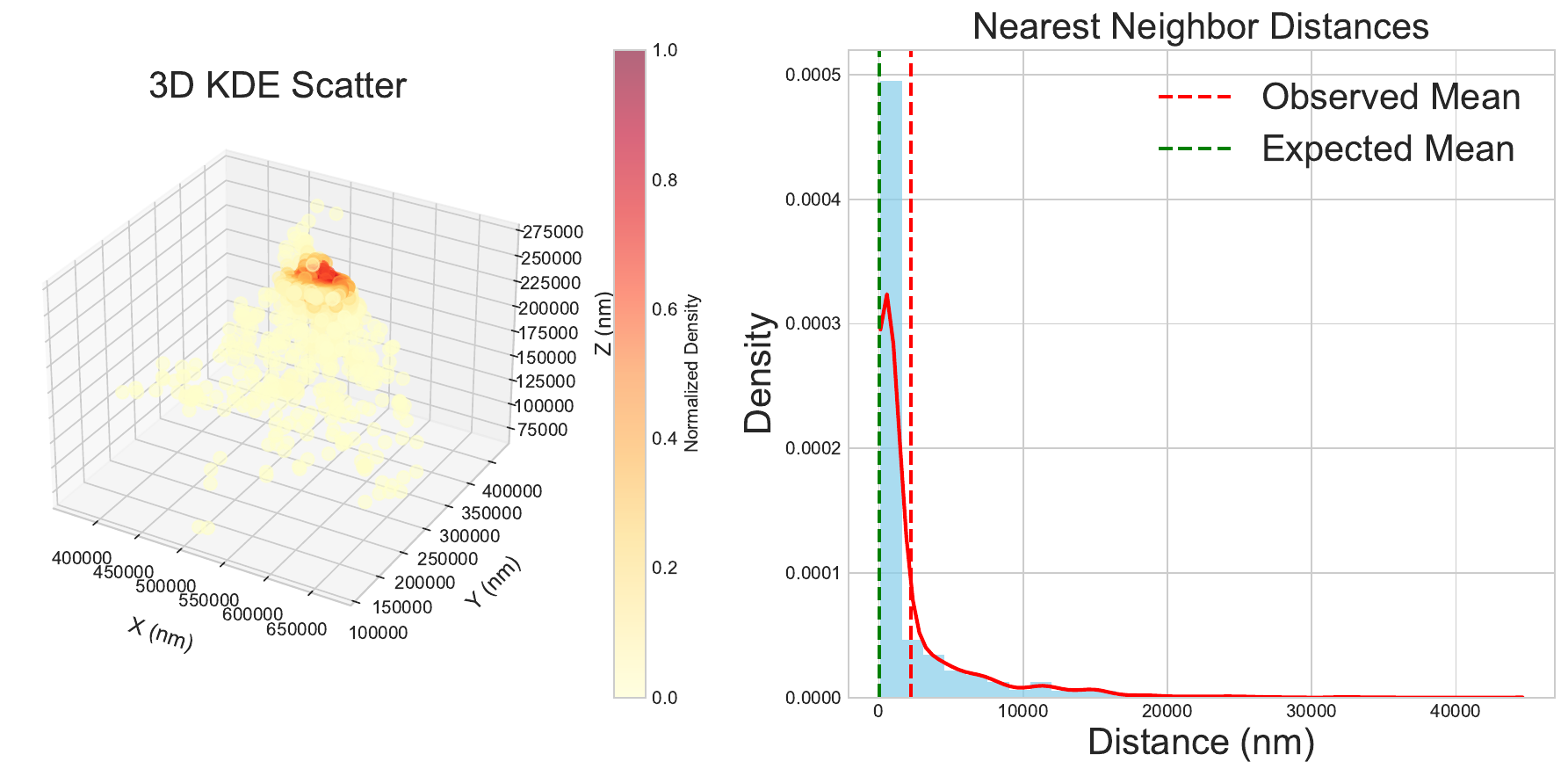}}  
	
	\vspace{-1em}  
	\subfloat[CX]{\includegraphics[width=0.19\textwidth]{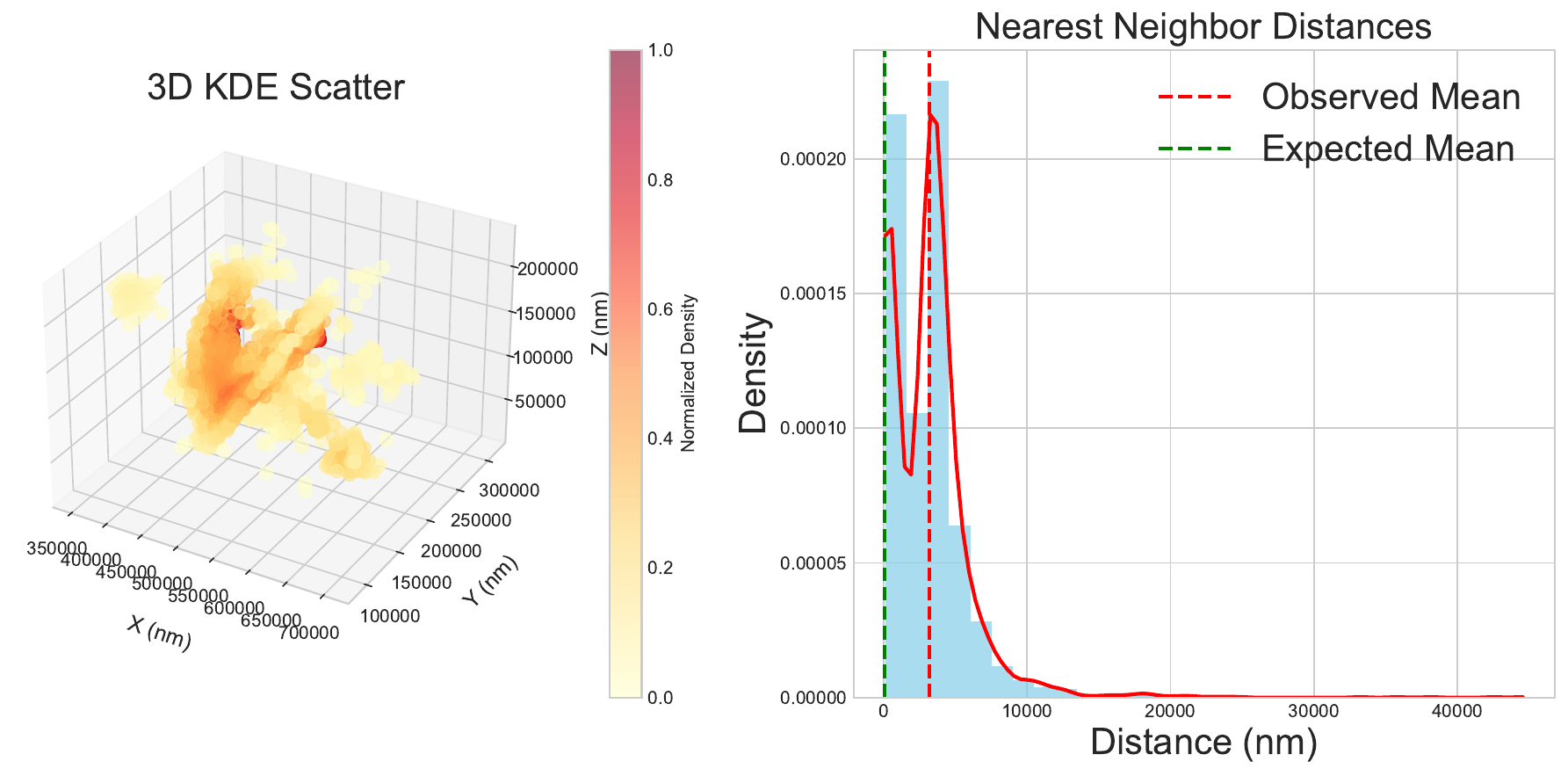}}  
	\hspace{-0.5em}  
	\subfloat[DAN]{\includegraphics[width=0.19\textwidth]{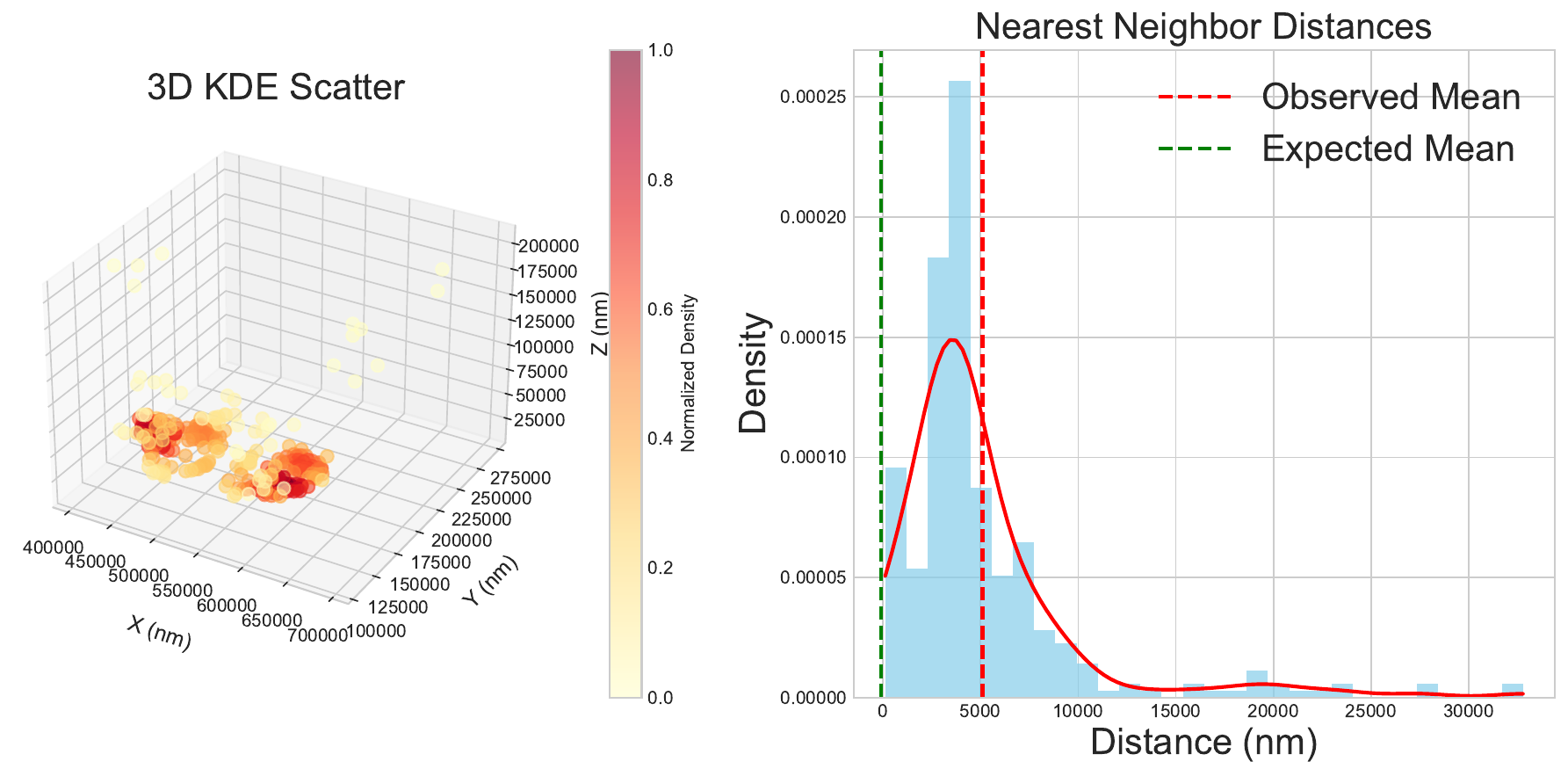}}  
	\hspace{-0.5em}  
	\subfloat[visual projection]{\includegraphics[width=0.19\textwidth]{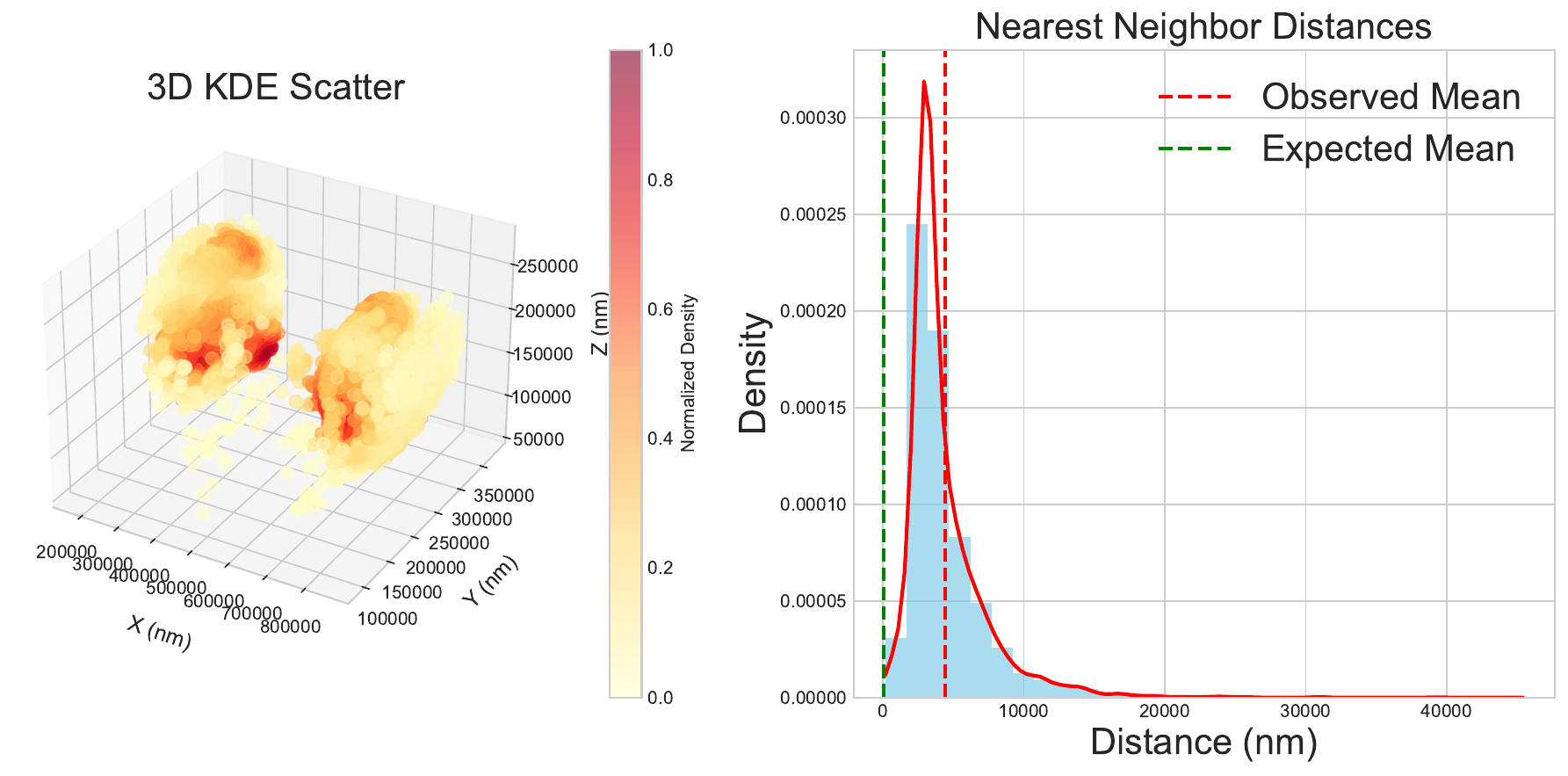}}  
	\hspace{-0.5em}  
	\subfloat[DN]{\includegraphics[width=0.19\textwidth]{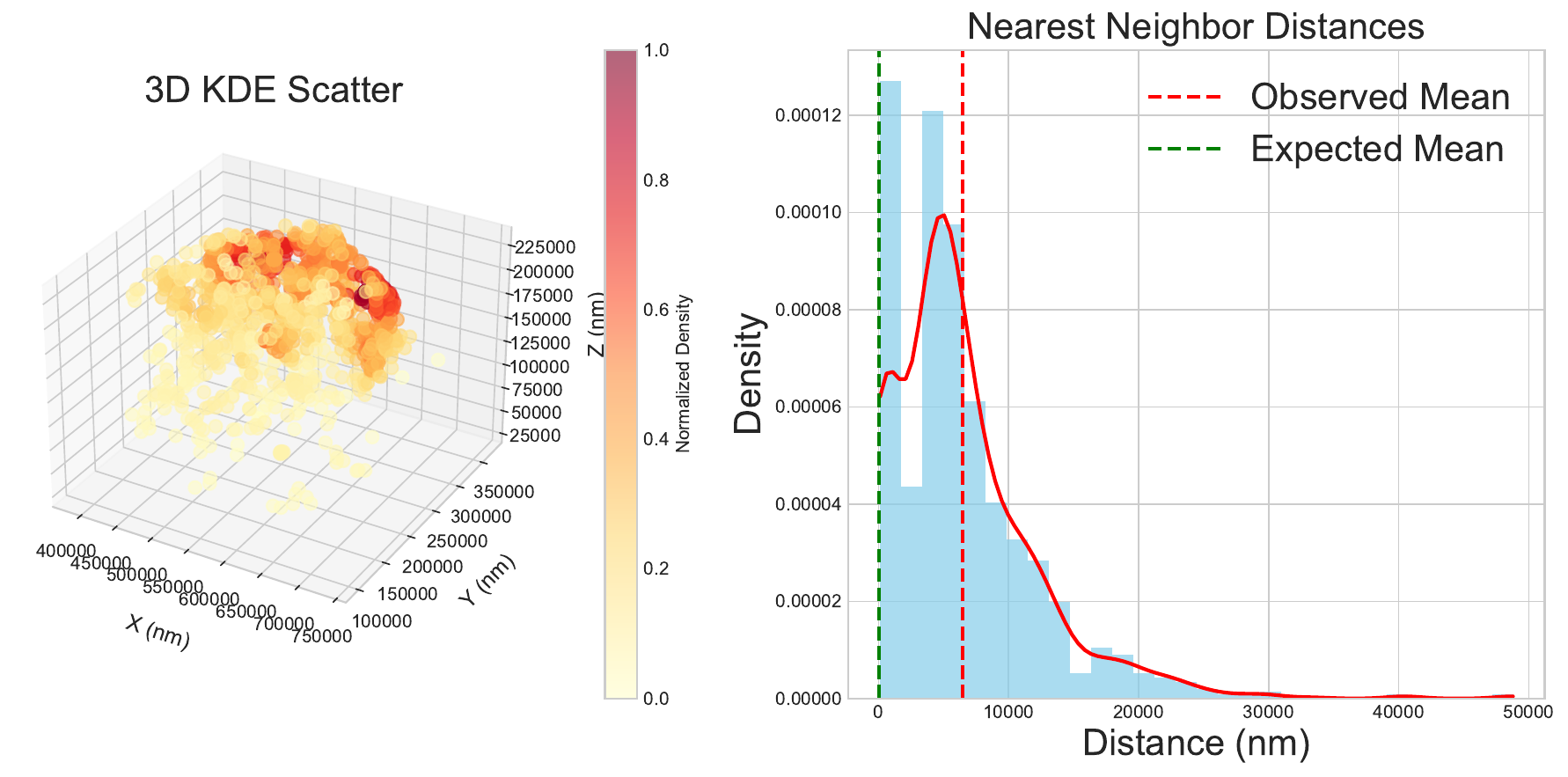}}  
	\hspace{-0.5em}  
	\subfloat[gustatory]{\includegraphics[width=0.19\textwidth]{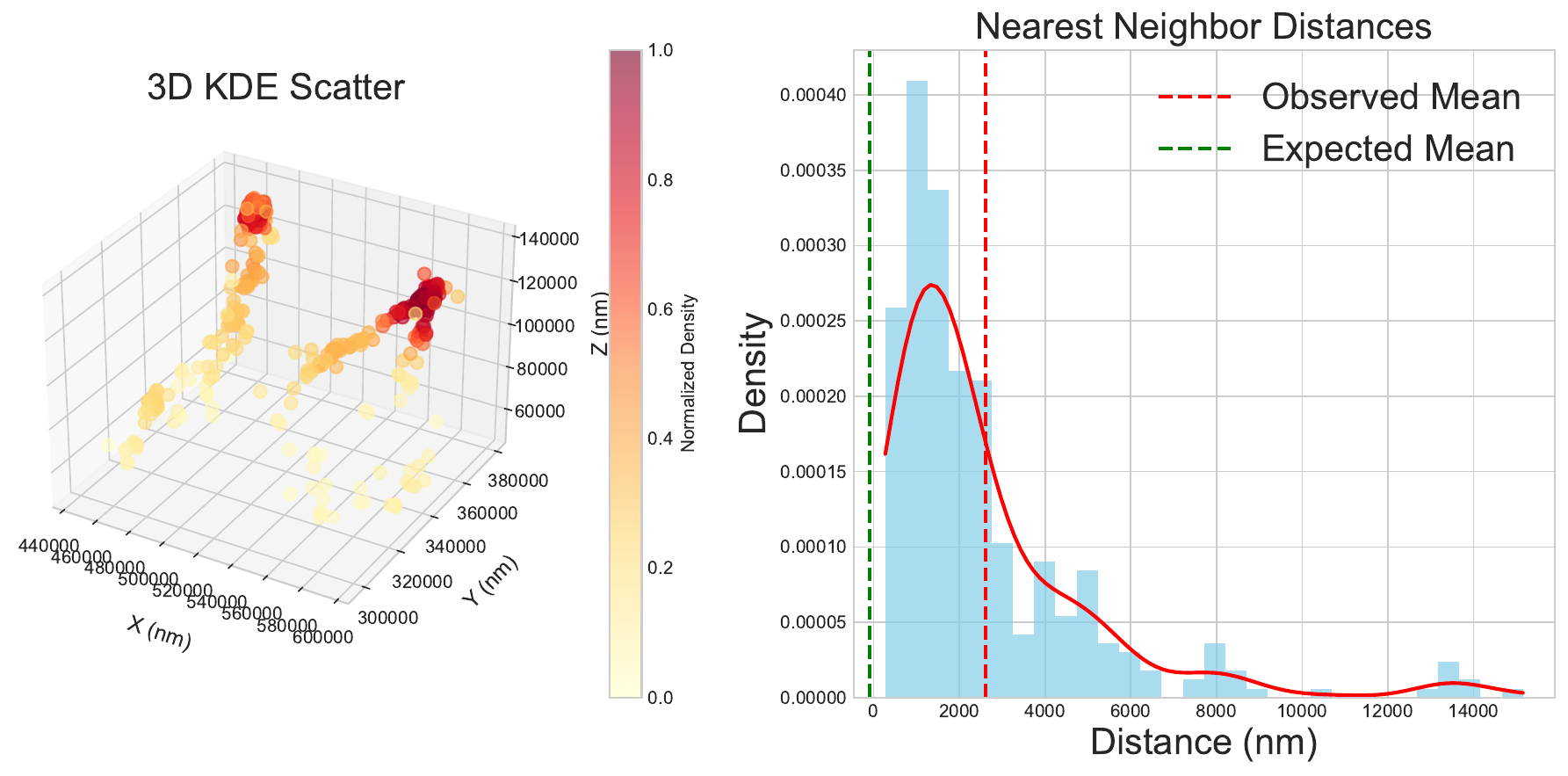}}  
	
	\vspace{-1em}  
	\subfloat[hygrosensory]{\includegraphics[width=0.19\textwidth]{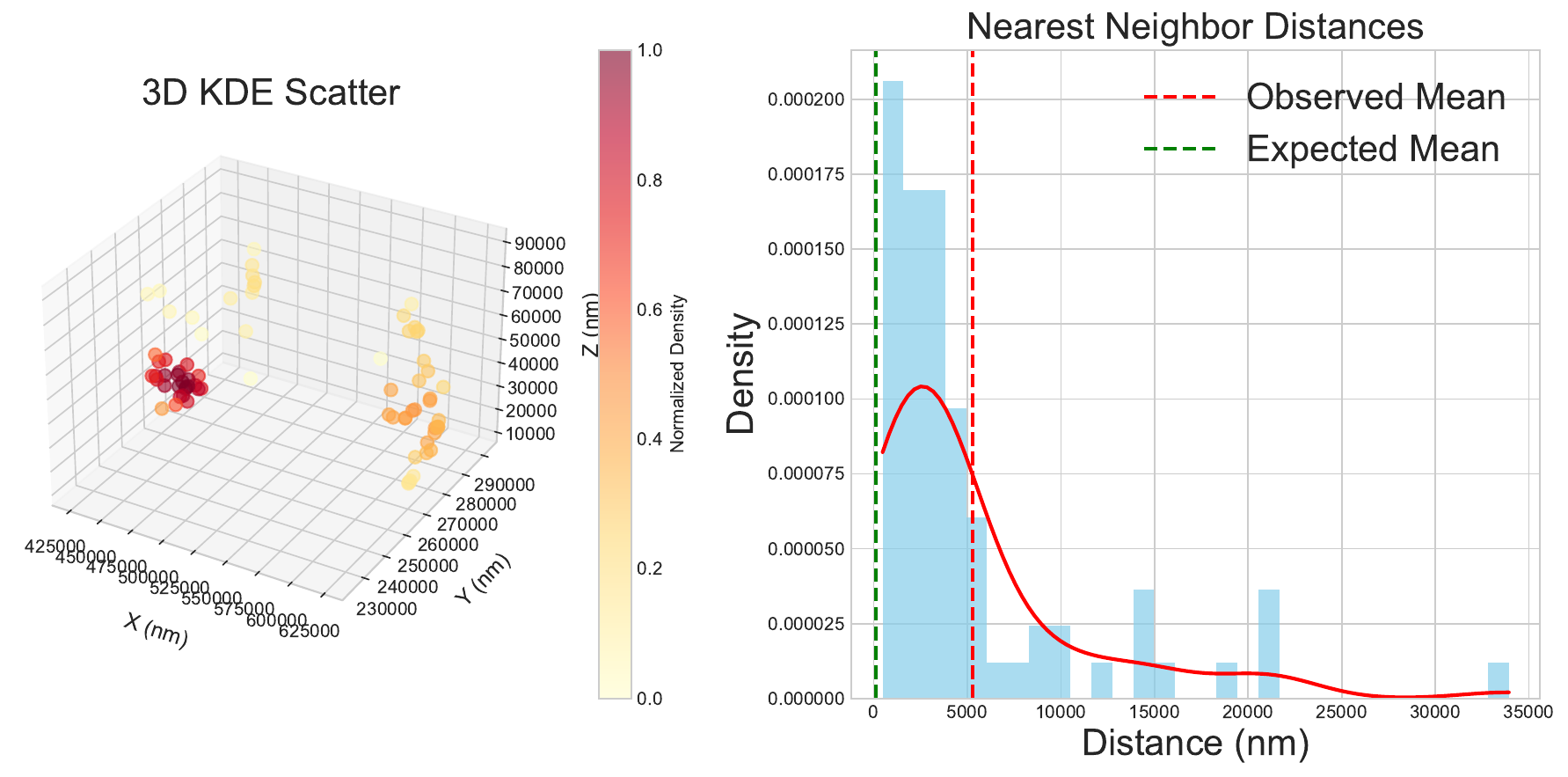}}  
	\hspace{-0.5em}  
	\subfloat[Kenyon Cell]{\includegraphics[width=0.19\textwidth]{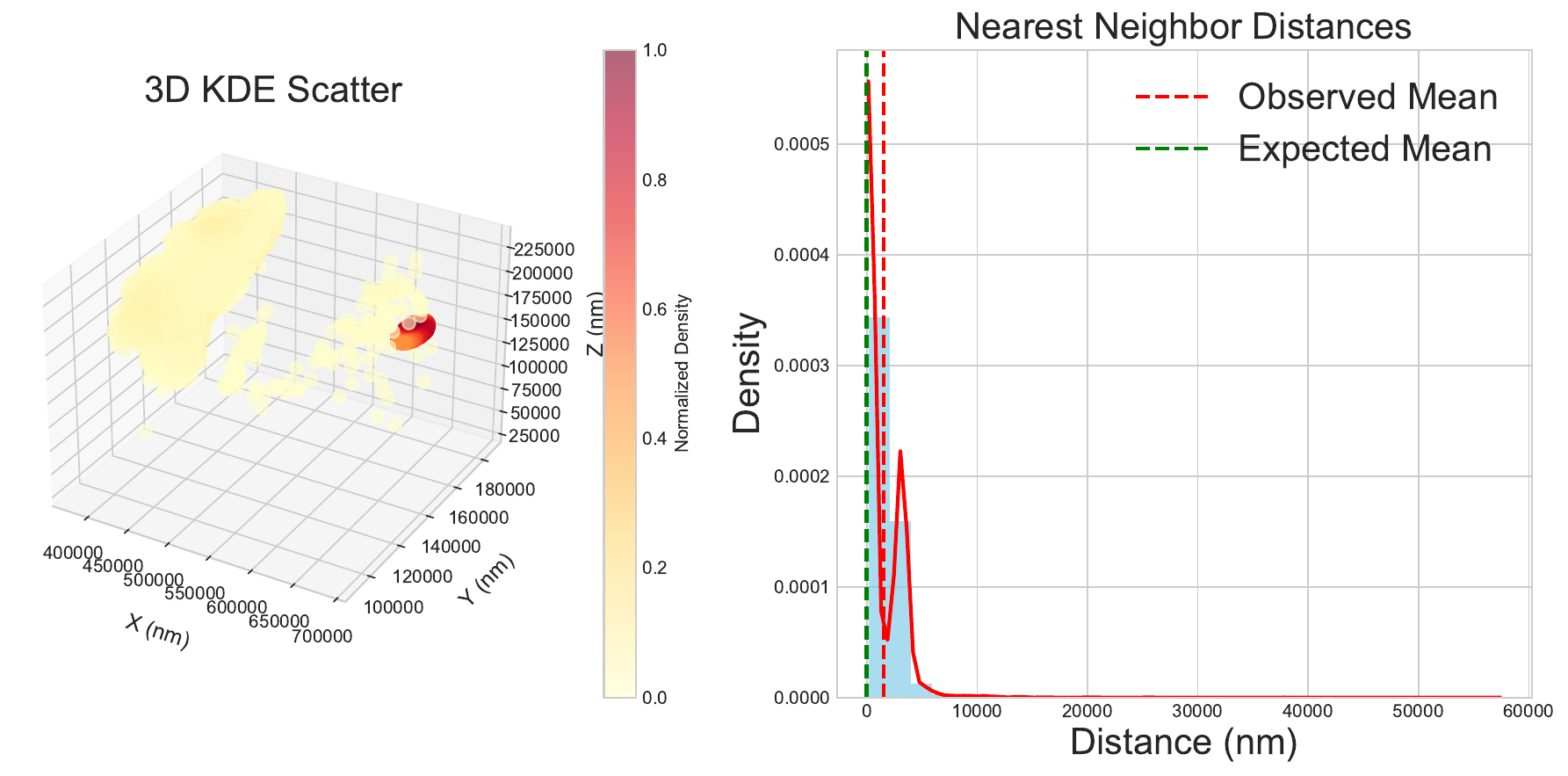}}  
	\hspace{-0.5em}  
	\subfloat[LHCENT]{\includegraphics[width=0.19\textwidth]{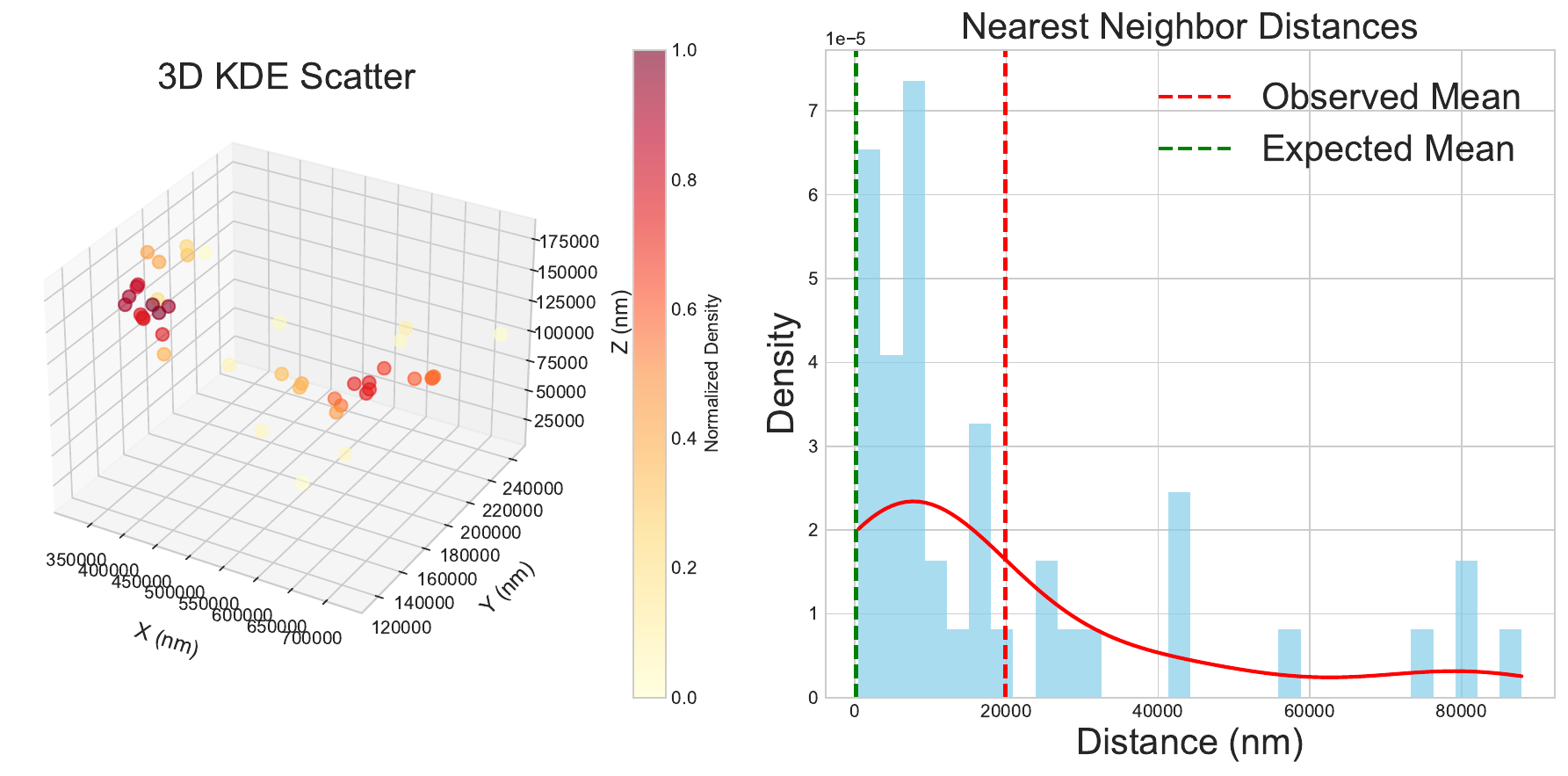}}  
	\hspace{-0.5em}  
	\subfloat[LHLN]{\includegraphics[width=0.19\textwidth]{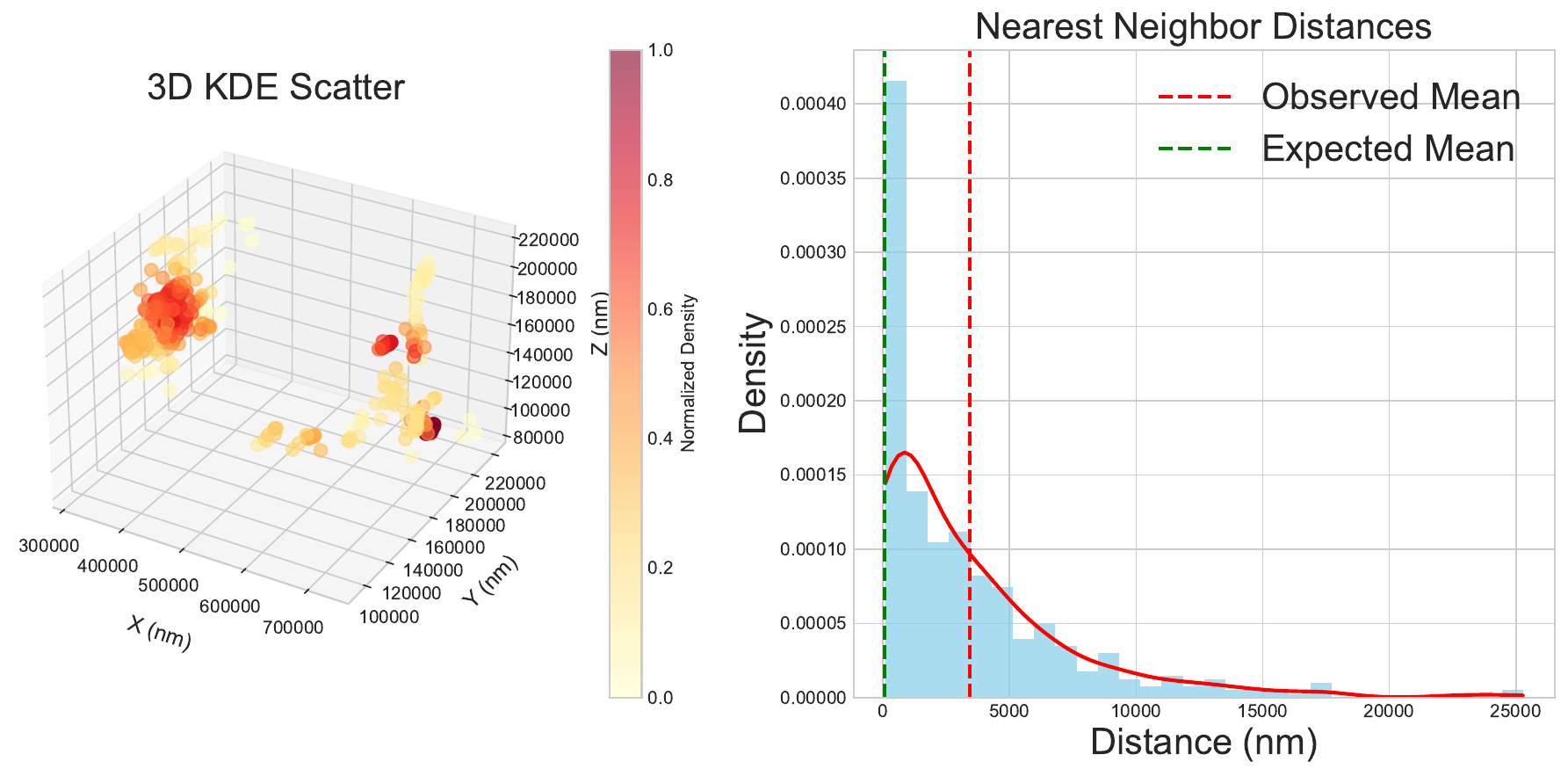}}  
	\hspace{-0.5em}  
	\subfloat[mAL]{\includegraphics[width=0.19\textwidth]{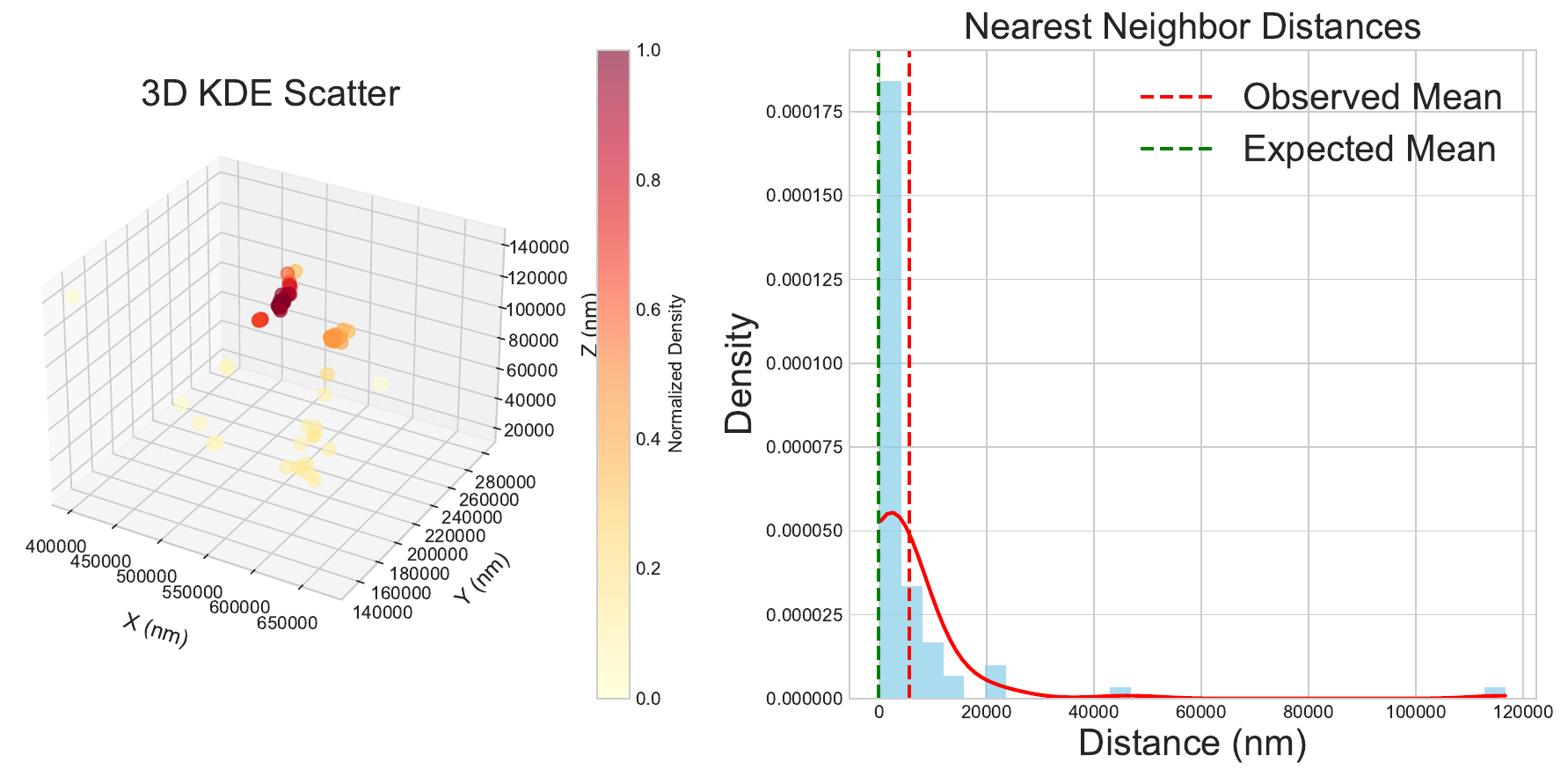}}  
	
	\vspace{-1em}  
	\subfloat[MBIN]{\includegraphics[width=0.19\textwidth]{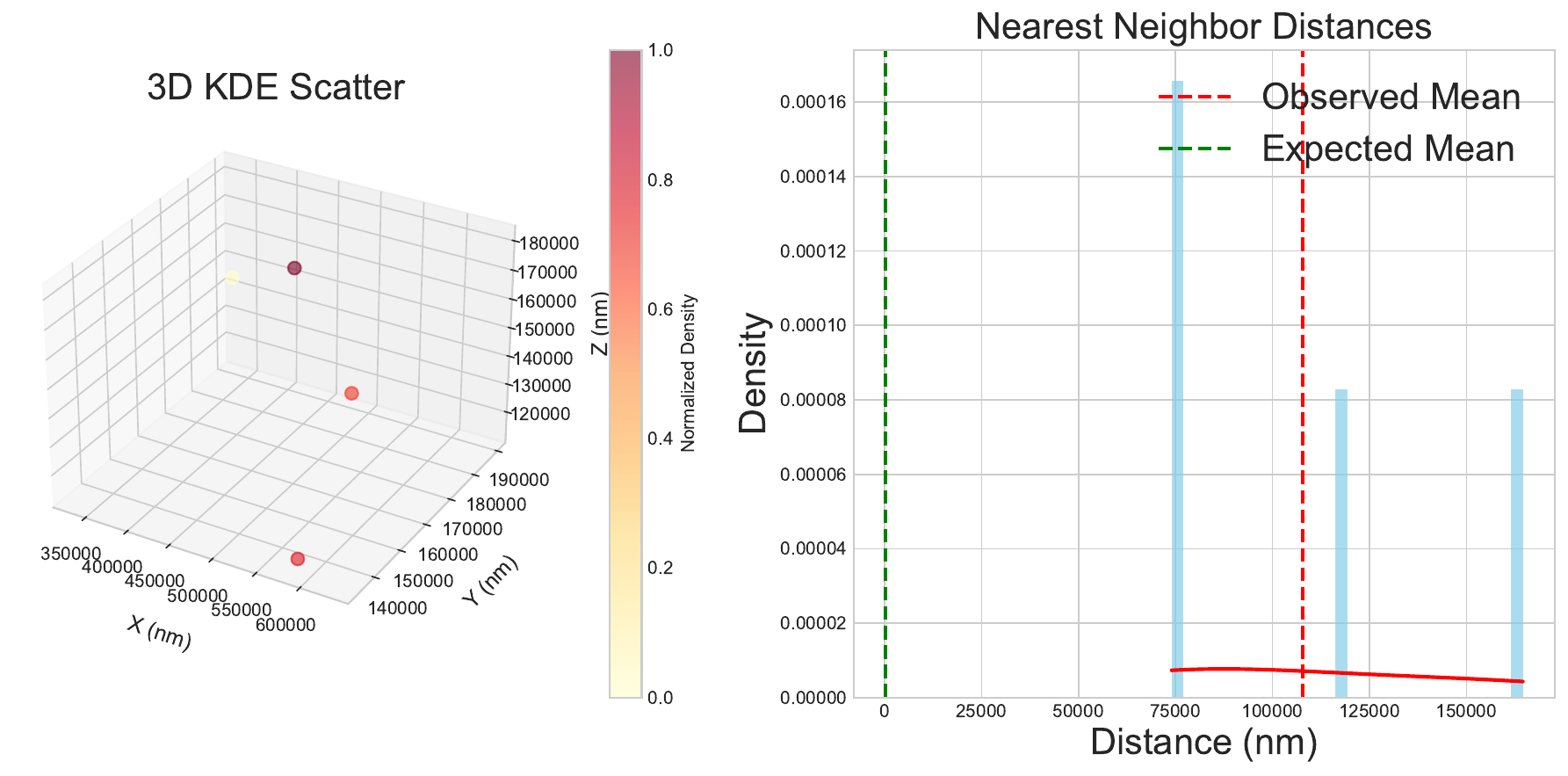}}  
	\hspace{-0.5em}  
	\subfloat[MBON]{\includegraphics[width=0.19\textwidth]{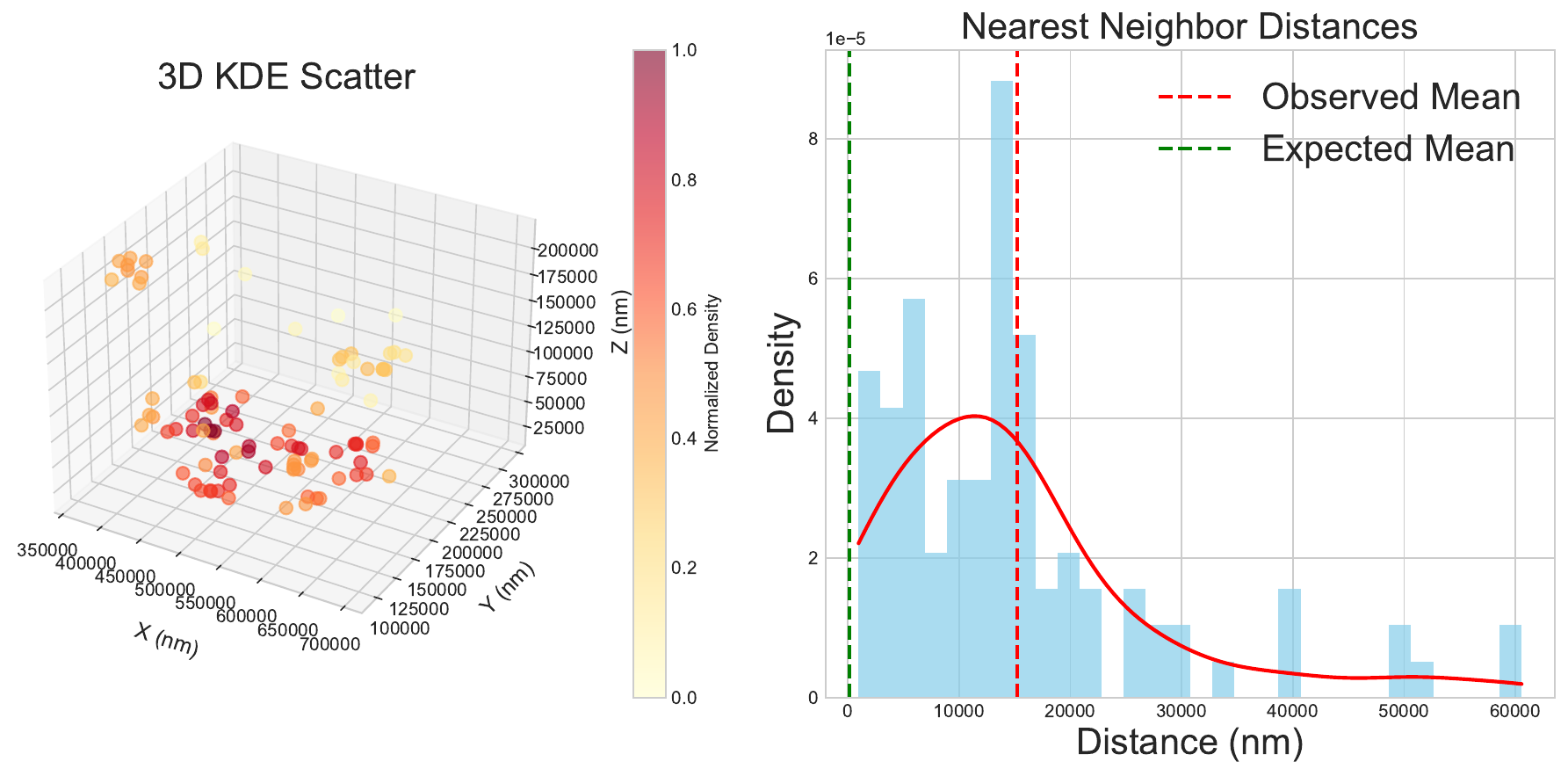}}  
	\hspace{-0.5em}  
	\subfloat[mechanosensory]{\includegraphics[width=0.19\textwidth]{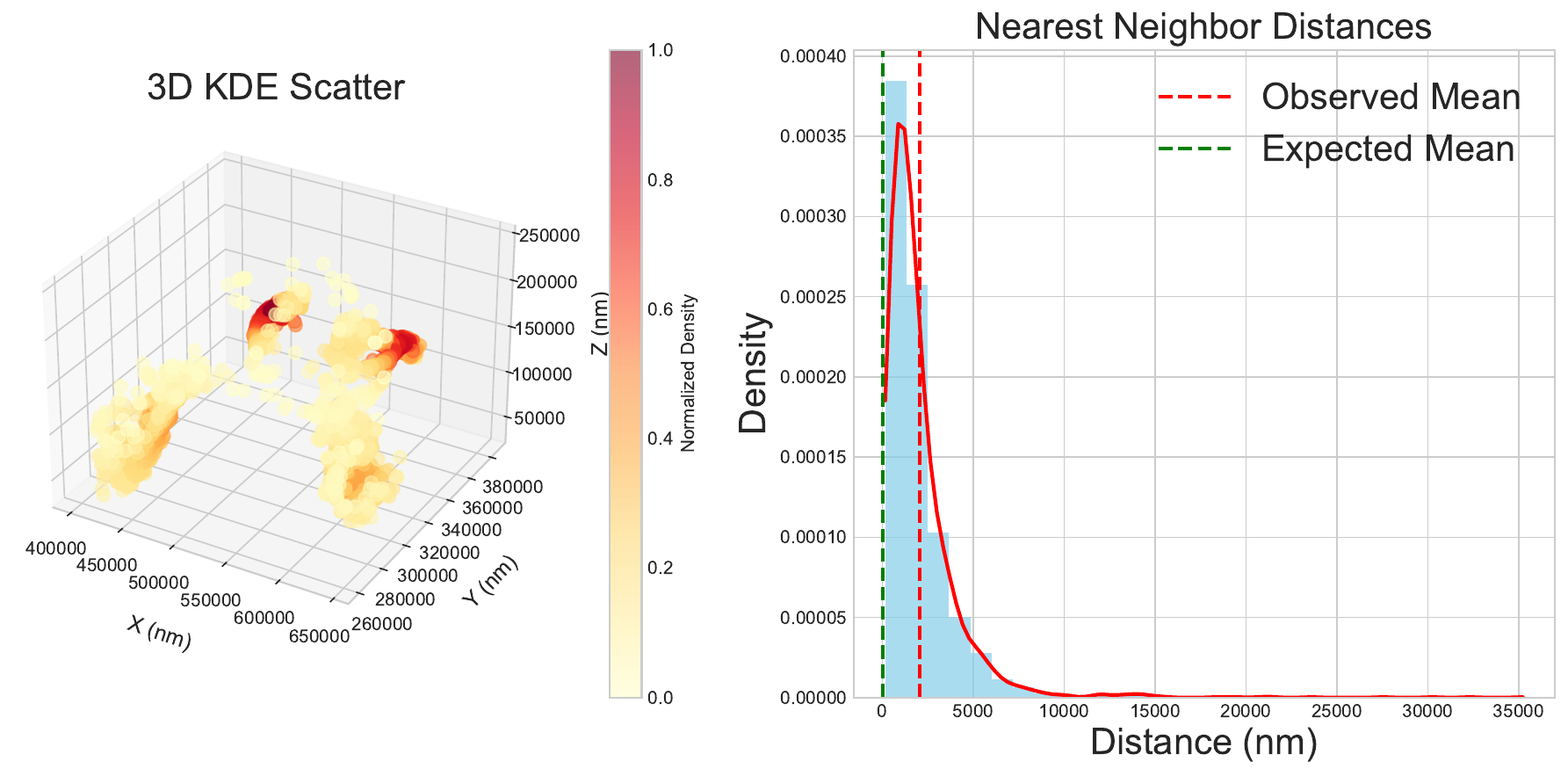}}  
	\hspace{-0.5em}  
	\subfloat[motor]{\includegraphics[width=0.19\textwidth]{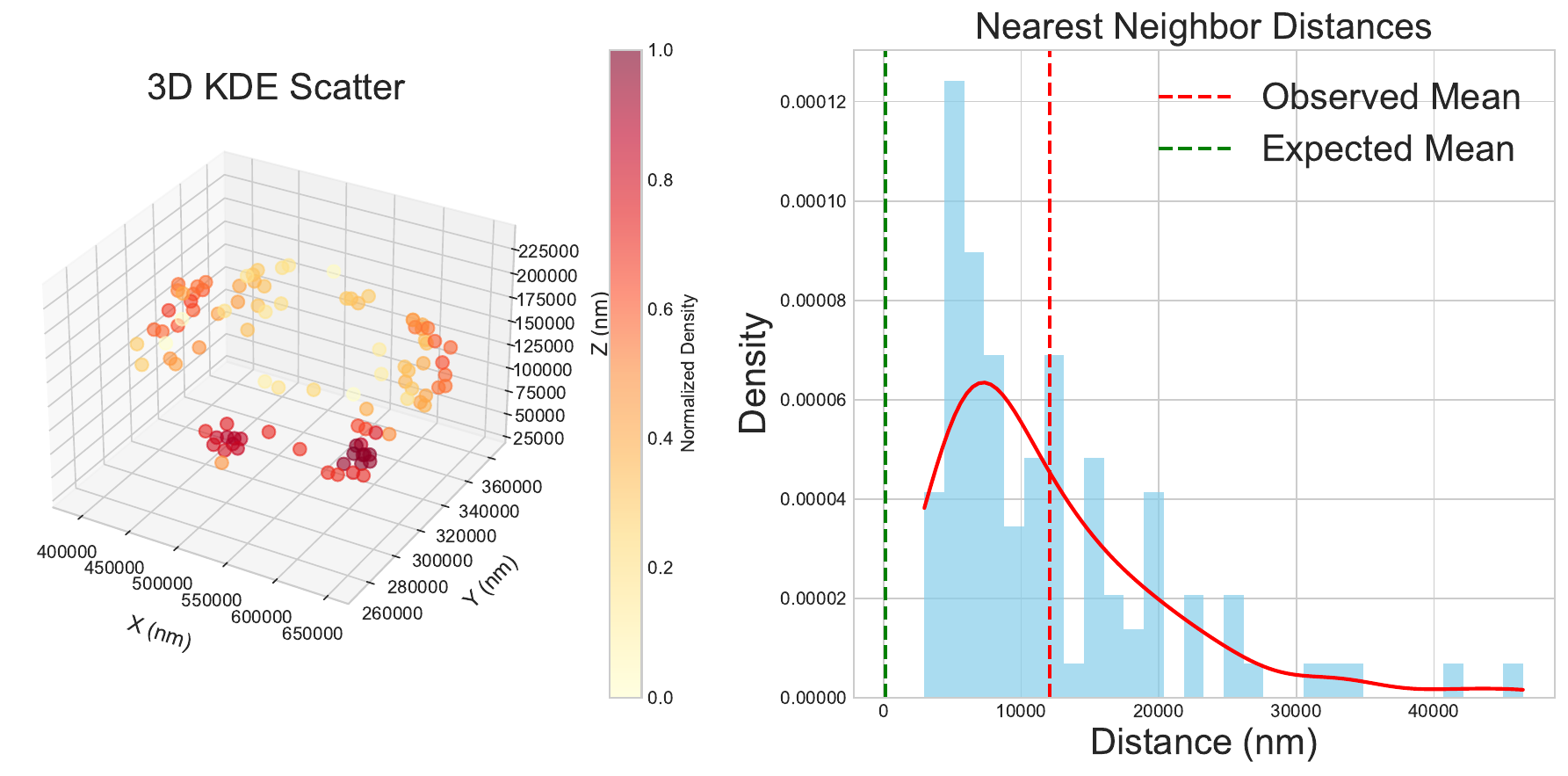}}  
	\hspace{-0.5em}  
	\subfloat[olfactory sensory]{\includegraphics[width=0.19\textwidth]{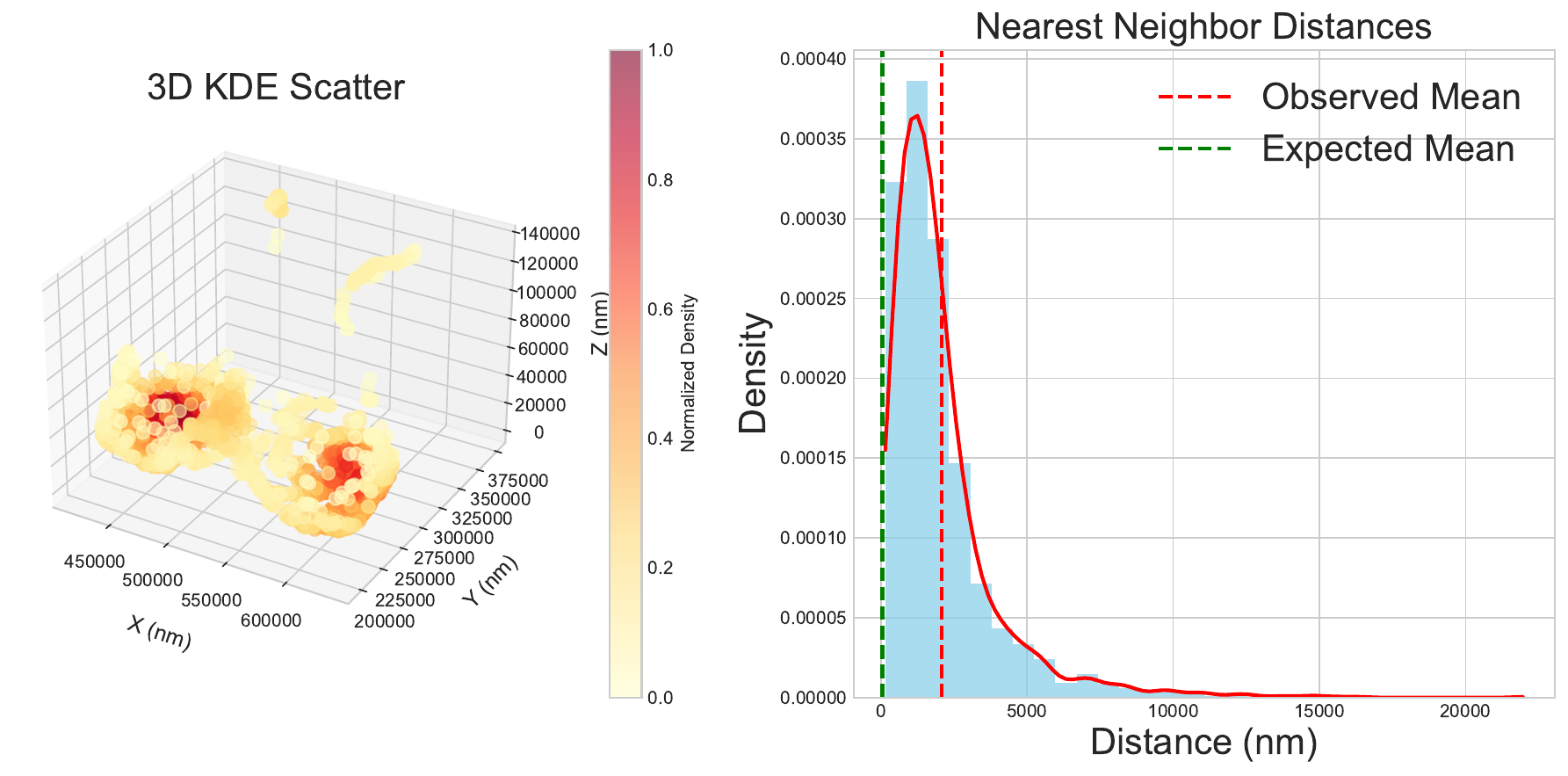}}  
	
	\vspace{-1em}  
	\subfloat[ocellar]{\includegraphics[width=0.19\textwidth]{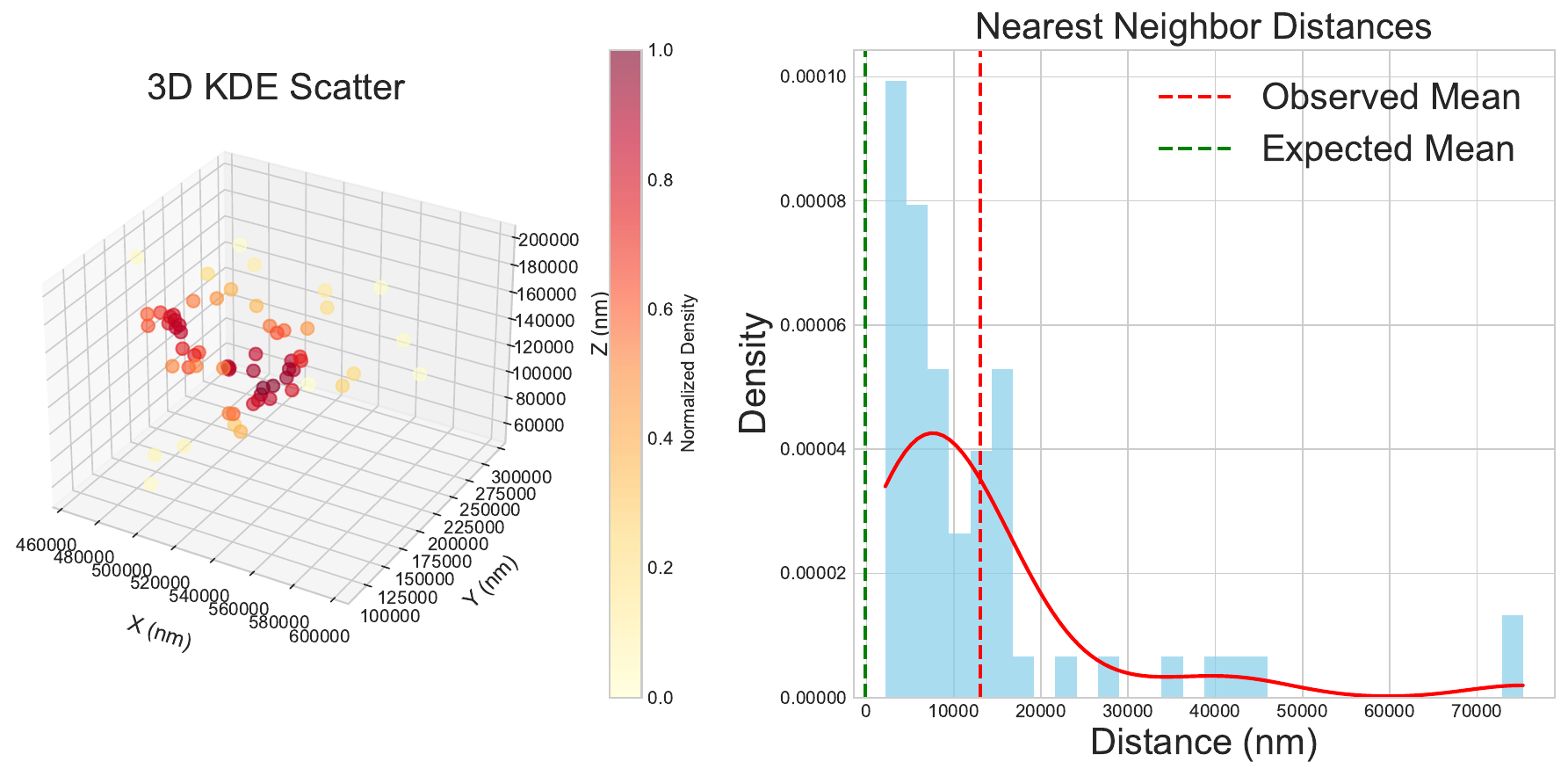}}  
	\hspace{-0.5em}  
	\subfloat[optic]{\includegraphics[width=0.19\textwidth]{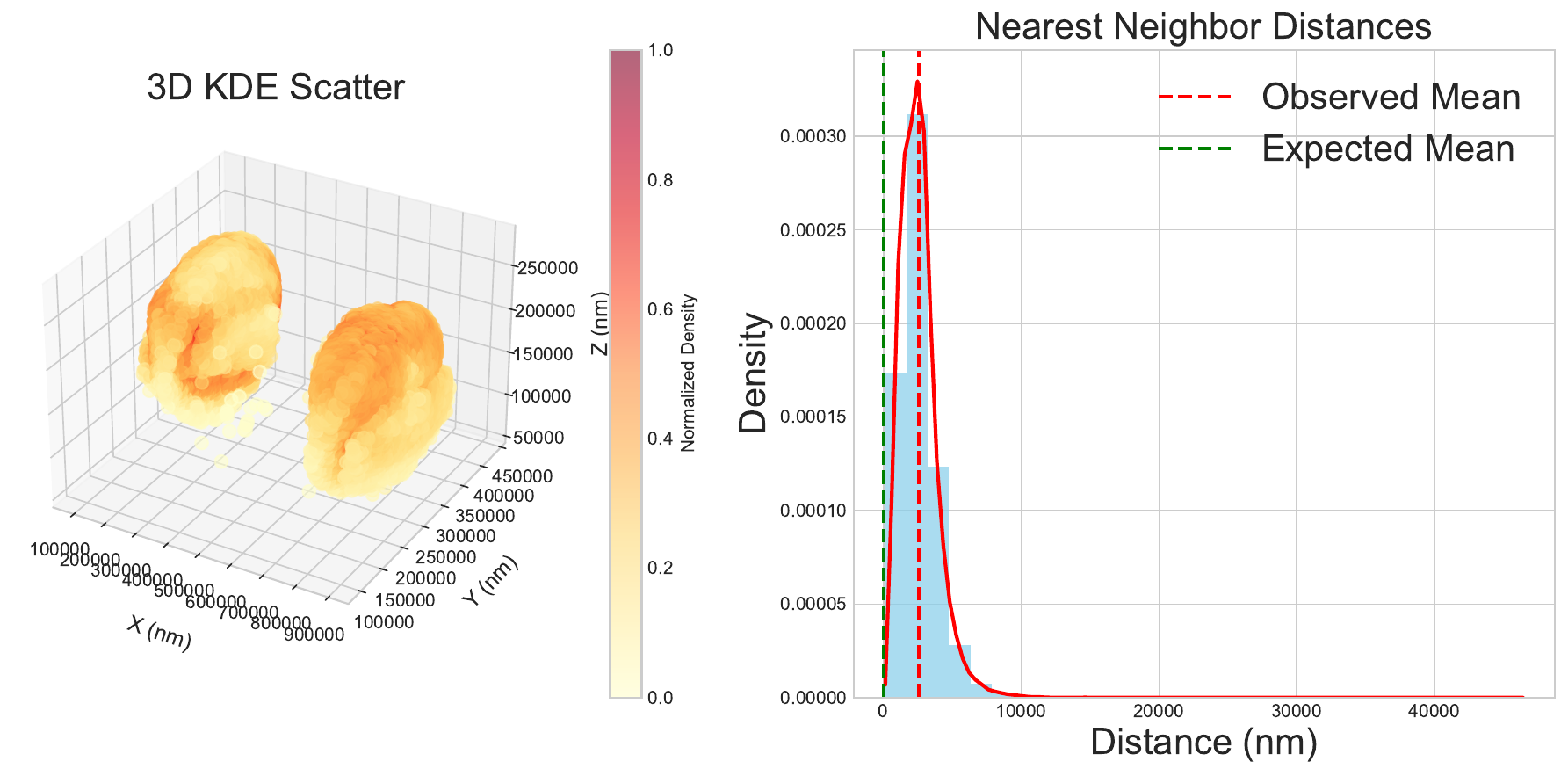}}  
	\hspace{-0.5em}  
	\subfloat[thermosensory]{\includegraphics[width=0.19\textwidth]{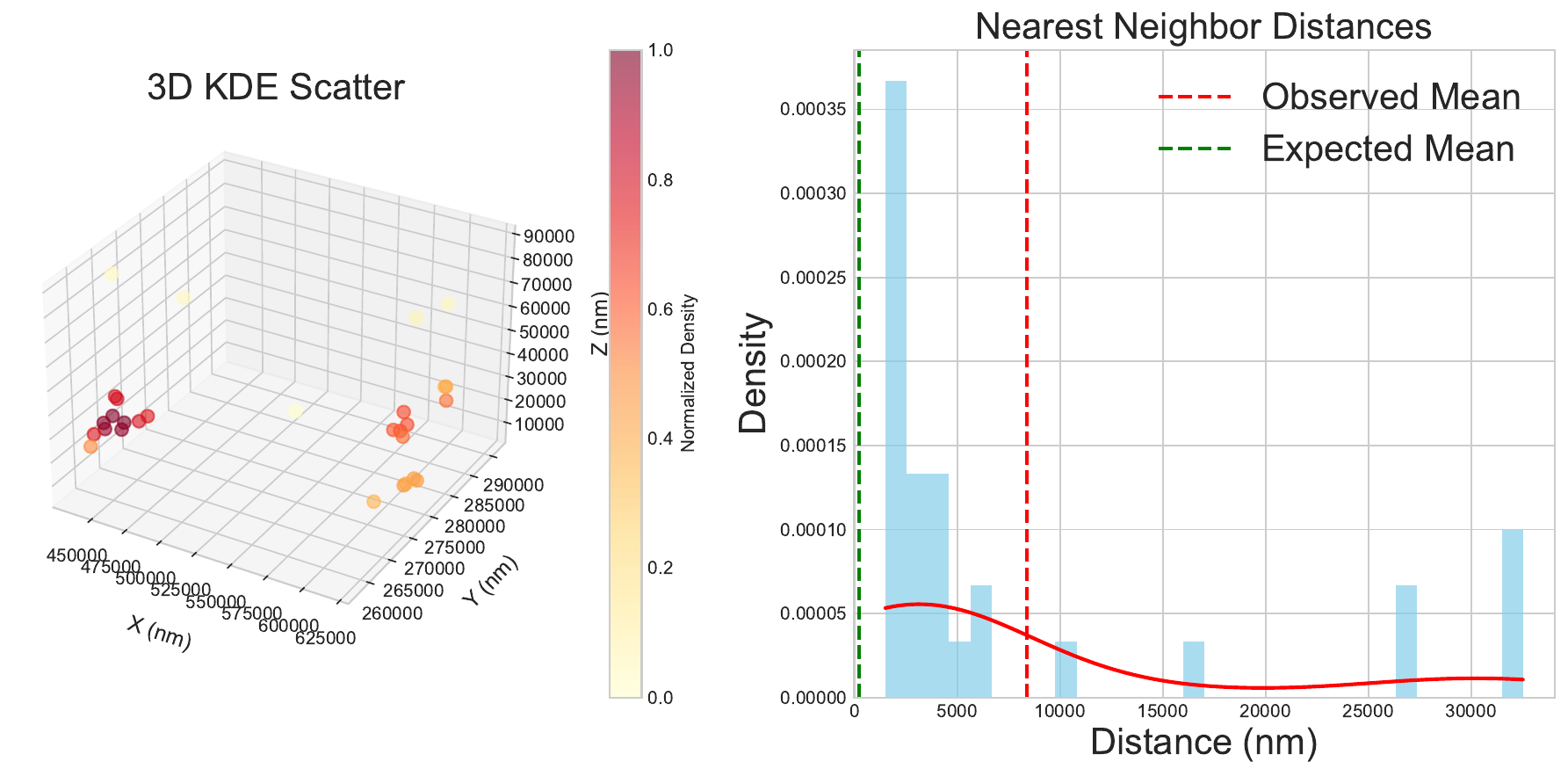}}  
	\hspace{-0.5em}  
	\subfloat[TuBu]{\includegraphics[width=0.19\textwidth]{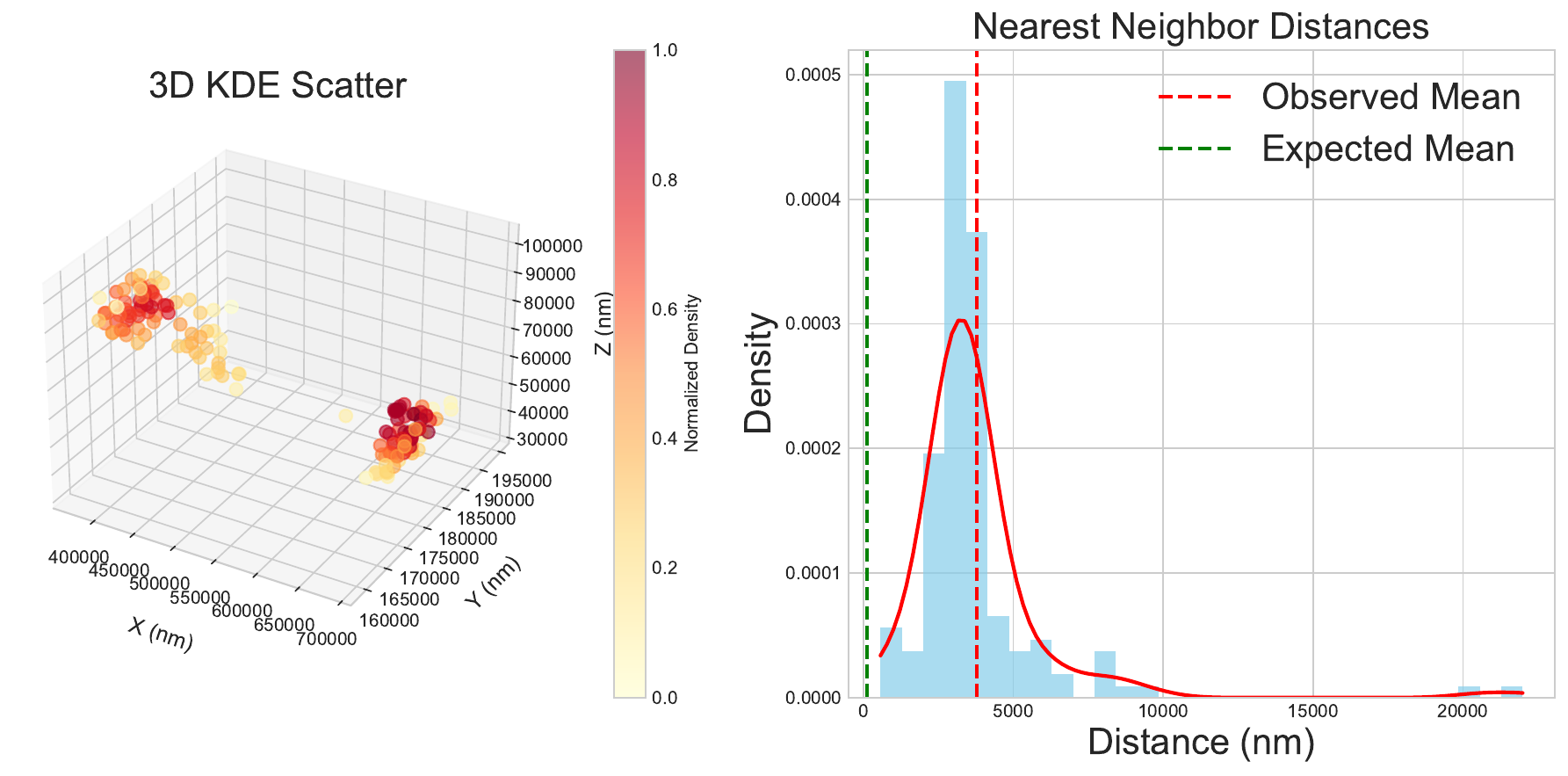}}  
	\hspace{-0.5em}  
	\subfloat[visual sensory]{\includegraphics[width=0.19\textwidth]{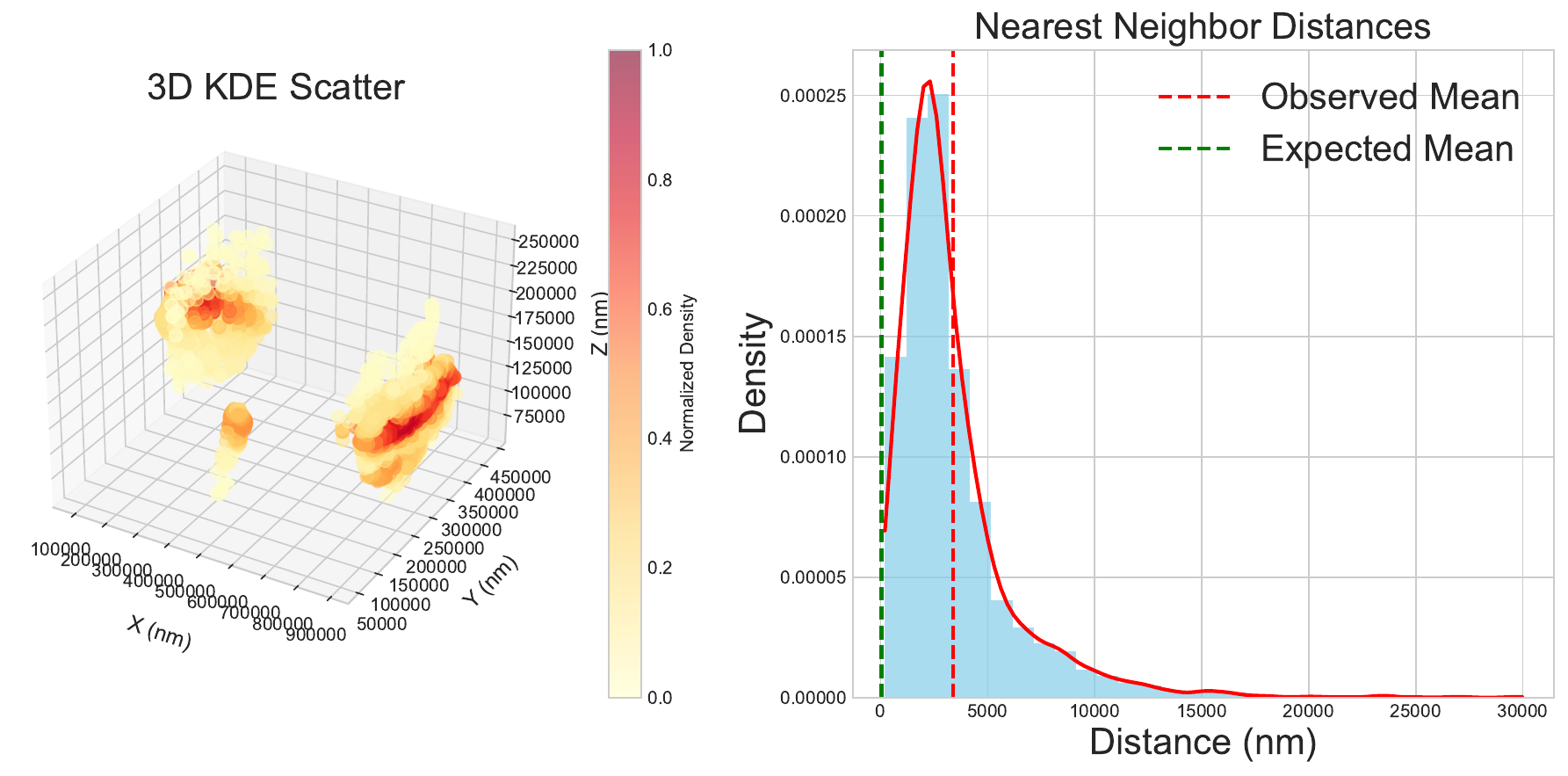}}  
	
	\vspace{-1em}  
	\subfloat[visual centrifugal]{\includegraphics[width=0.19\textwidth]{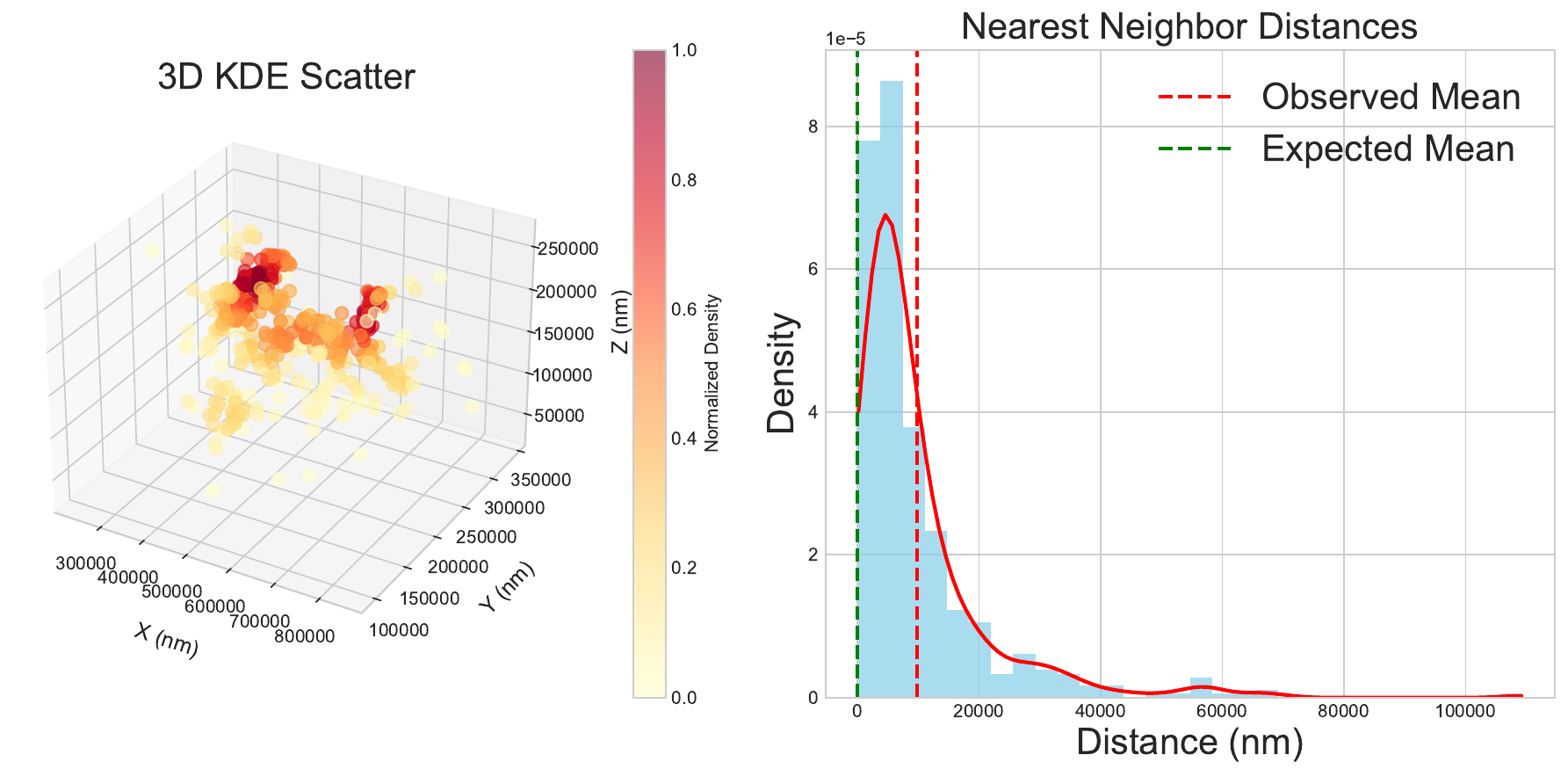}}  
	\hspace{-0.5em}  
	\subfloat[gustatory]{\includegraphics[width=0.19\textwidth]{./figures/spatial_distribution/gustatory_spatial_analysis_3d.pdf}}  
	\caption{Spatial agglomeration patterns across diverse regions}  
	\label{spatial_region}  
\end{figure*}

In order to further examine the spatial distribution patterns across specific functional regions, we present spatial clustering images for all regions (Fig. \ref{spatial_region}). These images elucidate the intra-regional variations in spatial agglomeration and spatial concentration patterns. All regions exhibited significant spatial agglomeration patterns, indicating that neurons within the same functional regions demonstrate a strong tendency for spatial clustering. However, notable heterogeneity in spatial aggregation was observed between bilateral hemispheres, particularly in the hygrosensory, Kenyon Cell, and thermosensory regions.

\begin{figure*}  
	\centering  
	\subfloat[]{  
		\begin{minipage}{4cm}  
			\centering  
			\includegraphics[width=\linewidth]{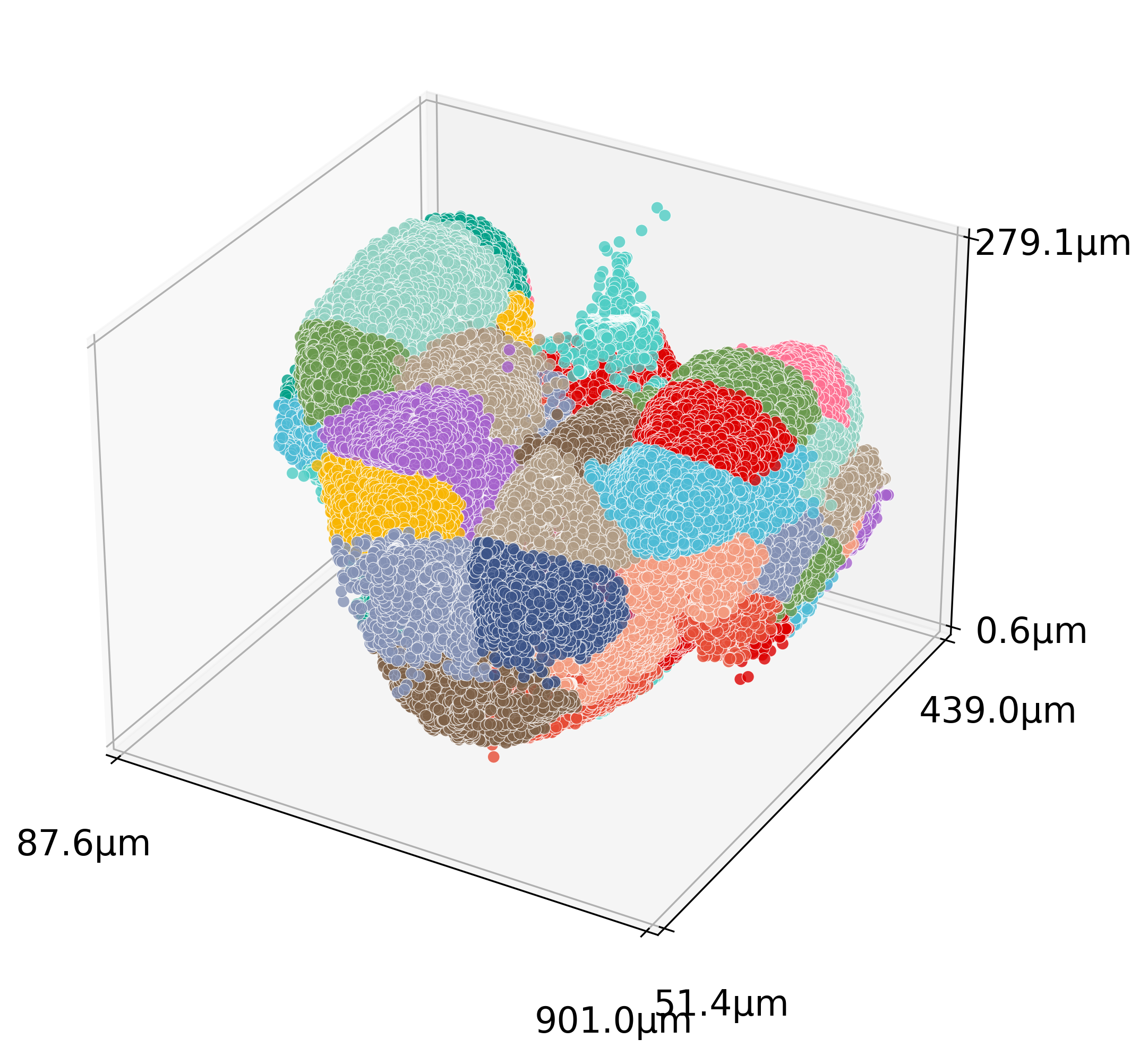}  
		\end{minipage}  
	}  
		\subfloat[]{  
		\begin{minipage}{4cm}  
			\centering  
			\includegraphics[width=\linewidth]{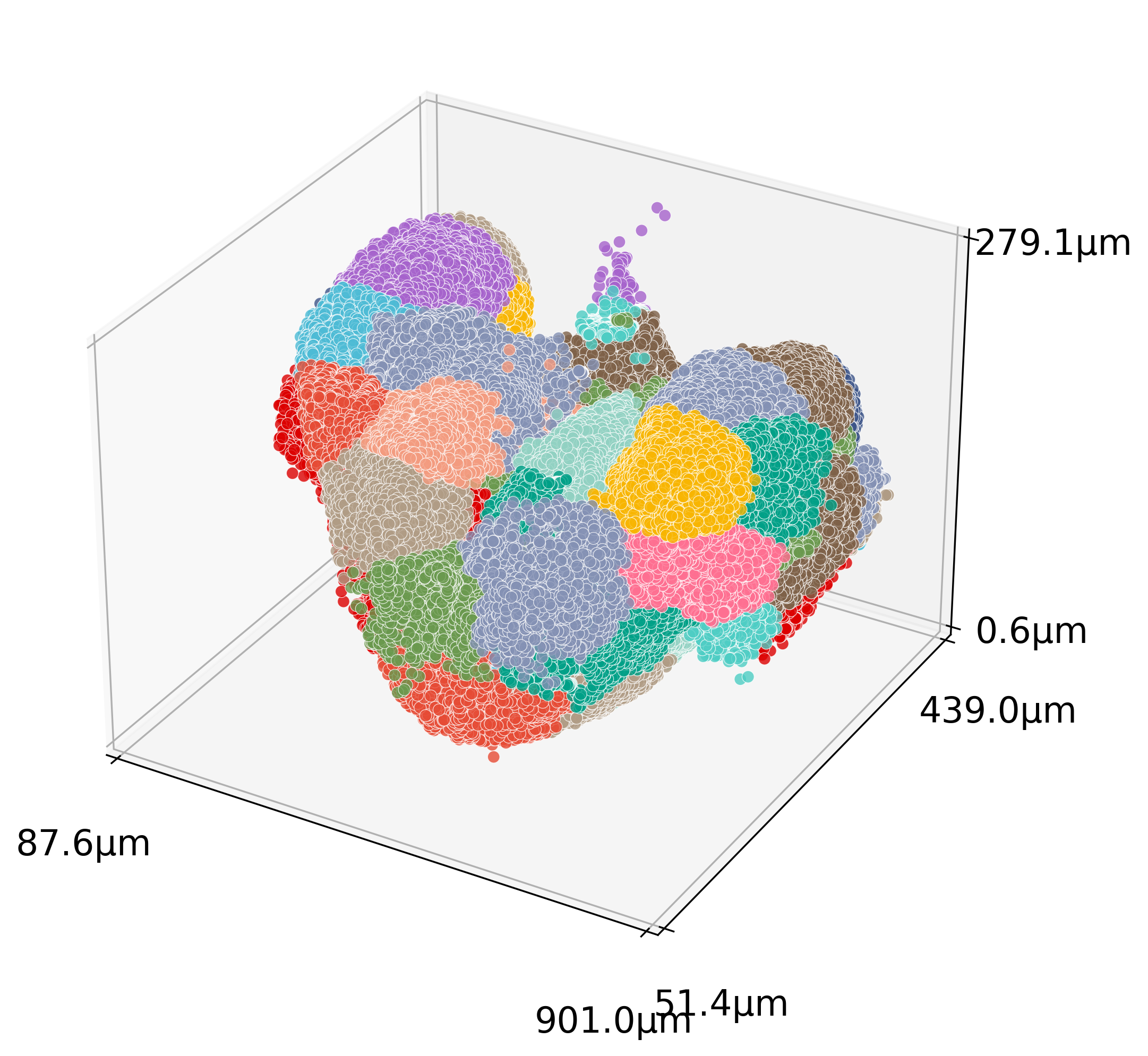}  
		\end{minipage}  
	}  
		\subfloat[]{  
		\begin{minipage}{4cm}  
			\centering  
			\includegraphics[width=\linewidth]{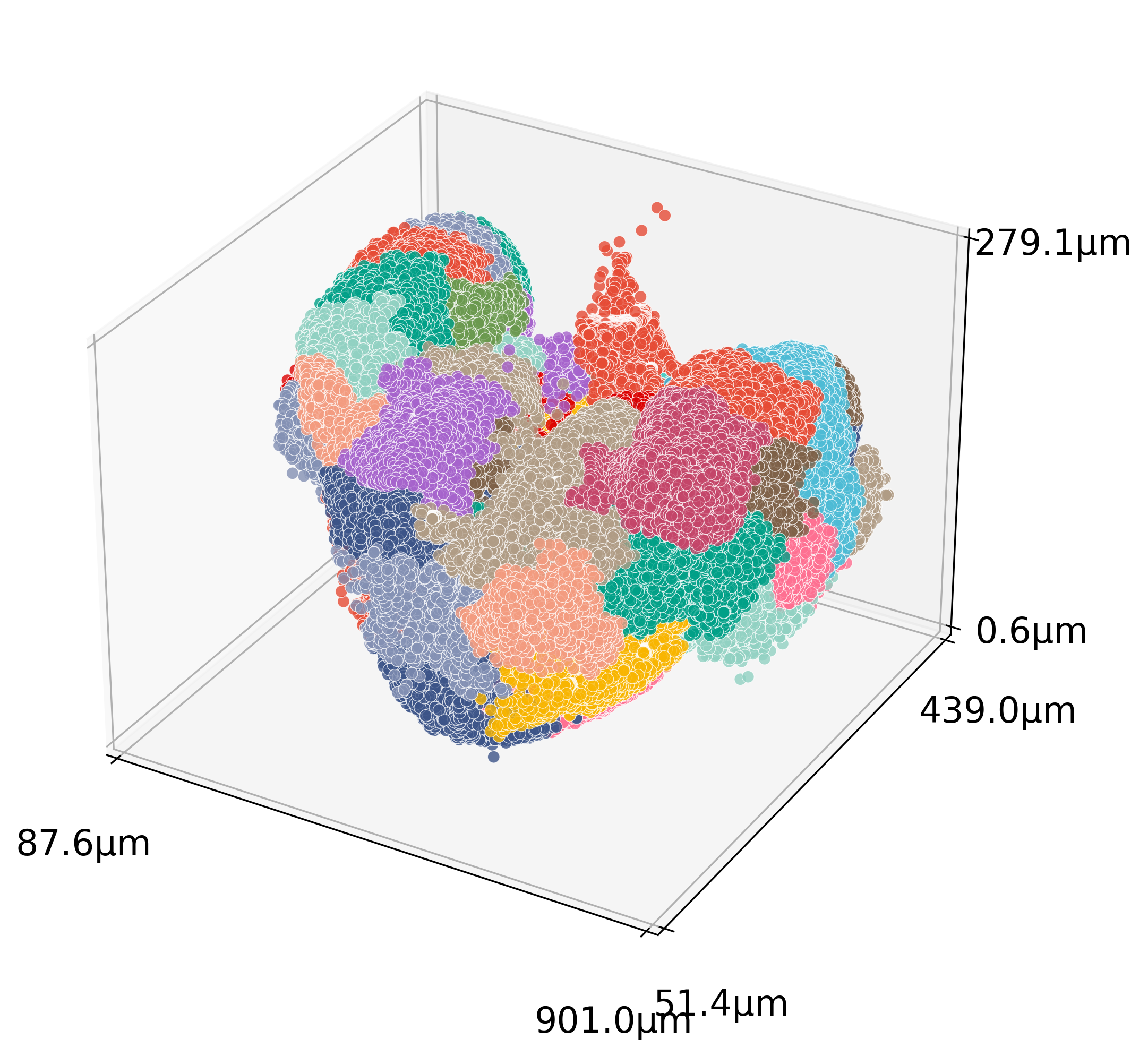}  
		\end{minipage}  
	}  
	\subfloat[]{  
		\begin{minipage}{4cm}  
			\centering  
			\includegraphics[width=\linewidth]{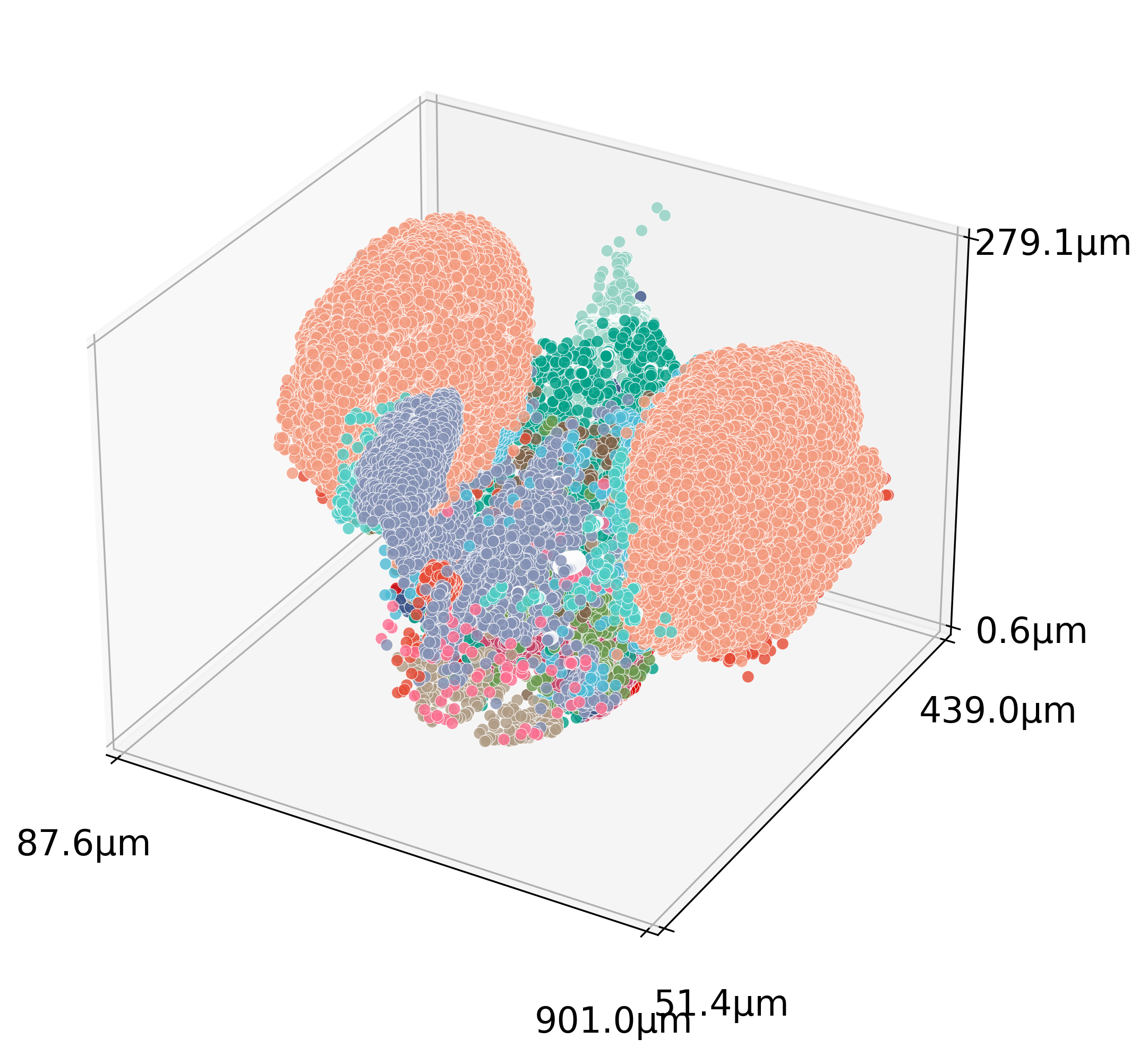}  
		\end{minipage}  
	}  
	\caption{Comparative Analysis of Spatial Clustering and Neuronal Typology. (a), (b) and (c) demonstrate the outcomes of spatial clustering analysis using k-means, spectral clustering, and hierarchical clustering respectively, segregating neurons into 58 distinct categories that numerically correspond to the established neuronal classifications. (d) Portrays the results of neuronal classification, employing a color-coded scheme to differentiate various neuronal subtypes. }
\label{kmeans}
\end{figure*}

Further, to investigate the correlation between the pervasive spatial clustering observed in cerebral networks and neuronal typology, we implemented the K-means clustering\cite{lloyd1982least}, Spectral clustering\cite{shi2000normalized} and Hierarchical clustering algorithm\cite{murtagh2012algorithms} to categorize the entire neuronal population into 58 distinct clusters, corresponding to the total number of identified neuronal types. Fig. \ref{kmeans} presents a comparative analysis between the outcomes of spatial clustering algorithms applied to the entire neuronal population and the actual neuronal classifications. Fig. \ref{kmeans}(a) and Fig. \ref{kmeans}(b) elucidates the inefficacy of spatial clustering algorithms in effectively partitioning diverse neuronal types. 
These results underscores the inadequacy of relying solely on spatial attributes for neuronal type discrimination.

\subsection{Validation of Hemispheric Functional Heterogeneity}
As a form of heterogeneity, previous analyses have reported structural symmetry in the left and right part of the Drosophila brain, yet numerous studies have demonstrated functional asymmetry\cite{kong2022mapping,10.1093/molbev/msad181} in neuronal systems. One hypothesis addressing functional heterogeneity suggests that neurons in the left and right hemispheres demonstrate intrinsic differences in their functional patterns\cite{kong2022mapping}. To test this hypothesis, we employed the Kuramoto model (for details, please refer to Appendix \ref{kuramoto}) to simulate neuronal activity, where the intrinsic frequencies of neurons in the left and right hemispheres were initialized using Gaussian distributions with distinct means and variances. Following random initialization, we conducted iterations over an extended time period and recorded the temporal evolution of the synchronization rate.

\begin{figure*}[t]
	\centering
	\subfloat[$\mu_L$=0,$\quad \sigma_L$=0.1,$\quad \mu_R$=0,$\quad \sigma_R$=0.5,global]{  
		\begin{minipage}{0.3\textwidth}  
			\centering  
			\includegraphics[width=\textwidth]{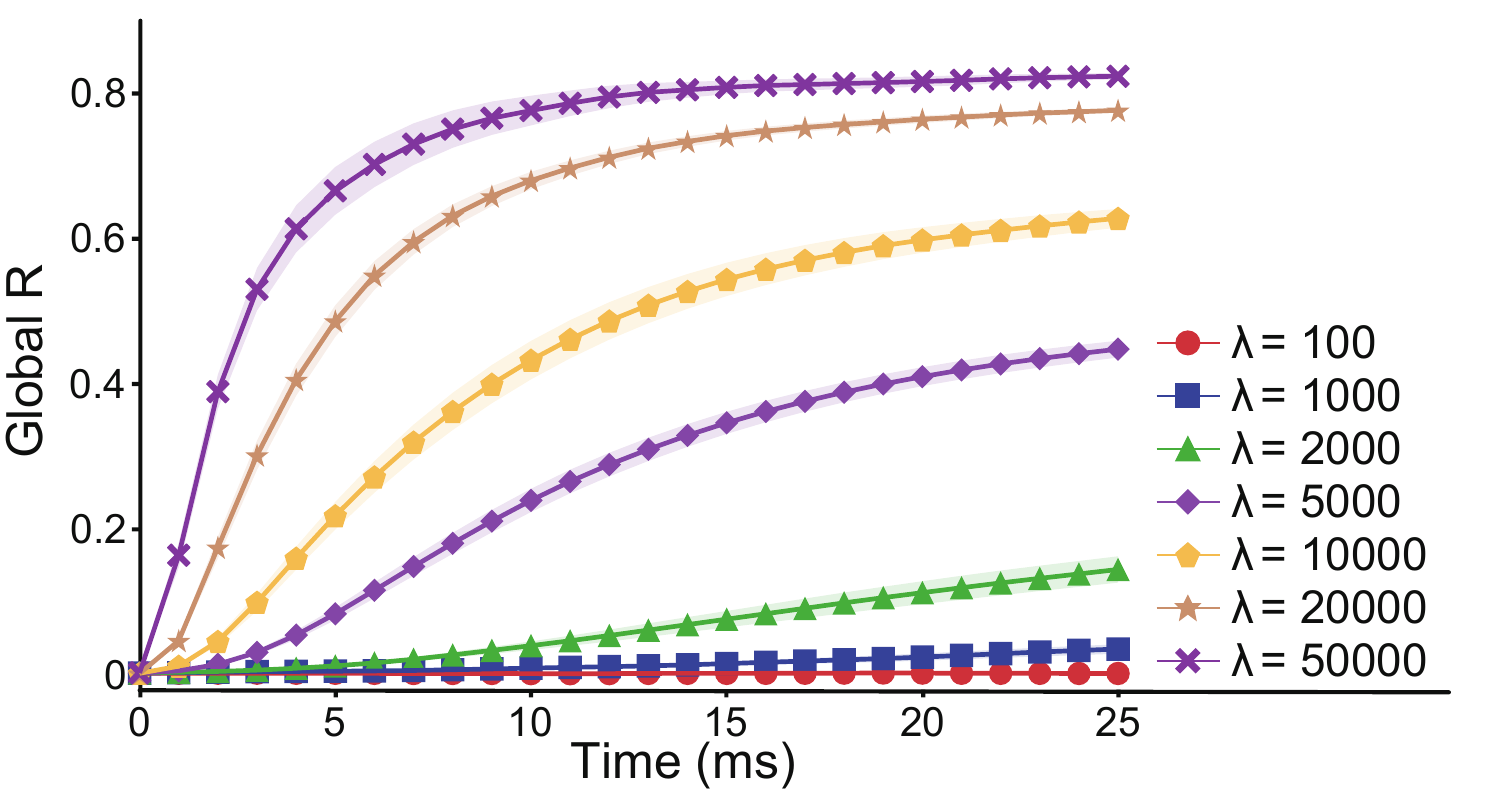}  \end{minipage}  	} 
	\subfloat[$\mu_L$=0,$\quad \sigma_L$=0.5,$\quad \mu_R$=0,$\quad \sigma_R$=0.1,global]{ 		
		\begin{minipage}{0.3\textwidth}  
			\centering  
			\includegraphics[width=\textwidth]{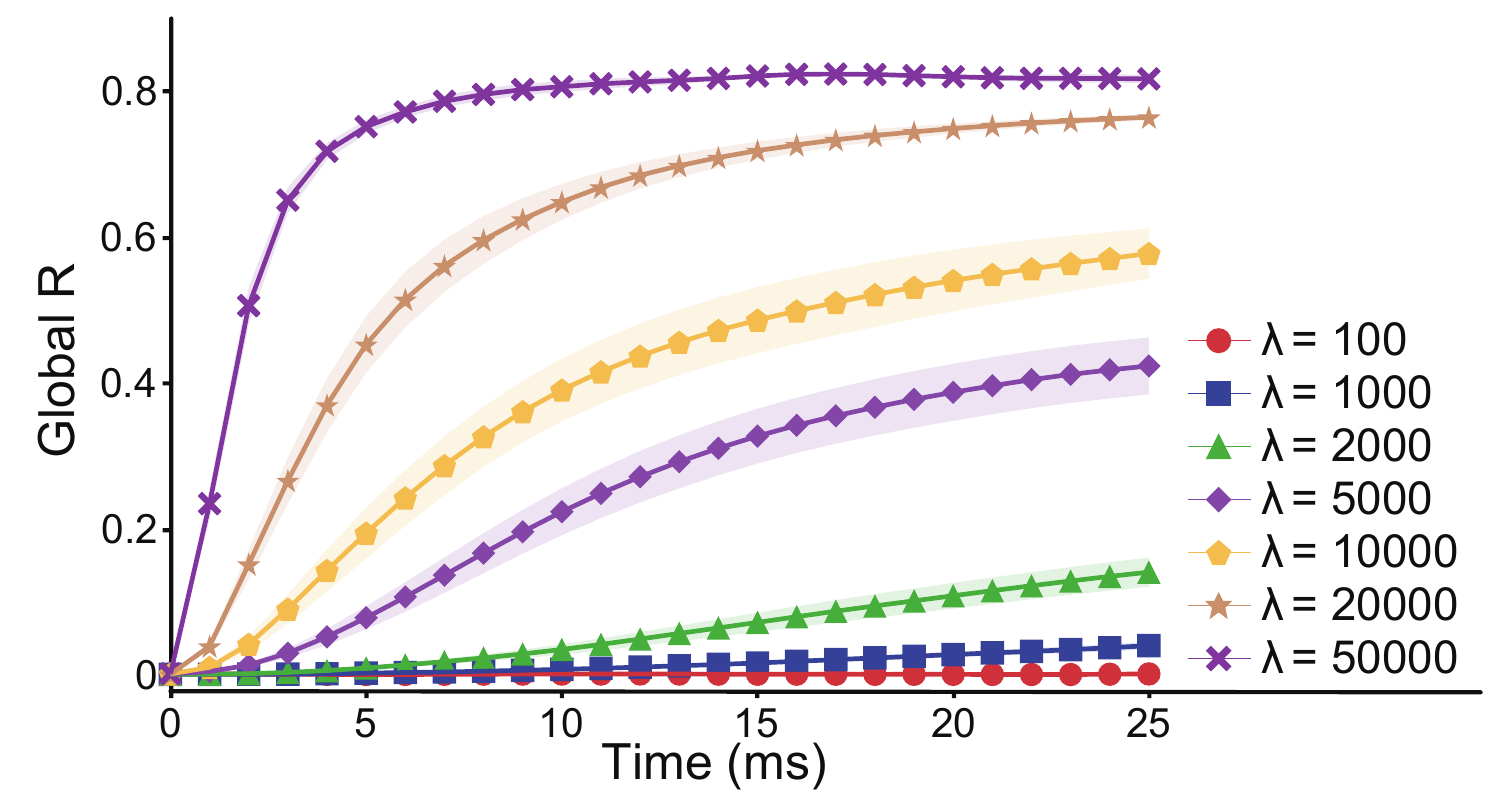}  \end{minipage}} 
	\subfloat[$\mu_L$=0,$\quad \sigma_L$=0.1,$\quad \mu_R$=0,$\quad \sigma_R$=0.5,left]{  
		\begin{minipage}{0.3\textwidth}  
			\centering  
			\includegraphics[width=\textwidth]{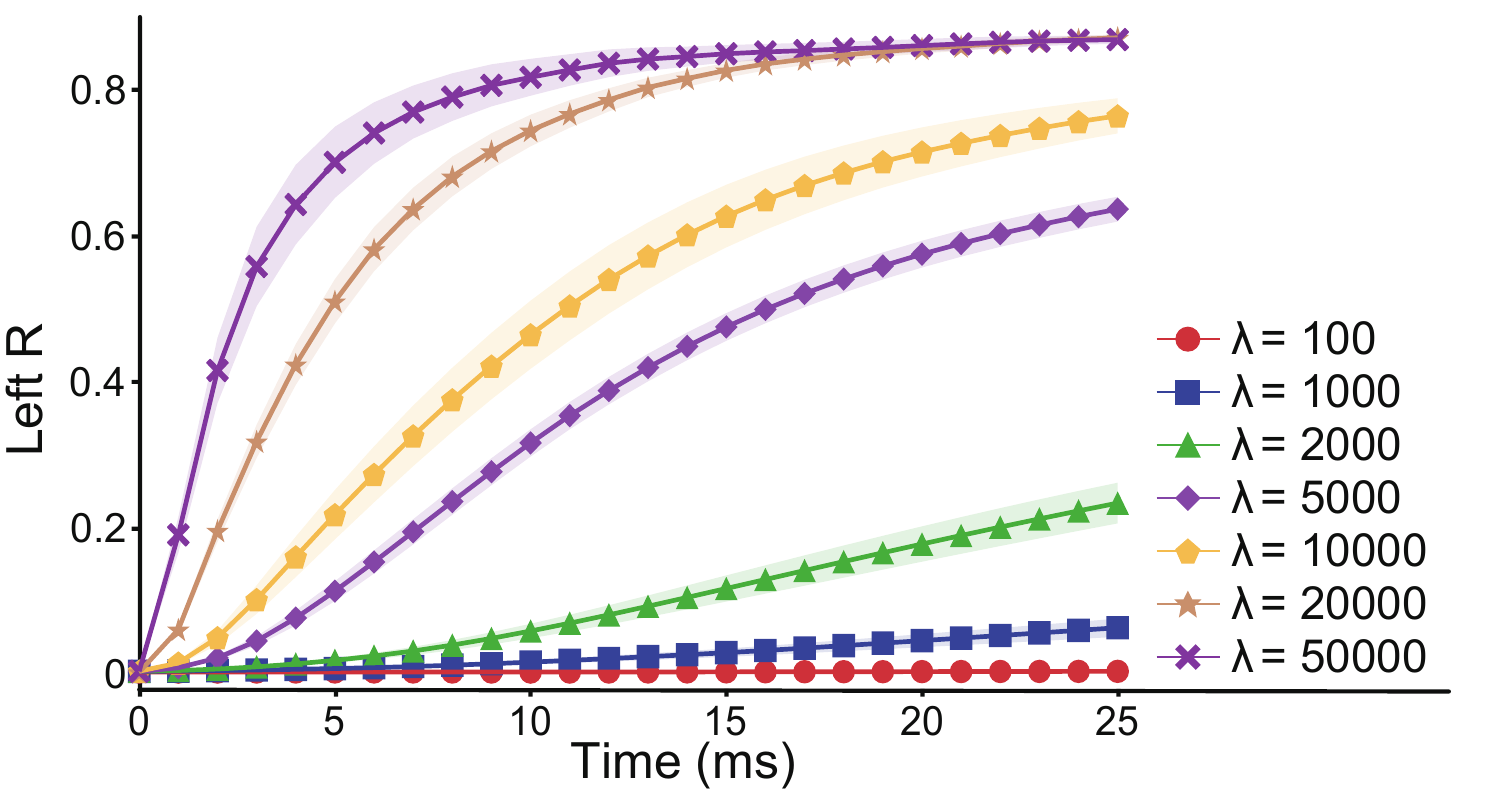}  \end{minipage}  	} 
		
	\subfloat[$\mu_L$=0,$\quad \sigma_L$=0.5,$\quad \mu_R$=0,$\quad \sigma_R$=0.1,left]{ 		
		\begin{minipage}{0.3\textwidth}  
			\centering  
			\includegraphics[width=\textwidth]{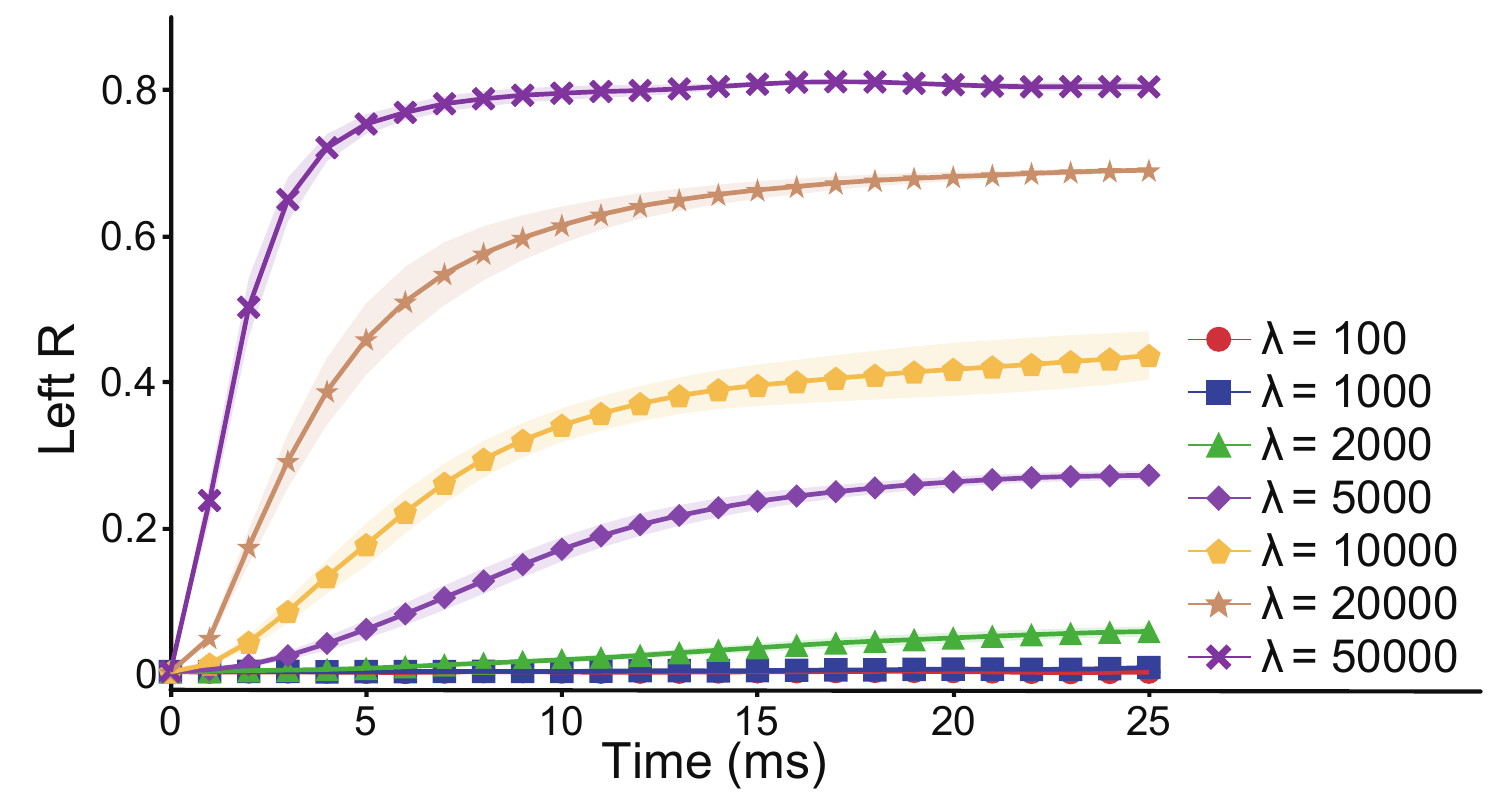}  \end{minipage}} 
	\subfloat[$\mu_L$=0,$\quad \sigma_L$=0.1,$\quad \mu_R$=0,$\quad \sigma_R$=0.5,right]{  
		\begin{minipage}{0.3\textwidth}  
			\centering  
			\includegraphics[width=\textwidth]{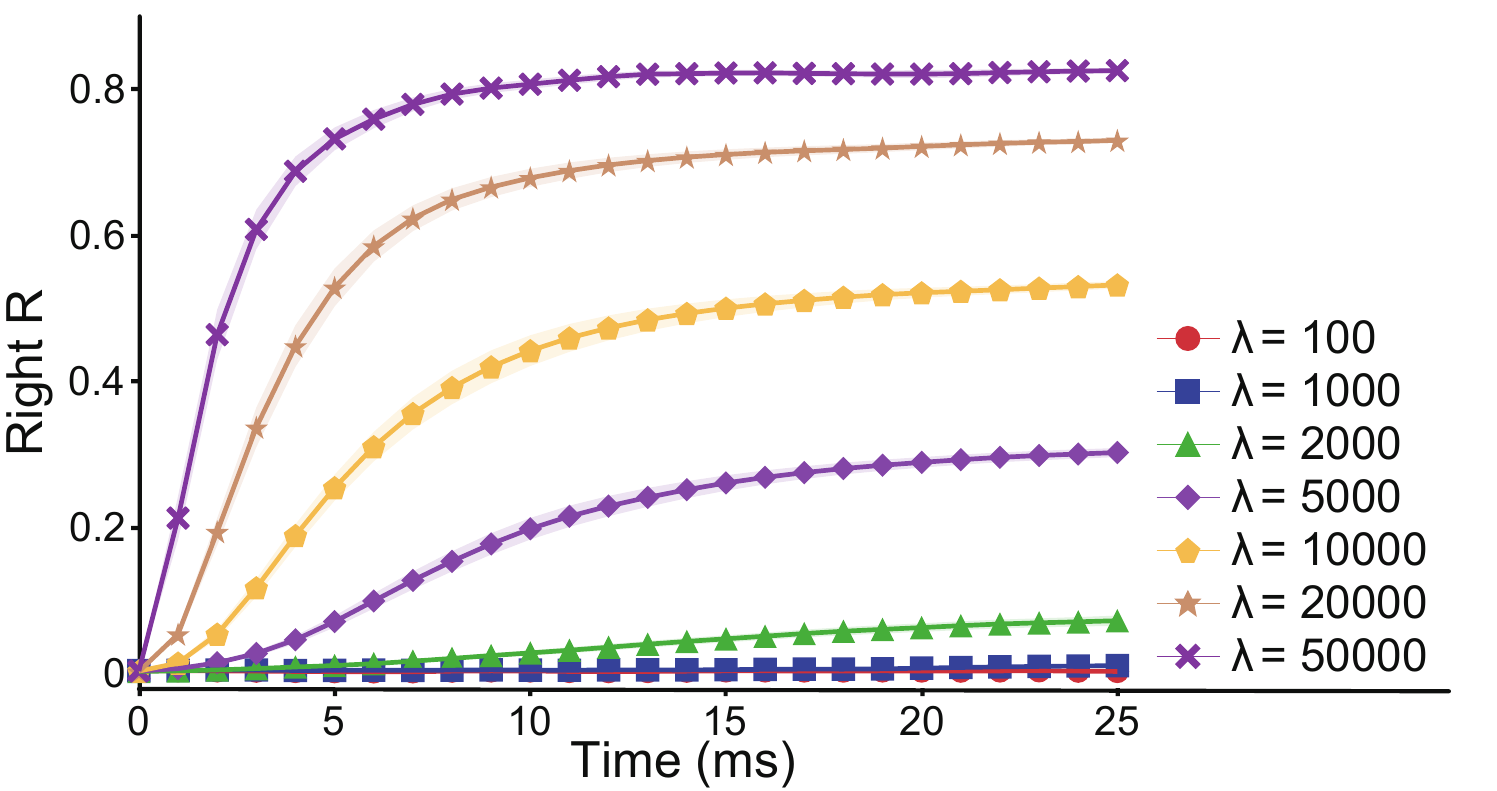}  \end{minipage}  	} 
	\subfloat[$\mu_L$=0,$\quad \sigma_L$=0.5,$\quad \mu_R$=0,$\quad \sigma_R$=0.1,right]{ 		
		\begin{minipage}{0.3\textwidth}  
			\centering  
			\includegraphics[width=\textwidth]{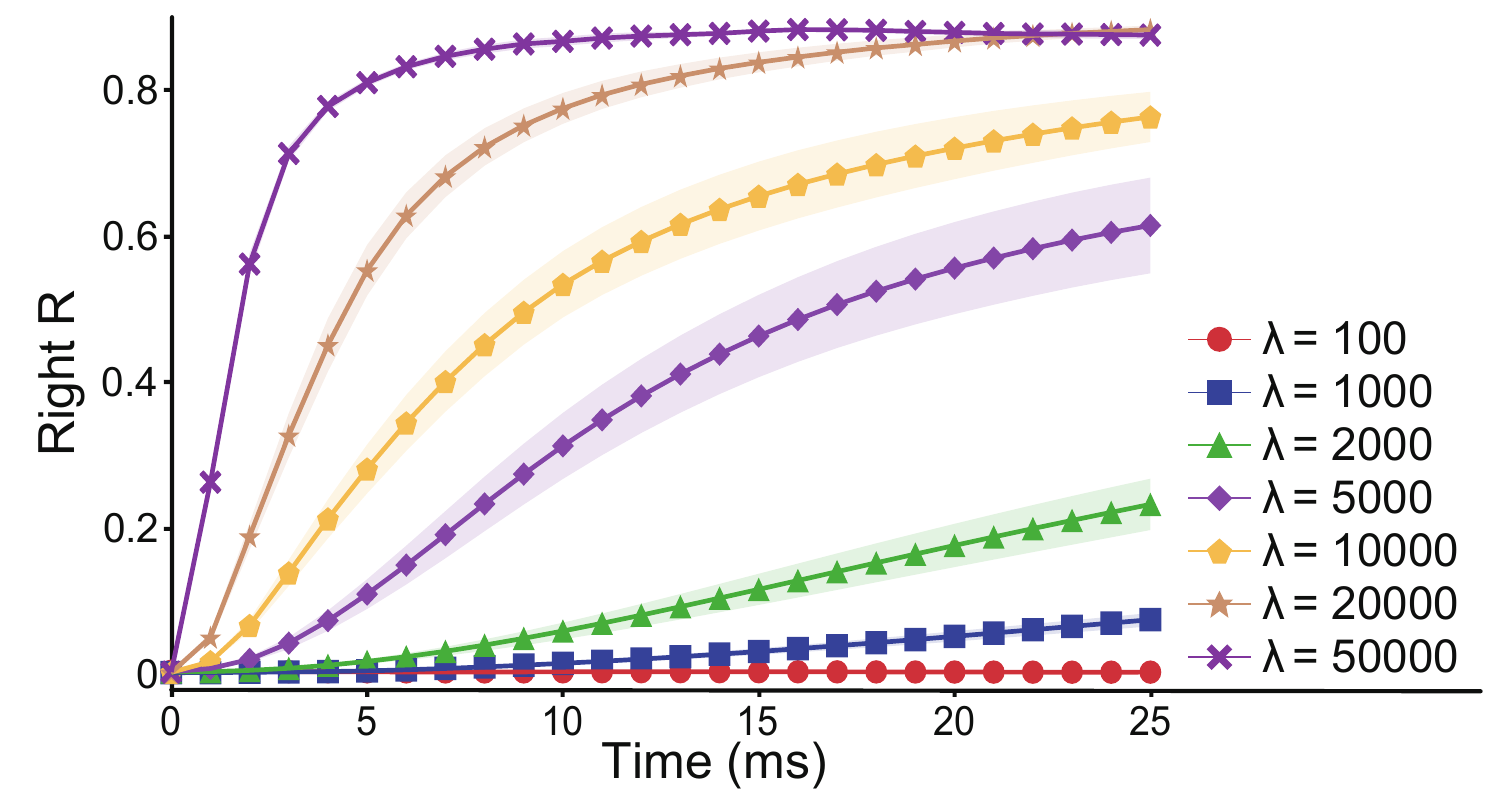}  \end{minipage}} 
 
	\caption{Temporal evolution of synchronization rates (R) for the whole brain, left hemisphere, and right hemisphere under different initialization parameters. The time step $\delta t = 0.05ms$ with a simulation period $T =  25ms$. Neural oscillator frequencies are characterized by $\mu_L$,$\quad \sigma_L$ (mean and S.D. for left hemisphere) and $\mu_R$,$\quad \sigma_R$ (mean and S.D. for right hemisphere). }
	\label{kuramoto1}

\end{figure*}

Notable differences in synchronization rates emerge (Fig. \ref{kuramoto1}). Specifically, at $\lambda=10000$ and $t$ = 25$ms$, the synchronization rates for the left and right hemispheres are $0.765$ (Fig. \ref{kuramoto1}(c)) and $0.532$ (Fig. \ref{kuramoto1}(e)), respectively. Similarly, at $\lambda=2000$ and $t$ = 25$ms$, the synchronization rates are $0.059$ and $0.233$, and at $\lambda=5000$, they are $0.273$ and $0.615$. These results indicate that for each hemisphere, a higher mean variance impedes its own synchronization capability, consequently leading to differences in overall activation levels (Fig. \ref{kuramoto1}(d) and (f)). 

Despite the relatively similar network structures between the left and right hemispheres (Fig. \ref{degree_region}), subtle structural disparities persist. Consequently, whether these architectural distinctions between hemispheric networks potentially contribute to differential synchronization rates between the left and right hemispheres remains an unresolved question warranting further investigation.


Accordingly, we investigated functional lateralization differences between the left and right hemispheres by maintaining identical intrinsic frequencies of neurons across both hemispheres while employing differential input signals, thereby examining hemispheric functional asymmetries in response to varied input conditions. The intrinsic frequencies of all neurons were initialized using Gaussian distributions with identical means (0.1) and standard deviations (0.1), and their initial phases were assumed to follow uniform distributions ranging from 0 to 2$\pi$. On this basis, we recognized all 270 ocellar retinula cells and adjusted their frequencies to simulate instantaneous frequency changes when neurons receive external stimulation (ocellar-type neurons, which is part of visual input neurons). These neurons respond to light stimuli and are activated relatively early in the visual processing of Drosophila. We conducted iterations over an extended time period and recorded changes in synchronization rates between left and right hemispheres. Considering that different neurons receive varying light intensities depending on their angles, we established a maximum activation frequency and adjusted the activation frequencies of stimulated neurons to any value between their intrinsic frequency and the maximum activation frequency. 

\begin{figure*}  
	\centering  
	\centering
	\subfloat[]{  
		\begin{minipage}{0.4\textwidth}  
			\centering  
			\includegraphics[width=\textwidth]{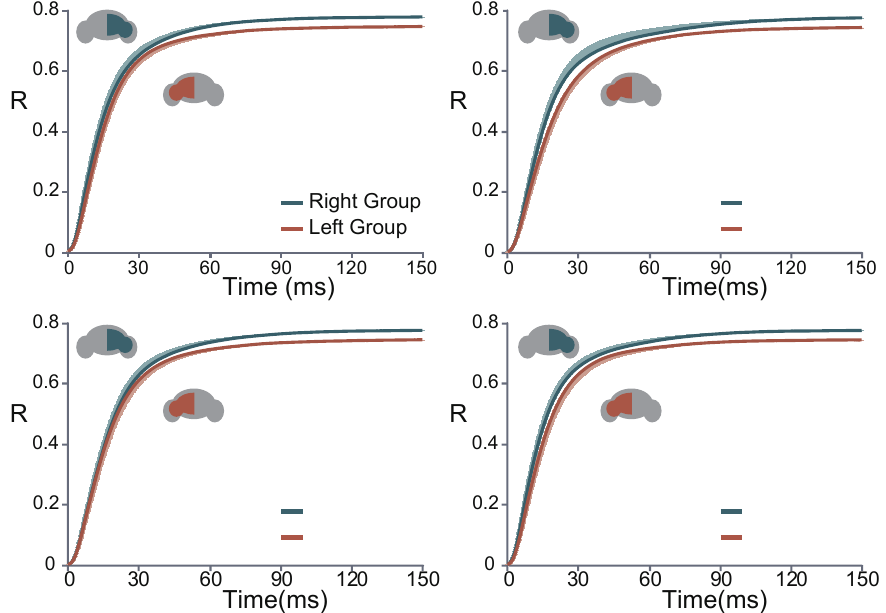}  \end{minipage}  	} 
	\subfloat[]{ 		
		\begin{minipage}{0.4\textwidth}  
			\centering  
			\includegraphics[width=\textwidth]{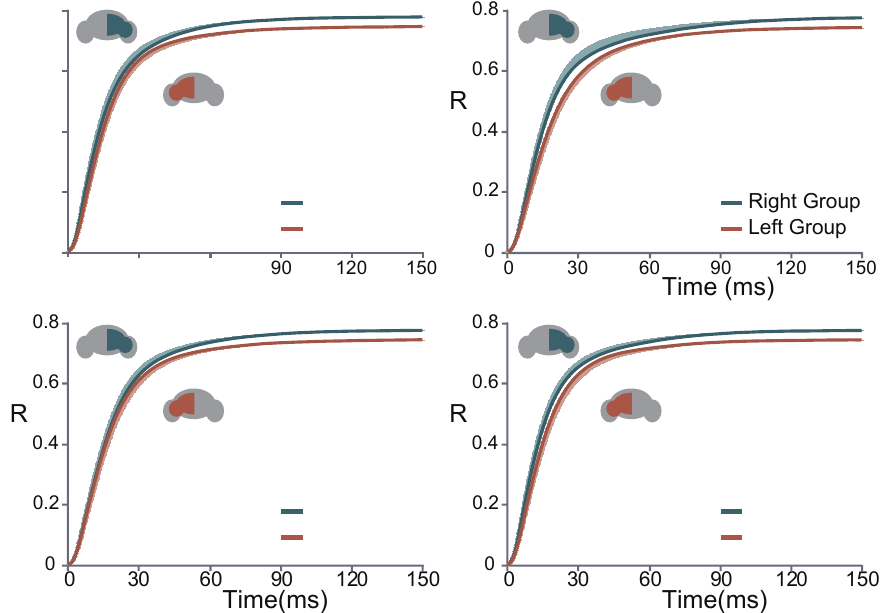}  \end{minipage}} 
	
	\subfloat[]{  
		\begin{minipage}{0.4\textwidth}  
			\centering  
			\includegraphics[width=\textwidth]{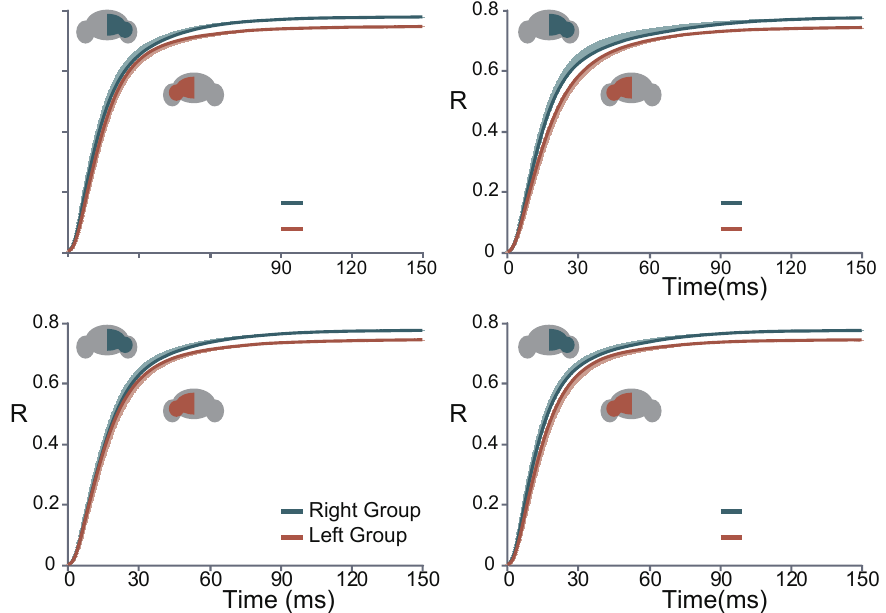}  \end{minipage}  	} 
	\subfloat[]{ 		
		\begin{minipage}{0.4\textwidth}  
			\centering  
			\includegraphics[width=\textwidth]{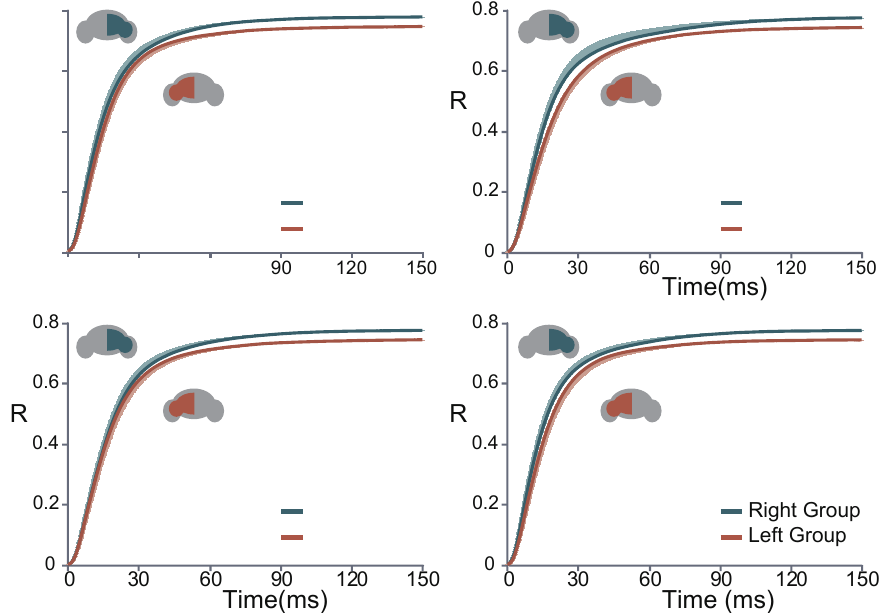}  \end{minipage}}  
	\caption{Temporal evolution of left-right brain synchronization rate (R) during external stimulation of optic superclass ocellar-type neurons. We set coupling strength $lambda$=5000, time step $\delta t$=0.25$ms$, simulation period $T$=150$ms$. $A$, $B$, $C$ and $D$ illustrate the temporal evolution of left-right brain synchronization rate (R) when stimulating ocellar-type neurons exclusively in the central brain, left brain, right brain and all ocellar-type neurons, respectively.}
	\label{out}
\end{figure*}

The average temporal evolution of synchronization rates between the left and right hemispheres across 10 experiments following external stimulation is shown in Fig. \ref{out}. The overall trend across all four experimental groups indicates that synchronization rates between the two hemispheres remain largely consistent during the initial phase. As time progresses, a divergence emerges: when ocellar-type neurons in the left hemisphere are selectively stimulated, the synchronization rate in the left hemisphere surpasses that of the right; in contrast, under all other stimulation conditions, the right hemisphere exhibits a higher synchronization rate than the left. In the resting state, no significant difference in synchronization rates between hemispheres is observed, regardless of the stimulation modality.

At 150 ms, the differences in activation rates between the two hemispheres are consistently small, with values of 0.032, 0.032, 0.031, and 0.032 in panels (a), (b), (c), and (d), respectively. These results suggest that, irrespective of input modality, the interhemispheric differences in synchronization dynamics remain minimal throughout the observed period.

\section{Data and Methods}
\subsection{Drosophila Connectome Data}
We utilize the recently released largest Drosophila brain connectome\cite{dorkenwald2023neuronal}, which represents the first complete connectome of an adult Drosophila brain. It encompasses approximately 120,000 reconstructed neurons and over 56 million mapped chemical synapses, covering both the central, afferent, and efferent neurons. Neuron type annotation in this data set mainly comes from expert knowledge. The annotation process is designed to be dynamic and continuously updated, allowing for the incorporation of new research findings. This comprehensive approach also features detailed synaptic connections and high-precision 3D coordinates (accurate to the nanometer level) for both cell bodies and synapses. For more detailed information,please refer to \cite{dorkenwald2023neuronal,schlegel2023consensus}.
\subsection{Network Analysis Metrics}  

\subsubsection{Graph Definition}  
In the Drosophila connectome data set, multiple directed connections may exist between neurons. To accurately represent this complex connectivity pattern, we employ a directed multigraph model to describe this data set. 

A directed multigraph $G$ is defined as an ordered pair $G = (V, E)$. Here, let $V$ represents a set of vertices, also known as nodes, which are the fundamental units of the network. Let $E$, on the other hand, is a set of ordered pairs of nodes, called directed edges or arcs. These edges represent the connections or relationships between the nodes, with the direction indicating the nature of the relationship. In a multigraph, multiple edges are allowed between the same pair of nodes, which can represent different types of relationships or interactions. In this paper, we represent neurons as nodes and synaptic connections between neurons as edges.

\subsubsection{Degree}  
The degree of a node is a fundamental measure of its connectivity in a graph. In a directed graph, we distinguish between two types of degree for a node $v$. The out-degree, denoted as $d^+(v)$, is the number of edges that originate from vertex $v$ and is formally defined as $d^+(v) = |\{(v,u) \in E, u \in V\}|$. This measures the number of connections that $v$ initiates with other vertices. Conversely, the in-degree, denoted as $d^-(v)$, is the number of edges that terminate at vertex $v$ and is defined as $d^-(v) = |\{(u,v) \in E, u \in V\}|$. This measures the number of connections that other vertices initiate with $v$. The total degree of a vertex is the sum of its in-degree and out-degree, that is:
\begin{equation}
	d(v) = d^+(v) + d^-(v).
\end{equation}
These measures provide insights into the importance or centrality of a node in the network.  

\subsubsection{Clustering Coefficient}  
The clustering coefficient is a measure of the degree to which nodes in a graph tend to cluster together. It provides insights into the local structure of networks and is often used to identify tightly-knit groups within the network.  

\paragraph{Undirected Networks}  
For undirected networks, the local clustering coefficient for a node $i$ with $k_i$ neighbors is defined as:  
\begin{equation}  
	C_i = \frac{2|\{e_{jk}: v_j, v_k \in N_i, e_{jk} \in E\}|}{k_i(k_i-1)}  
\end{equation}  
where $N_i$ is the neighborhood of vertex $i$. This coefficient represents the fraction of pairs of $i$'s neighbors that are connected to each other, indicating how close the vertex and its neighbors are to forming a clique.  

\paragraph{Directed Networks}  
In directed networks, the local clustering coefficient for a node $i$ is calculated as:  
\begin{equation}  
	C_i = \frac{|\{e_{jk}: v_j, v_k \in N_i, e_{jk} \in E\}|}{k_i(k_i-1)}  
\end{equation}  
This definition takes into account the directionality of the edges, providing a more nuanced measure of clustering in directed graphs.  

\paragraph{Weighted Networks}  
For weighted networks, the clustering coefficient is adapted to incorporate edge weights. The weighted clustering coefficient for a vertex $i$ is defined as:  
\begin{equation}  
	C_i^w = \frac{1}{s_i(k_i-1)} \sum_{j,k} \frac{(w_{ij} + w_{ik})}{2} a_{ij} a_{ik} a_{jk}  
\end{equation}  
where $s_i$ is the strength of vertex $i$, $k_i$ is its degree, $w_{ij}$ is the weight of the edge between $i$ and $j$, and $a_{ij}$ is 1 if an edge exists between $i$ and $j$, and 0 otherwise. This weighted version provides a more comprehensive measure of clustering in networks where edge weights are significant.

\subsubsection{Weight}  
In network analysis, the weight $w_{ij}$ of an edge between nodes $i$ and $j$ represents the intensity or capacity of the connection. The incorporation of weights allows for a more detailed and realistic representation of complex systems, where not all connections are equal. In this study, we operate under the assumption that all synaptic connections are equally weighted, that is to say:
\begin{equation}
	W_{ijl} = 1,
\end{equation} 
where $l$ represents the $l$-th edge of node $i$ and $j$. 
So the total weight between nodes $i$ and $j$ are:
\begin{equation}
	W_{ij} = Num(i,j).
\end{equation} 
Where $Num(i,j)$ represents the number of synapses between neuron $i$ and $j$.

\subsubsection{Strength}  
In weighted networks, the concept of degree is extended to strength. The strength $S_i$ of a node $i$ is defined as the sum of the weights of all edges connected to it:  
\begin{equation}  
	S_i = \sum_{j \in N_i} W_{ij}  ,
\end{equation}  
where $N_i$ is the set of neighbors of node $i$ and $W_{ij}$ is the weight of the edge between $i$ and $j$. This measure provides a more nuanced view of a node's importance in weighted networks, taking into account not just the number of connections, but also their intensities.  

%
%

\subsection{Spatial Clustering Method}  
Spatial clustering refers to the tendency of similar elements to be located near each other in physical or abstract space. In analysis, it can provide insights into the underlying structure and organization of point patterns.  
\paragraph{Distance of Points}
In this study, the "distance" between two points refers to their Euclidean distance, unless otherwise specified. The equation of Euclidean distance is expressed as follows:
\begin{equation}  
	d(p,q) = \sqrt{\sum_{i=1}^n (q_i - p_i)^2}  ,
\end{equation} 
where $d(p,q)$ is the Euclidean distance between points $p$ and $q$. $n$ is the number of dimensions. $p_i$ and $q_i$ are the coordinates of $p$ and $q$ in the
$i$-th dimension, respectively.

\paragraph{Nearest Neighbor Distance}  
The nearest neighbor distance is a fundamental concept in spatial analysis. For a point $i$, the nearest neighbor distance $d_i$ is defined as the distance to its closest neighbor:  
\begin{equation}  
	d_i = \min_{j \neq i} \{d_{ij}\}  ,
\end{equation}  
where $d_{ij}$ is the distance between points $i$ and $j$. This measure is useful for understanding the local spatial structure around individual points.  

\paragraph{R-index}  
The R-index is a measure used to quantify the degree of spatial clustering in a point pattern. It is defined as:  
\begin{equation}  
	R = \frac{\bar{d}_O}{\bar{d}_E}  ,
\end{equation}  
where $\bar{d}_O$ is the observed mean nearest neighbor distance and $\bar{d}_E$ is the expected mean nearest neighbor distance in a random distribution. An R-index less than 1 indicates clustering, while an R-index greater than 1 suggests dispersion. This index provides a simple yet effective way to characterize the overall spatial distribution of points in a network.  

\subsection{Similarity Metric: Earth Mover's Distance (EMD)}  

The Earth Mover's Distance (EMD), also known as the Wasserstein metric, is a measure of the distance between two probability distributions over a region. Conceptually, it can be understood as the minimum cost of transforming one distribution into the other, where the cost is the amount of probability mass that needs to be moved, multiplied by the distance it needs to be moved.  

Formally, for two distributions $P$ and $Q$ defined over a space $X$ with a ground distance $d$, the EMD is defined as:  

\begin{equation}  
	EMD(P, Q) = \inf_{\gamma \in \Gamma(P, Q)} \int_{X \times X} d(x, y) d\gamma(x, y)  ,
\end{equation}  
where $\Gamma(P, Q)$ is the set of all joint distributions $\gamma(x, y)$ whose marginals are $P$ and $Q$.  

In practical applications, when dealing with discrete distributions or empirical samples, the EMD is often computed using linear programming techniques. For two discrete distributions $P = \{(p_i, w_{p_i})\}_{i=1}^m$ and $Q = \{(q_j, w_{q_j})\}_{j=1}^n$, where $p_i$ and $q_j$ are the locations and $w_{p_i}$ and $w_{q_j}$ are the weights, the EMD can be formulated as:  

\begin{equation}  
	EMD(P, Q) = \min_{f_{ij}} \frac{\sum_{i=1}^m \sum_{j=1}^n f_{ij} d_{ij}}{\sum_{i=1}^m \sum_{j=1}^n f_{ij}}, 
\end{equation}  
which subject to the following constraints:  
\begin{equation}
	\begin{aligned}  
		f_{ij} &\geq 0, & 1 \leq i \leq m, 1 \leq j \leq n \\
		\sum_{j=1}^n f_{ij} &\leq w_{p_i}, & 1 \leq i \leq m \\
		\sum_{i=1}^m f_{ij} &\leq w_{q_j}, & 1 \leq j \leq n \\
		\sum_{i=1}^m \sum_{j=1}^n f_{ij} &= \min(\sum_{i=1}^m w_{p_i}, \sum_{j=1}^n w_{q_j}),  
	\end{aligned}  
\end{equation}
where $f_{ij}$ represents the flow from $p_i$ to $q_j$, and $d_{ij}$ is the ground distance between $p_i$ and $q_j$.  


The interpretation of EMD values depends on the context and the ground distance used. Generally, an EMD value of 0 indicates that the two distributions are identical. Low EMD values suggest high similarity between the distributions, while high EMD values indicate significant differences. As EMD approaches infinity, it represents maximally dissimilar distributions.  


\subsection{Clustering Algorithms}  

The K-means spatial clustering algorithm\cite{macqueen1967some} aims to partition $n$ observations into $k$ clusters in space, where each observation belongs to the cluster with the nearest mean (centroid). The algorithmic procedure is delineated as follows:

\begin{enumerate}  
	\item Initialize: Select $k$ points as initial centroids $\{\mu_1, \mu_2, ..., \mu_k\}$ in the $d$-dimensional space.  
	
	\item Assign each point $x_i$ to the nearest centroid, forming clusters $S_j$:  
	\begin{equation}
		S_j^{(t)} = \{x_i : \|x_i - \mu_j^{(t)}\|^2 \leq \|x_i - \mu_l^{(t)}\|^2 \quad \forall l, 1 \leq l \leq k\}.
	\end{equation}

	\item Update centroids to the mean of assigned points:  
	\begin{equation}
		\mu_j^{(t+1)} = \frac{1}{|S_j^{(t)}|} \sum_{x_i \in S_j^{(t)}} x_i
	\end{equation}

	\item Repeat steps 2-3 until convergence or a maximum number of iterations is reached.  
	
	\item Minimize the within-cluster sum of squares (WCSS):  
	\begin{equation}
		\text{WCSS} = \sum_{j=1}^{k} \sum_{x_i \in S_j} \|x_i - \mu_j\|^2  
	\end{equation}

	\item Evaluate clustering quality using the silhouette coefficient:  
	\begin{equation}
		s(i) = \frac{b(i) - a(i)}{\max\{a(i), b(i)\}}  
	\end{equation}
	
	where $a(i)$ is the mean distance between $i$ and points in the same cluster,  
	and $b(i)$ is the mean distance between $i$ and points in the nearest different cluster.  
\end{enumerate}  
\subsection{Kuramoto Model} \label{kuramoto}
The Kuramoto model stands as a paradigmatic framework for studying synchronization phenomena in systems of coupled oscillators, originally proposed by Yoshiki Kuramoto\cite{kuramoto1975self}. This mathematical model has emerged as a fundamental tool for understanding collective behavior in diverse complex systems, ranging from neural networks and chemical oscillators to power grids and biological rhythms. The model describes a population of $N$ coupled phase oscillators, where each oscillator is characterized by its natural frequency drawn from a given probability distribution and is coupled to all other oscillators through sine functions of their phase differences. The governing equation for the $i$-th oscillator is given by
\begin{equation}
	\frac{d\theta_i}{dt} = \omega_i + \frac{\lambda}{N}\sum_{j=1}^N \sin(\theta_j - \theta_i),
\end{equation}
where $\theta_i$ represents the phase of the $i$-th oscillator, $\omega_i$ is its natural frequency, $\lambda$ is the coupling strength, and $N$ is the total number of oscillators. 

The order parameter $R_{\lambda}$ in the Kuramoto model quantifies the degree of synchronization among oscillators through a complex-valued mean field. The instantaneous order parameter is calculated as  

\begin{equation}  
	z(t) = R_{\lambda}(t)e^{i\psi(t)} = \frac{1}{N}\sum_{j=1}^N e^{i\theta_j(t)},  
\end{equation}  	
where $z(t)$ is the complex order parameter, $R_{\lambda}(t)$ is its magnitude, and $\psi(t)$ is the average phase. The practical calculation decomposes this into real and imaginary parts as:
\begin{equation}  
	x(t) = \frac{1}{N}\sum_{j=1}^N \cos(\theta_j(t))  
\end{equation}  
and  
\begin{equation}  
	y(t) = \frac{1}{N}\sum_{j=1}^N \sin(\theta_j(t)),  
\end{equation}  
from which we obtain  
\begin{equation}  
	R_{\lambda}(t) = \sqrt{x(t)^2 + y(t)^2}  
\end{equation}  
and  
\begin{equation}  
	\psi(t) = \arctan\left(\frac{y(t)}{x(t)}\right).  
\end{equation}  
We use the time-averaged order parameter: 
\begin{equation}  
	\langle R_{\lambda} \rangle = \frac{1}{T}\int_{t_0}^{t_0+T} R_{\lambda}(t)dt \approx \frac{1}{M}\sum_{k=1}^M R_{\lambda}(t_k),  
\end{equation}  
where $T$ is the averaging time interval and $M$ is the number of discrete time points. The magnitude of $R_{\lambda}$ ranges from 0 (complete desynchronization) to 1 (perfect synchronization), with intermediate values indicating partial synchronization states.

\section{Discussion}

In this study, we conducted a comprehensive analysis of the most extensive Drosophila brain network available to date, with a particular focus on the heterogeneity of its connectome. This heterogeneity is reflected in both the large number of neurons and their intricate connectivity patterns, especially in the highly uneven distribution of synaptic weights between neuronal pairs. Region-specific analyses further reveal substantial variability in network properties across distinct functional areas, indicating that the pronounced structural differences among these regions contribute significantly to the overall complexity of the Drosophila brain.

We also examined spatial heterogeneity within the brain network by performing global and local analyses of neuronal clustering. The results show that neuronal somata are significantly clustered throughout the brain. Moreover, most brain regions exhibit lower spatial agglomeration coefficients compared to the whole-brain average, suggesting that intra-regional clustering is more pronounced than inter-regional clustering.

Previous studies have classified neurons based on morphological, functional, and neurotransmitter-related characteristics, often leading to a single neuron being assigned to multiple functional domains\cite{zeng2017neuronal}. To further explore this issue, we applied spatial clustering techniques for neuron classification. However, our findings suggest that spatial proximity alone is insufficient for delineating functional neuron categories. Functional regions often comprise spatially intermixed neurons—including both intrinsic neurons and those originating from anatomically distinct areas—which complicates attempts to infer function from spatial localization alone. This observation underscores a fundamental limitation of traditional spatial parcellation approaches\cite{eickhoff2018imaging}, as neurons within the same anatomically defined region may not share common functional properties.

A notable finding of our study is the strong bilateral structural symmetry between the left and right hemispheres, observed in both network topology and spatial distribution. Homologous regions across hemispheres display similar neuron and synapse counts, as well as comparable topological features. Spatial clustering characteristics are also closely matched between corresponding regions. Despite this structural symmetry, numerous studies have demonstrated lateralized functional differences in the Drosophila brain, with important implications for behavior and cognition\cite{pascual2004brain,abubaker2024asymmetric,linneweber2020neurodevelopmental,lapraz2023asymmetric}. Such functional lateralization appears to be essential for the modulation of adaptive behaviors and other complex neurobiological processes.

While we confirmed this structural symmetry, our findings further reveal that differences in intrinsic neuronal dynamics may underlie the emergence of functional asymmetry. Specifically, initializing the left and right hemispheres with distinct intrinsic frequency distributions leads to divergent steady-state eigenoscillation patterns. In contrast, when intrinsic frequencies are identical, differential external inputs alone do not produce significant asymmetries in activation patterns. These results support the hypothesis that functional asymmetry is driven more by internal neuronal dynamics than by structural differences.

Taken together, our findings provide new insights into the heterogeneous structural and functional organization of the Drosophila brain. By revealing the interplay between network topology, spatial distribution, and intrinsic dynamics, this work contributes to a deeper understanding of how brain structure supports function. Future research should investigate the developmental and evolutionary mechanisms shaping these complex patterns, particularly the balance between bilateral symmetry and functional specialization.

Looking ahead, further exploration of Drosophila brain heterogeneity from a functional perspective could illuminate its causal relationship with specific cognitive and behavioral functions. Understanding how structural diversity shapes neural computation may help uncover fundamental principles of intelligence. Moreover, the principles of heterogeneity identified in this study may inform the design of biologically inspired artificial neural networks. Investigating how structural diversity influences learning, adaptation, and efficiency in artificial systems—such as large language models—could aid the development of more robust and biologically plausible intelligent architectures.

\section*{Acknowledgement}
This work was supported by the Major Program of Xiangjiang Laboratory (No. 24XJJCYJ01001), the National Natural Science Foundation of China (72025405, 72421002, 92467302, 72088101, 72301285, 72474223), the National Social Science Foundation of China (22ZDA102), the Natural Science Foundation of Hunan Province (2023JJ40685, 2024RC3133). We thank M. Murthy for data set support. 
\bibliography{ref} 
\bibliographystyle{unsrt}
\clearpage
\appendix 

\setcounter{figure}{0}
\renewcommand{\thefigure}{S\arabic{figure}}

\end{document}